\documentclass[11pt,a4paper]{article}
\usepackage{epsfig}
\usepackage{a4wide}
\usepackage{latexsym}
\usepackage{graphicx}
\usepackage{amsmath}
\usepackage{slashed}
\usepackage{amsfonts}
\usepackage{amssymb}
\usepackage{mathrsfs}
\usepackage{appendix}
\usepackage[usenames,dvips]{color}
\usepackage[colorlinks=true,urlcolor=PineGreen,anchorcolor=PineGreen,citecolor=PineGreen,filecolor=PineGreen,linkcolor=PineGreen,menucolor=PineGreen,pagecolor=PineGreen,linktocpage=true,pdfproducer=medialab]{hyperref}
\usepackage{layout}

\allowdisplaybreaks[4] 
\numberwithin{equation}{section}

\begin{document}

\begin{titlepage}

\begin{center}
{\Large \textbf{Two-loop QED radiative corrections to the decay $\pi^0\rightarrow e^+ e^-$:\\  The virtual corrections and soft-photon bremsstrahlung}}\\[1.5 cm%
]
\textbf{Petr Va\v{s}ko and Ji\v{r%
}\'{\i} Novotn\'y}\footnote{%
for emails use: \textit{surname\/} at ipnp.troja.mff.cuni.cz},  \\[1 cm]
\textit{Institute of Particle and Nuclear Physics, Faculty of
Mathematics
and Physics,}\\[0pt]
\textit{Charles University, V Hole\v{s}ovi\v{c}k\'ach 2, CZ-180 00
Prague 8,
Czech Republic} \\[0.5 cm]
\end{center}

\vspace*{1.0cm}

\vspace*{1.0cm}

\begin{abstract}
This  paper is devoted to the  two-loop QED radiative corrections
to the decay $\pi^0\rightarrow e^+ e^-$. We compute the virtual
corrections without using any approximation and we take into
account all the relevant graphs with the inclusion of those
omitted in the previous approximative calculations. The
bremsstrahlung is then treated within the soft photon
approximation. We concentrate on the technical aspects of the
calculation and discuss in detail the UV renormalization and the
treatment of IR divergences within the dimensional regularization.
As a result we obtain the $O(\alpha^3 p^2)$ contribution in closed
analytic form. We compare the exact two-loop results with existing
approximative calculations of QED corrections and find significant
disagreement in the kinematical region relevant for the KTeV
experiment.

\end{abstract}

\end{titlepage}

\setcounter{footnote}{0} \newpage\tableofcontents
%%%%%%%%%%%%%%%%%%%%%%%%%%%%%%%%%%%%%%%%%%%%%%%%%%%%%%%%%%%%%%%%%%%%

\section{Introduction}

The rare decay of the neutral pion into the electron-positron pair provides
an interesting tool to test the nonperturbative low-energy dynamics of the
Standard Model (SM). While the possible contributions of the weak sector of
the SM are tiny and can be safely neglected, the leading order QED
contribution is described by two virtual photon exchange diagram and is
therefore tightly connected to the doubly off-shell pion transition form
factor $F_{\pi^0\gamma^*\gamma^*}$ for the subprocess $\pi^0\rightarrow
\gamma^*\gamma^*$. Better understanding of this form factor which is not
known from the first principles is important \emph{e.g.} for the
determination of the light-by-light hadronic contribution to the muon
anomalous magnetic moment $g-2$. On the other hand the rareness of the decay
which is suppressed with respect to the $\pi^0\rightarrow \gamma\gamma$
decay by a factor of $2(\alpha m/M_{\pi^0})^2$ within the SM (here $m$ is
the electron mass which enters here as a consequence of the approximate
helicity conservation) makes it also a promising process possibly sensitive
to the physics beyond the SM.

The systematical theoretical treatment of the process dates back to 1959
when the first prediction of the decay rate \cite{Drell:1959} was published
by Drell. From that time, numerous attempts to model the form factor $F_{\pi
^{0}\gamma ^{\ast }\gamma ^{\ast }}$ and to get the predictions of the
leading order decay rate within various approaches have been made \cite%
{Berman:1960zz,Bergstrom:1983ay,Savage:1992ac,Ametller:1993we,Knecht:1999gb,Knecht:2001xc}%
. Recently this decay has attracted a renewed theoretical interest in
connection with a new precise branching ratio measurement by KTeV-E799-II
experiment at Fermilab \cite{Abouzaid:2006kk} with the result
\begin{equation}
B(\pi ^{0}\rightarrow e^{+}e^{-}(\gamma ),x_{D}>0.95)=(6.44\pm 0.25\pm
0.22)\times 10^{-8}.  \label{KTeV inclusive}
\end{equation}%
Here the Dalitz variable
\begin{equation}
x_{D}=\frac{m_{e^{+}e^{-}}^{2}}{M_{\pi ^{0}}^{2}}=1-2\frac{E_{\gamma }}{%
M_{\pi ^{0}}}  \label{x_D}
\end{equation}%
(where $E_{\gamma }$ is the real photon energy) has been bounded from below
in order to pick up the region where the final state radiation is soft and
where the contribution of the Dalitz decay $\pi ^{0}\rightarrow
e^{+}e^{-}\gamma $ which dominates at low $x_{D}$ is suppressed. Subsequent
comparison with theoretical predictions of the SM based on the dispersive
approach and various models for the pion transition form factor (including
the CELLO \cite{Behrend:1990sr} and CLEO \cite{Gronberg:1997fj} data) has
been done in \cite{Dorokhov:2007bd}. The necessary ingredient of such an
analysis is a good understanding of the QED radiative corrections to the
process. The KTeV analysis used the early calculation of Bergstr\"{o}m \cite%
{Bergstrom:1982wk} to extrapolate the full radiative tail beyond $x_{D}>0.95$
and to scale the result by the overall radiative corrections to get the
lowest order rate with the final state radiation removed with the result
\begin{equation}
B_{KTeV}^{\mathrm{no-rad}}(\pi ^{0}\rightarrow e^{+}e^{-})=(7.48\pm 0.29\pm
0.25)\times 10^{-8}.  \label{KTeV no rad}
\end{equation}%
This should be compared with the SM theoretical prediction of \cite%
{Dorokhov:2007bd} which has been found to be almost insensitive to the model
dependent part within the relevant class of models for the form factor $%
F_{\pi ^{0}\gamma ^{\ast }\gamma ^{\ast }}$. Using the CLEO+OPE they
obtained
\begin{equation}
B_{SM}^{\mathrm{no-rad}}(\pi ^{0}\rightarrow e^{+}e^{-})=(6.23\pm
0.09)\times 10^{-8}.  \label{Dorokhov prediction}
\end{equation}%
The result of the analysis can be interpreted as a 3.3$\sigma $ discrepancy
between the theory and the experiment. This discrepancy initiated further
theoretical investigation of its possible sources. Aside from the attempts
to find the corresponding mechanism within the physics beyond the SM \cite%
{Kahn:2007ru,Dorokhov:2008uk,Chang:2008np,McKeen:2008gd} also the possible
revision of the SM predictions has been taken into account. The theoretical
estimate of the mass corrections to the decay width using the Mellin-Barnes
representation has been made in \cite{Dorokhov:2008cd,Dorokhov:2009xs} and
this effect has been found to be negligible (the central value of the SM
prediction is shifted by 0.5\%). Also the incorporation of the new BABAR
data \cite{Aubert:2009mc} on the semi-off-shell form factor $F_{\pi
^{0}\gamma \gamma ^{\ast }}$ in the time-like region into the analysis \cite%
{Dorokhov:2009jd} has not influenced the SM prediction (\ref{Dorokhov
prediction}).

The QED radiative corrections as a possible source of the discrepancy have
been revisited calculating the contributions of the vertex-, box-type and self energy
two-loop graphs in the double logarithm approximation \cite{Dorokhov:2008qn}%
. The result has occasionally confirmed quantitatively the old Bergstr\"{o}m
calculation \cite{Bergstrom:1982wk} which used a different type of
approximation based on shrinking of the one-loop leading order graph into a
local $\pi^0 e^{+}e^{-}$ vertex.

The aim of our paper is to present a more detailed analysis of the two-loop
QED radiative corrections without using any approximation in order to check
the validity of the previous approximative results. The natural formalism to
treat the problem systematically is the Chiral perturbation theory ($\chi $%
PT) \cite{Weinberg:1978kz,Gasser:1983yg,Gasser:1984gg} enriched by photons and leptons \cite%
{Urech:1994hd,Knecht:1999ag}. The leading order amplitude which is $O(\alpha
^{2}p^{2})$ within the chiral power counting has been calculated in \cite%
{Savage:1992ac} and the matching of the relevant low energy
constant to the QCD in the leading order of the large $N_{C}$
expansion has been done in \cite{Knecht:1999gb}. The
next-to-leading order contributions have not been calculated
within this formalism yet. They can be divided into two groups.
The first group corresponds to the additional strong higher order
corrections to the $\pi ^{0}\gamma ^{\ast }\gamma ^{\ast }$ vertex
with pions inside the loops and it counts as $O(\alpha ^{2}p^{4})$
while the second one collects the pure QED corrections of the
order $O(\alpha ^{3}p^{2})$. It is the latter group we will
concentrate on in this paper. The relevant contributions consist
of the six two-loop Feynman diagrams, namely the one box-type, two
vertex-type, the electron self-energy insertion (these have been
approximately investigated in \cite{Dorokhov:2008qn}) and two
vacuum polarization insertions. In order to renormalize the
one-loop UV sub-divergences the corresponding one-loop counterterm
diagrams have to be taken into account. Finally the remaining
superficial UV divergence has to be renormalized by tree
counterterm graph. The box-type diagram suffers further from the
IR divergence, this is cancelled within the inclusive $\pi
^{0}\rightarrow e^{+}e^{-}(\gamma )$ width.

In this paper we address the technical aspects of the calculation of the six
two-loop Feynman diagram contributions mentioned above. The standard
strategy consists of their reduction to the dimensionally regularized scalar
integrals which will be subsequently expressed in terms of the eighteen
Master Integrals. This can be done using the Laporta-Remiddi algorithm \cite%
{Laporta:1996mq,Laporta:2001dd} which is based on the integration by parts
identities \cite{Tkachov:1981wb,Chetyrkin:1981qh} and Lorentz invariance
identities \cite{Gehrmann:1999as}. We then calculate the Master Integrals
using the technique of differential equations \cite{Kotikov:1990kg,Kotikov:1991hm,Kotikov:1991pm,Remiddi:1997ny,Caffo:1998yd}
and expand them up to and including the order $O(\varepsilon )$ (where $%
\varepsilon =2-d/2$) in terms of the harmonic polylogarithms \cite%
{Remiddi:1999ew}. Some of the Master Integrals has been already published in
the existing literature, we either take them over \cite{Fleischer:1999hp,Argeri:2002wz} or
make independent calculations in alternative bases within individual
topology classes and afterwards check the results \cite{Davydychev:2000na,Davydychev:2003mv,Bonciani:2003hc,Bonciani:2003te,Anastasiou:2006hc,Czakon:2004wm,Czakon:2004tg}.
This re-calculation found agreement with the formulae published earlier. We
have also added new yet unpublished parts of some of the Master
Integrals (typically the $O(\varepsilon )$ terms of their $\varepsilon -$%
expansion) in the closed form for the first time. We also discuss in detail
the aspects of the UV renormalization of the two-loop graphs including the
counterterm graphs described above and the treatment of the IR divergences
within the soft-photon approximation. We give also the numerical analysis
and discuss the various approximation to the exact two-loop expression.

The paper is organized as follows. In Section 2 we summarize our notation
and discuss the general structure of the amplitude. The third section is
devoted to the general aspects of the systematic chiral expansion of the
amplitude. Here we also give a list of the two-loop and one-loop Feynman
diagrams contributing to the next-to-leading order pure QED corrections. In
Section 4 we discuss the general strategy of the renormalization of the
one-loop and two-loop contributions and the treatment of IR divergences
within dimensional regularization in detail. Section 5 is devoted to the
calculation of the one-loop graphs with one one-loop counterterm vertex and
also our renormalization scheme is specified there. In Section 6 we
calculate the two-loop graphs and in Section 7 we discuss the soft-photon
bremsstrahlung. In Section 8 we put all the ingredients together and give
the final result for the virtual and real QED radiative corrections. We also
discuss large logarithm approximation and relate our result to the Bergstr%
\"{o}m's calculation. Some preliminary phenomenological applications are discussed in
Section 9. In Section 10 we give a brief summary and
conclusion. Some technical details are postponed to the
Appendices.  The relevant part of the $\chi PT$ Lagrangian with
virtual photons and leptons is summarized in Appendix
\ref{Lagrangian_appendix}. The reduction of the six
two-loop graphs to the scalar integrals is presented in Appendix \ref%
{Reduction_appendix}. In Appendix \ref{IPB_appendix} we list the
integration-by-parts identities for the scalar integrals and in Appendix \ref%
{MI appendix} we summarize the results of our (re-)calculation of the relevant Master
Integrals and give a comparison with existing literature.

\section{Basic properties of the amplitude}

In this section we discuss the basic features of the amplitude. We set the
notation and kinematics and then briefly comment on the general properties
of the lowest order amplitude which corresponds to $O(\alpha ^{2})$ order in
electromagnetic interaction (and all orders in QCD for the pion transition
form factor, which is here the only nonperturbative ingredient).

\subsection{Notation and kinematics}

The invariant amplitude $\mathcal{M}_{\pi ^{0}\rightarrow e^{+}e^{-}}$ for
the decay is defined by means of the matrix element
\begin{equation}
\langle e^{+}(q_{+},s_{+})e^{-}(q_{-},s_{-});out|\pi ^{0}(Q);in\rangle =%
\mathrm{i}(2\pi )^{4}\delta ^{(4)}(Q-q_{+}-q_{-})\mathcal{M}_{\pi
^{0}\rightarrow e^{+}e^{-}}
\end{equation}%
which is supposed to be calculated in the presence of strong and
electromagnetic interactions. According to the Lorentz covariance we can
further write
\begin{equation}
\mathcal{M}_{\pi ^{0}\rightarrow e^{+}e^{-}}=\overline{u}(q_{-},s_{-})\Gamma
_{\pi ^{0}e^{+}e^{-}}(q_{-},q_{+})v(q_{-},s_{-})
\end{equation}%
where $\Gamma _{\pi ^{0}e^{+}e^{-}}(q_{-},q_{+})$ is a one particle
irreducible $\pi ^{0}e^{+}e^{-}$ vertex. Off shell it can be conveniently
decomposed introducing four scalar form factors $P$, $A_{\pm }$ and $T$
defined as\footnote{%
In what follows we use the convention $\varepsilon ^{0123}=1$ and $\gamma
_{5}=\mathrm{i}\gamma ^{0}\gamma ^{1}\gamma ^{2}\gamma ^{3}$.}
\begin{equation} \begin{split}
\mathrm{i}\Gamma _{\pi ^{0}e^{+}e^{-}}(q_{-},q_{+})
&=P(q_{-}^{2},q_{+}^{2},Q^{2})\gamma ^{5}+(\slashed{q}_{-}-m)\gamma
^{5}A_{-}(q_{-}^{2},q_{+}^{2},Q^{2}) \\
&+A_{+}(q_{-}^{2},q_{+}^{2},Q^{2})\gamma
^{5}(\slashed{q}_{+}+m)
+T(q_{-}^{2},q_{+}^{2},Q^{2})(\slashed{q}_{-}-m)\gamma ^{5}(\slashed{q}%
_{+}+m).
\end{split} \end{equation}%
Here $Q=q_{+}+q_{-}$ and the charge conjugation invariance implies
\begin{eqnarray}
A_{-}(q_{-}^{2},q_{+}^{2},Q^{2}) &=&-A_{+}(q_{+}^{2},q_{-}^{2},Q^{2})  \notag
\\
P(q_{-}^{2},q_{+}^{2},Q^{2}) &=&P(q_{+}^{2},q_{-}^{2},Q^{2})  \notag \\
T(q_{-}^{2},q_{+}^{2},Q^{2}) &=&T(q_{+}^{2},q_{-}^{2},Q^{2}).
\end{eqnarray}%
For the electron-positron pair on shell we get then
\begin{equation}
\mathrm{i}\mathcal{M}_{\pi ^{0}\rightarrow e^{+}e^{-}}=\overline{u}%
(q_{-},s_{-})\gamma ^{5}v(q_{-},s_{-})P(m^{2},m^{2},Q^{2}),
\end{equation}%
and, as a consequence, the total decay rate is given solely in terms of the
on-shell form factor $P(m^{2},m^{2},M_{\pi ^{0}}^{2})$ as
\begin{equation}
\Gamma _{\pi ^{0}\rightarrow e^{+}e^{-}}=\frac{M_{\pi ^{0}}}{8\pi }\beta
(M_{\pi ^{0}}^{2})\left\vert P(m^{2},m^{2},M_{\pi ^{0}}^{2})\right\vert ^{2}
\label{rate}
\end{equation}%
where
\begin{equation}
\beta (Q^{2})=\sqrt{1-\frac{4m^{2}}{Q^{2}}}
\end{equation}%
is the velocity of the electron-positron pair in the CM frame.

Note that the semi-on-shell form factor $P(m^{2},m^{2},Q^{2})$ can be
extracted from the one particle irreducible vertex $\Gamma _{\pi
^{0}e^{+}e^{-}}(q_{-},q_{+})$ by means of the following projection
\begin{equation}
P(m^{2},m^{2},Q^{2})=-\lim_{q_{\pm }^{2}\rightarrow m^{2}}\frac{1}{2Q^{2}}%
\mathrm{Tr}\left[ (\slashed{q}_{-}+m)\Gamma _{\pi
^{0}e^{+}e^{-}}(q_{-},q_{+})(\slashed{q}_{+}-m)\gamma ^{5}\right] \mathrm{.}
\label{projection_P}
\end{equation}
This dimensionless form factor is an analytical function of the variable $%
s=Q^{2}$ in the complex plain with a cut $[0,\infty )$ where the unphysical
threshold $Q^{2}=0$ corresponds to the two-photon intermediate state. For
further convenience we introduce two dimensionless kinematical variables,
namely
\begin{equation}
y=\frac{Q^{2}}{4m^{2}}
\end{equation}
and
\begin{equation}
x=\frac{\beta (Q^{2})-1}{\beta (Q^{2})+1}=\frac{\sqrt{1-\frac{1}{y}}-1}{%
\sqrt{1-\frac{1}{y}}+1}  \label{x_definition}
\end{equation}
which map the unphysical threshold to $y=0$ and $x=1$ respectively.

In what follows we assume perturbative expansion of the amplitude in the QED
coupling $\alpha $. Consequently we can write for the form factor $P$
\begin{equation}
P(m^{2},m^{2},M_{\pi ^{0}}^{2})=P^{LO}(m^{2},m^{2},M_{\pi
^{0}}^{2})+P^{NLO}(m^{2},m^{2},M_{\pi ^{0}}^{2})+O(\alpha ^{4}),
\end{equation}%
where $P^{LO}=O(\alpha ^{2})$ and $P^{NLO}=O(\alpha ^{3})$, and for the
decay rate
\begin{equation}
\Gamma (\pi ^{0}\rightarrow e^{+}e^{-})=\Gamma ^{LO}(\pi ^{0}\rightarrow
e^{+}e^{-})+\Gamma ^{NLO}(\pi ^{0}\rightarrow e^{+}e^{-})+O(\alpha ^{6}).
\end{equation}%
In order to cancel the infra red (IR) divergences present in $\Gamma ^{NLO}$
we have to add also the real photon bremsstrahlung contribution and consider
inclusive decay rate of the process\footnote{%
Note that the same final state has also the Dalitz decay $\pi
^{0}\rightarrow \gamma \gamma ^{\ast }\rightarrow \gamma e^{+}e^{-}$, which
is however dominant in different region of the phase space corresponding to
small $x_{D}$ (cf. (\ref{x_D})). For large enough $x_{D}$ the Dalitz decay
contribution is tiny, however, for the experimentally used cut on $x_{D}$
this contribution should be also included.} $\pi ^{0}\rightarrow
e^{+}e^{-}(\gamma )$. \ The size of the NLO real and virtual \ QED radiative
corrections can be then described by means of the factor $\delta (x_{D}^{%
\mathrm{cut}})$ defined as
\begin{equation}
\Gamma ^{NLO}(\pi ^{0}\rightarrow e^{+}e^{-}(\gamma ),x_{D}>x_{D}^{\mathrm{%
cut}})=\delta (x_{D}^{\mathrm{cut}})\Gamma ^{LO}(\pi ^{0}\rightarrow
e^{+}e^{-})  \label{delta_definition}
\end{equation}%
where $x_{D}$ is the Dalitz variable (\ref{x_D}) and where all the $\pi
^{0}\rightarrow e^{+}e^{-}(\gamma )$ events with $x_{D}>x_{D}^{\mathrm{cut}}$
are included.

\subsection{The amplitude at the order \texorpdfstring{$O(\protect\alpha ^{2})$}{alpha2}}

\begin{figure}[t]
\begin{center}
\epsfig{width=0.35\textwidth,file=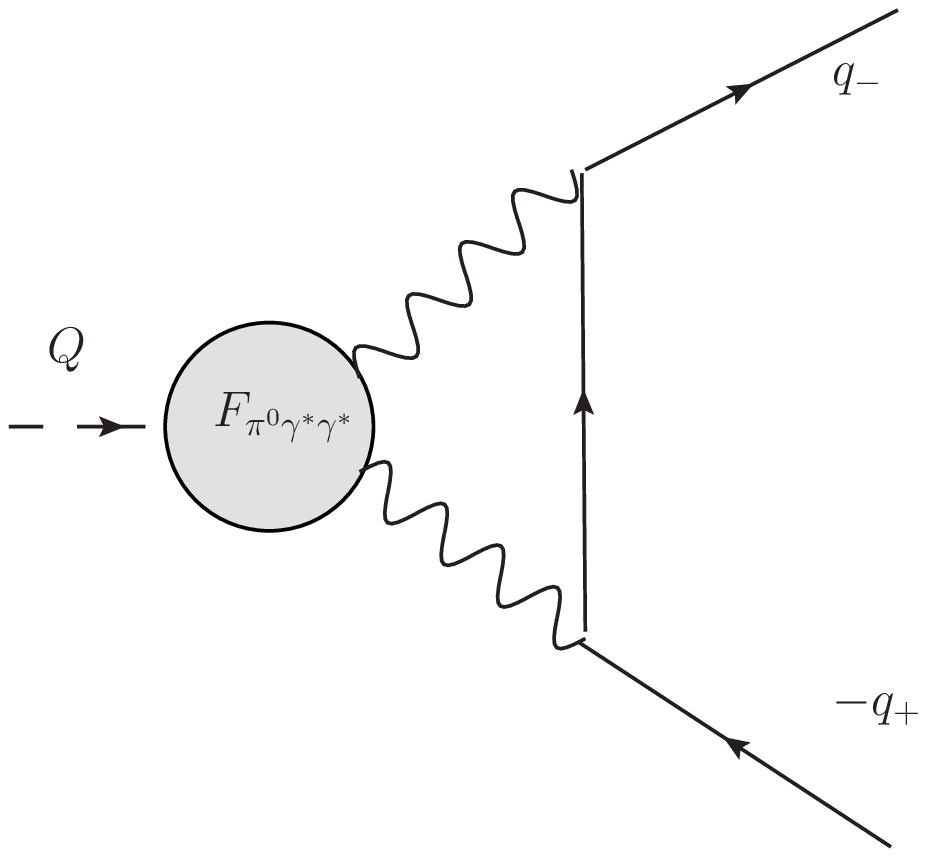}
\end{center}
\caption{The leading order $O(\protect\alpha ^{2})$ contribution to the
amplitude. The blob corresponds to the pion transition form factor. }
\label{LO_QED}
\end{figure}

Within the QED the leading order contribution $P^{LO}(m^{2},m^{2},M_{\pi
^{0}}^{2})$ to $P(m^{2},m^{2},M_{\pi ^{0}}^{2})$ is of the order $O(\alpha
^{2})$ and comes from the diagram depicted in Fig. \ref{LO_QED}. The bubble
there corresponds to the strong matrix element
\begin{equation}
-e^{2}\int \mathrm{d}^{4}x~e^{{i}k\cdot x}\langle 0|T(j^{\mu }(x)j^{\nu
}(0))|\pi ^{0}(Q)\rangle =\mathrm{i}e^{2}\varepsilon ^{\mu \nu \alpha \beta
}k_{\alpha }Q_{\beta }F_{\pi ^{0}\gamma ^{\ast }\gamma ^{\ast
}}(k^{2},(Q-k)^{2})
\end{equation}
where $j^{\mu }(x)$ is the hadronic part of the electromagnetic current
\begin{equation}
j^{\mu }=\frac{2}{3}\overline{u}\gamma ^{\mu }u-\frac{1}{3}\overline{d}%
\gamma ^{\mu }d
\end{equation}
and $F_{\pi ^{0}\gamma ^{\ast }\gamma ^{\ast }}(k^{2},l^{2})$ is the pion
transition form factor.

The explicit formula reads then
\begin{eqnarray}
\mathrm{i}\Gamma _{\pi ^{0}e^{-}e^{+}}^{LO}(q_{-},q_{+}) &=&-\mathrm{i}%
e^{4}\varepsilon ^{\mu \nu \alpha \beta }\int \frac{\mathrm{d}^{4}l}{(2\pi
)^{4}}\,F_{\pi ^{0}\gamma ^{\ast }\gamma ^{\ast
}}((l-q_{-})^{2},(l+q_{+})^{2})  \notag \\
&&\times \frac{(l+q_{+})_{\alpha }(l-q_{-})_{\beta }}{((l-q_{-})^{2}+\mathrm{%
i}0)[(l+q_{+})^{2}+\mathrm{i}0]}\,\gamma _{\mu }\,\frac{\mathrm{i}}{%
\slashed{l}-m+\mathrm{i}0}\,\gamma _{\nu }\,,  \label{Gamma_integral}
\end{eqnarray}
and using the projection (\ref{projection_P}) we get (cf. \cite%
{Bergstrom:1982zq})
\begin{equation}
P^{LO}(m^{2},m^{2},Q^{2})=-\mathrm{i}\frac{e^{4}m}{Q^{2}}\int \frac{\mathrm{d%
}^{4}l}{(2\pi )^{4}}\frac{F_{\pi ^{0}\gamma ^{\ast }\gamma ^{\ast
}}(D^{(-)},D^{(+)})}{D^{(-)}D^{(+)}D^{(0)}}\lambda (Q^{2},D^{(-)},D^{(+)}).
\label{P_integral}
\end{equation}
where we abbreviated
\begin{align}
D^{(\pm )}&=(l\pm q_{\pm })^{2}+\mathrm{i}0, \\
D^{(0)}&=l^{2}-m^{2}+\mathrm{i}0
\end{align}
and where
\begin{equation}
\lambda (a,b,c)=a^{2}+b^{2}+c^{2}-2ab-2ac-2bc
\end{equation}
is the triangle function.

The pion transition form factor represents the unknown nonperturbative QCD
ingredient of the above formula, therefore the evaluation of the integral (%
\ref{P_integral}) is model dependent. However some general features of (\ref%
{P_integral}) can be deduced in a model independent way. We will discuss
them in the rest of this section.

Let us first briefly remind the properties of $F_{\pi ^{0}\gamma
^{\ast }\gamma ^{\ast }}$. It has the following short distance
asymptotics which is a consequence of OPE. For $Q$ fixed and
$\lambda \rightarrow \infty $ we have \cite{Cornwall:1966zz} (see
also \cite{Knecht:2001xc})
\begin{equation}
F_{\pi ^{0}\gamma ^{\ast }\gamma ^{\ast }}((\lambda l)^{2},(Q-\lambda
l)^{2})=-\frac{1}{(\lambda l)^{2}}\frac{2}{3}F_{\pi }\left( 1+O(\alpha
_{s},\lambda ^{-1})\right)  \label{UV_asymptotics}
\end{equation}
(where $F_{\pi }$ is the pion decay constant) and therefore we can conclude
that the integrals (\ref{Gamma_integral}) and (\ref{P_integral}) are
convergent. Also the long distance asymptotics of $F_{\pi ^{0}\gamma ^{\ast
}\gamma ^{\ast }}(l^{2},(Q-l)^{2})$ is known from the first principles being
fixed by the QCD chiral anomaly. Namely in the chiral limit
\begin{equation}
F_{\pi ^{0}\gamma ^{\ast }\gamma ^{\ast }}^{\chi -\lim }(0,0)=\frac{1}{4\pi
^{2}F_{\pi }},
\end{equation}
and therefore the low energy behavior is expected to be given by the chiral
expansion of the form
\begin{equation}
F_{\pi ^{0}\gamma ^{\ast }\gamma ^{\ast }}(l^{2},(Q-l)^{2})=\frac{1}{4\pi
^{2}F_{\pi }}\left( 1+O\left( \frac{m_{q}}{\Lambda _{H}},\frac{l^{2}}{%
\Lambda _{H}^{2}},\frac{Q\cdot l}{\Lambda _{H}^{2}}\right) \right) .
\label{IR_asymptotics}
\end{equation}
where $\Lambda _{H}\sim 1GeV$ is the hadronic scale limiting the
applicability \ of $\chi PT$.

Another useful model independent property is that the two-photon
intermediate state contribution to the imaginary part of the leading order
form factor (\ref{P_integral}) with $Q^{2}\equiv s$ extended off the pion
mass shell\footnote{%
Note that $F_{\pi ^{0}\gamma ^{\ast }\gamma ^{\ast }}(k^{2},l^{2})$ can be
obtained by means of the LSZ reduction formula from the three-point
correlator \cite{Knecht:1999gb}
\begin{equation*}
\int \mathrm{d}^{4}x\mathrm{~{d}}^{4}ye\mathrm{^{{i}k\cdot x}}e\mathrm{^{{i}%
(Q-k)\cdot x}}\langle 0|T(j^{\mu }(x)j^{\nu }(y)P^{3}(0))|0\rangle =\frac{2}{%
3}\varepsilon ^{\mu \nu \alpha \beta }k_{\alpha }Q_{\beta
}H(k^{2},(Q-k)^{2},Q^{2})
\end{equation*}
where
\begin{equation*}
P^{a}=\frac{1}{2}\overline{q}\lambda ^{a}\gamma ^{5}q
\end{equation*}
is the pseudoscalar density. Namely for $Q^{2}\rightarrow M_{\pi ^{0}}^{2}$
we have
\begin{eqnarray*}
H(k^{2},(Q-k)^{2},Q^{2}) &=&\frac{3}{2}\frac{\langle \pi
^{0}(Q)|P^{3}(0)|0\rangle }{(Q^{2}-M_{\pi ^{0}}^{2})}F_{\pi ^{0}\gamma
^{\ast }\gamma ^{\ast }}(k^{2},(Q-k)^{2}) \\
&&+O\left( (Q^{2}-M_{\pi ^{0}}^{2})^{0}\right) .
\end{eqnarray*}
This offers the natural possibility to extend the form factor $F_{\pi
^{0}\gamma ^{\ast }\gamma ^{\ast }}(k^{2},l^{2})$ off the mass shell
\begin{equation*}
F_{\pi ^{0}\gamma ^{\ast }\gamma ^{\ast }}(k^{2},(Q-k)^{2},Q^{2})=\frac{2}{3}%
\frac{(Q^{2}-M_{\pi ^{0}}^{2})}{\langle \pi ^{0}(Q)|P^{3}(0)|0\rangle }%
H(k^{2},(Q-k)^{2},Q^{2})
\end{equation*}%
} is uniquely fixed by the value $F_{\pi ^{0}\gamma ^{\ast }\gamma ^{\ast
}}(0,0)$. Indeed, using the Cutkosky rules and cutting the two internal
photon lines we get \cite{Drell:1959},\cite{Berman:1960zz}
\begin{equation}
\mathrm{Im}\text{ }P^{LO}(m^{2},m^{2},s)|_{2\gamma }=\theta (s)\pi \alpha
^{2}m\frac{1}{\beta (s)}\ln \left( \frac{1-\beta (s)}{1+\beta (s)}\right)
F_{\pi ^{0}\gamma ^{\ast }\gamma ^{\ast }}(0,0).
\end{equation}
For $s<(M_{\pi ^{0}}+2M_{\pi^{+}})^{2}$ this contribution saturates the imaginary part of $%
P^{LO}(m^{2},m^{2},s)$. As a consequence, the known imaginary part $\mathrm{%
Im}$ $P^{LO}(m^{2},m^{2},M_{\pi ^{0}}^{2})$ and the known width $\Gamma
_{2\gamma }^{LO}$ of the $2\gamma $ pion decay in the leading order in the
QED expansion,
\begin{equation}
\Gamma _{2\gamma }^{LO}=\frac{1}{4}\pi \alpha ^{2}M_{\pi ^{0}}^{3}\left\vert
F_{\pi ^{0}\gamma ^{\ast }\gamma ^{\ast }}(0,0)\right\vert ^{2},
\end{equation}
can be further used to get another exact result concerning the branching
ratio
\begin{equation}
R=\frac{B^{LO}(\pi ^{0}\rightarrow e^{+}e^{-})}{B^{LO}(\pi ^{0}\rightarrow
\gamma \gamma )}.
\end{equation}
Namely, using $\mathrm{Im}P^{LO}|_{2\gamma }$ instead of $P^{LO}$ in (\ref%
{rate}), we get the following model independent unitarity bound \cite%
{Berman:1960zz}
\begin{equation}
R\geq \frac{1}{2}\left( \frac{\alpha m}{M_{\pi ^{0}}}\right) ^{2}\frac{1}{%
\beta (M_{\pi ^{0}}^{2})}\ln ^{2}\left( \frac{1-\beta (M_{\pi ^{0}}^{2})}{%
1+\beta (M_{\pi ^{0}}^{2})}\right) =4.75\times 10^{-8}.
\end{equation}

The explicit knowledge of $\mathrm{Im}$ $P^{LO}|_{2\gamma }$ allows also to
pinpoint the most important nonanalytic contribution to $%
P^{LO}(m^{2},m^{2},s)$, namely that stemming from the two-photon
intermediate state. The latter is given by means of the following once
subtracted dispersive representation \cite{Bergstrom:1983ay}
\begin{equation}
P^{LO}(m^{2},m^{2},s)|_{2\gamma }=P^{LO}(m^{2},m^{2},0)|_{2\gamma }+\frac{s}{%
\pi }\int_{0}^{\infty }\frac{\mathrm{d}s\mathrm{^{^{\prime }}}}{s^{^{\prime
}}}\frac{\mathrm{{Im}\text{ }}P^{LO}(m^{2},m^{2},s^{^{\prime }})|_{2\gamma }%
}{s^{^{\prime }}-s},
\end{equation}
and it is therefore fixed uniquely up to one unknown subtraction constant $%
P^{LO}(m^{2},m^{2},0)|_{2\gamma }$. The dispersion integral for
the physically relevant region $\ s=Q^{2}>4m^{2}$ reads
\begin{equation} \begin{split}
P^{LO}(m^{2},m^{2},s)|_{2\gamma ,\mathrm{disp}} &=\frac{Q^{2}}{\pi }%
\int_{0}^{\infty }\frac{ds^{^{\prime }}}{s^{^{\prime }}}\frac{\mathrm{{Im}%
\text{ }}P^{LO}(m^{2},m^{2},s^{^{\prime }})|_{2\gamma }}{s^{^{\prime
}}-Q^{2}-\mathrm{i}0}   \\
&=\alpha ^{2}mF_{\pi ^{0}\gamma ^{\ast }\gamma ^{\ast }}(0,0)\frac{1}{\beta
(s)}\left[ \mathrm{Li}_{2}\left( x\right) -\mathrm{Li}_{2}\left( \frac{1}{x}%
\right) +\mathrm{i}\pi \ln \left( -x\right) \right].
\end{split} \end{equation}
In this formula $x<0$ is given by (\ref{x_definition}) and $\mathrm{Li}_{2}$
is the dilogarithm defined as
\begin{equation}
\mathrm{Li}_{2}\left( z\right) =-\int_{0}^{z}\frac{{d}t}{t}\ln (1-t)
\end{equation}
which is analytic in the complex plain with cut $[1,\infty )$. The unknown
explicit form of the form factor $F_{\pi ^{0}\gamma ^{\ast }\gamma ^{\ast }}$
in the intermediate energy region can influence only the subtraction
constant $P^{LO}(m^{2},m^{2},0)|_{2\gamma }$.

As a consequence, we can split the leading order amplitude $%
P^{LO}(m^{2},m^{2},s)$ into two parts, namely
\begin{eqnarray}
P^{LO}(m^{2},m^{2},s) &=&P^{LO}(m^{2},m^{2},s)|_{2\gamma ,\mathrm{disp}}
\notag \\
&&+2\alpha ^{2}mF_{\pi ^{0}\gamma ^{\ast }\gamma ^{\ast }}(0,0)\left[ \frac{3%
}{2}\ln \left( \frac{m^{2}}{\Lambda ^{2}}\right) -\frac{5}{2}+\chi \left(\frac{s}{%
\Lambda ^{2}},\frac{m^{2}}{\Lambda ^{2}}\right)\right] .
\label{P_LO_split}
\end{eqnarray}%
The first part corresponding to the two photon intermediate state is
completely independent on the details of the pion transition formfactor. The
second part \ (where $\Lambda $ is an intrinsic scale characteristic for the
formfactor $F_{\pi ^{0}\gamma ^{\ast }\gamma ^{\ast }}$, \emph{i.e.} the
scale at which is the integral (\ref{P_integral}) effectively cut off)
accumulates the above subtraction constant $P^{LO}(m^{2},m^{2},0)|_{2\gamma }
$ as well as the contributions of higher intermediate states which appear
when also the bubble in the Fig. \ref{LO_QED} is cut. The explicit form of the $\chi (%
\frac{s}{\Lambda ^{2}},\frac{m^{2}}{\Lambda ^{2}})$ as a functional of
unknown transition formfactor $F_{\pi ^{0}\gamma ^{\ast }\gamma ^{\ast }}$
has been derived in \cite{Dorokhov:2009xs} with use of the Mellin-Barnes
representation.

We have explicitly kept apart the logarithmic term in the square
brackets of (\ref{P_LO_split}). The reason is that this term
corresponds to the leading dependence of $P^{LO}(m^{2},m^{2},s)$
on the effective cut-off scale
$\Lambda $. The origin of this term is easy to understand \cite%
{Bergstrom:1983ay}. \ In the case of point-like pion (\emph{i.e.} when $%
F_{\pi ^{0}\gamma ^{\ast }\gamma ^{\ast }}(k^{2},l^{2})=F_{\pi
^{0}\gamma ^{\ast }\gamma ^{\ast }}(0,0)=const.$), the integral
(\ref{P_integral}) is logarithmically divergent. Using the sharp
cut-off at the scale $\Lambda $ instead of effective cut-off
provided by $F_{\pi ^{0}\gamma ^{\ast }\gamma ^{\ast
}}(l^{2},(Q-l)^{2})$ we get for $\Lambda \rightarrow \infty $
\begin{equation}
P^{LO}(m^{2},m^{2},s)|_{\mathrm{point-like}}^{\mathrm{sharp\ cut-off}%
}=-3\alpha ^{2}mF_{\pi ^{0}\gamma ^{\ast }\gamma ^{\ast }}(0,0)\ln \left(
\frac{\Lambda ^{2}}{m^{2}}\right) +O(1)
\end{equation}%
where the $O(1)$ part includes terms that are finite or suppressed
for $\Lambda \rightarrow \infty $. Thus we expect the same
behavior also for the full
amplitude $P^{LO}$ and therefore \ $\chi =O(1)$ for $\Lambda \rightarrow \infty $%
.

Because the discontinuities of $\chi $ as a function of $s$ start at $s\sim
\Lambda ^{2}$, $\chi $ is analytic in the physical region $s<\Lambda ^{2}$
and can be therefore expanded in this region in the power series of the
variable $s/\Lambda ^{2}$. This suggests that $\chi (\frac{s}{\Lambda ^{2}},%
\frac{m^{2}}{\Lambda ^{2}})$ can be approximated at the leading order of
this expansion by $\Lambda $-independent constant.

\section{Systematic chiral expansion}

Within the $SU(2)\times SU(2)$ variant of $\chi PT$ supplemented
with dynamical photons and electrons (cf. \cite{Knecht:1999ag})
the formfactor $P$ is given in terms of the systematic
simultaneous expansion in powers of the momenta ($s$), the quark
and electron masses and the fine structure constant $\alpha $, to
which the chiral orders are formally assigned according to
\begin{equation}
s,~m_{q},m^{2},\alpha =O(p^{2}).  \label{chiral_orders_assignment}
\end{equation}%
The relevant parts of the enlarged $\chi PT$ Lagrangian are listed in the
Appendix \ref{Lagrangian_appendix}. The hierarchy of the various
contributions is then controlled by the Weinberg power-counting formula \cite%
{Weinberg:1978kz}. As is usual in $\chi PT$, for the regularization of UV as well
as IR divergences we use the dimensional regularization (DR) in what
follows. In order to avoid problems with intrinsically four-dimensional
objects like Levi-Civita pseudo-tensor and $\gamma _{5}$ we use here the
variant known as Dimensional Reduction. For calculation of $P$ this means
that we first project out the form factor from the amplitude by means of (%
\ref{projection_P}) using four-dimensional Dirac algebra and only then we
dimensionally regularize the resulting scalar integrals.

\subsection{The leading order of the chiral expansion}

\begin{figure}[t]
\begin{center}
\epsfig{width=0.95\textwidth,file=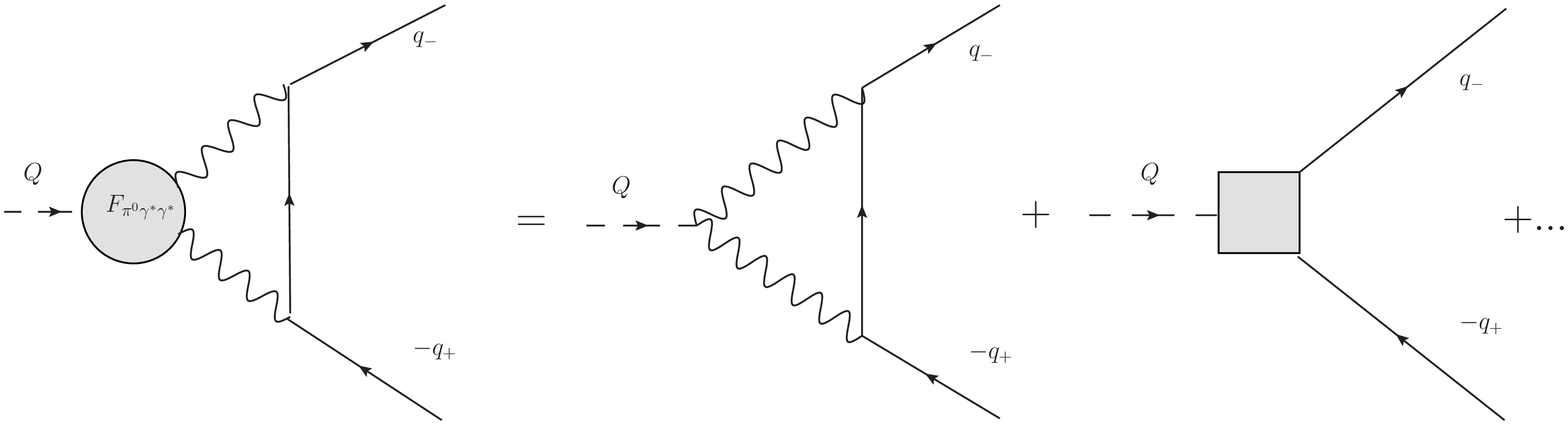}
\end{center}
\caption{The leading order $O(\protect\alpha ^{2}p^2)$
contribution to the amplitude. The shaded box corresponds to the
$\protect\pi e^+ e^-$ counterterm contribution. } \label{LO_ChPT}
\end{figure}

The leading order contribution $P^{\chi LO}(m^{2},m^{2},s)$ represents the
chiral order\footnote{%
Note, that for the amplitude $\mathcal{M}_{\pi ^{0}\rightarrow e^{+}e^{-}}$%
we have $\mathcal{M}_{\pi ^{0}\rightarrow e^{+}e^{-}}=O(\alpha ^{2}p^{2})$,
because the fermion wave functions are counted as $O(p^{1/2})$.} $O(\alpha
^{2}p)$ and can be divided into two parts (see Fig. \ref{LO_ChPT}). The
first one is a logarithmically divergent one-loop graph with local $\pi
^{0}\gamma ^{\ast }\gamma ^{\ast }$ vertex stemming form the $O(\alpha
p^{2}) $ Wess-Zumino-Witten Lagrangian (here and in what follows we write
down only the relevant vertices)
\begin{equation}
\mathcal{L}_{WZW}^{\alpha p^{2}}=\frac{1}{8}\left( \frac{\alpha }{\pi }%
\right) \frac{\pi ^{0}}{F_{0}}\varepsilon _{\mu \nu \alpha \beta }F^{\mu \nu
}F^{\alpha \beta }+\ldots  \label{WZW}
\end{equation}%
($F_{0}$ is the pion decay constant in the chiral limit) and the second one
corresponds to a tree-level counterterm graph originating in the $O(\alpha
^{2}p^{2})$ Lagrangian
\begin{equation}
\mathcal{L}_{\pi ee}^{\alpha ^{2}p^{2}}=-\mu ^{-2\varepsilon }\frac{1}{4}%
\left( \frac{\alpha }{\pi }\right) ^{2}\left[ \chi ^{r}(\mu )+\frac{3}{2}%
\left( \frac{1}{\varepsilon }+\ln 4\pi -\gamma \right) \right] \overline{e}%
\gamma ^{\mu }\gamma _{5}e\frac{\partial _{\mu }\pi ^{0}}{F_{0}}+\ldots ,
\label{CT_alpha^2p^2}
\end{equation}%
where $\chi ^{r}(\mu )$ is a renormalized counterterm coupling at a scale $%
\mu $. One can think of the loop part of $P^{\chi LO}$ as approximating the
leading order formfactor $P^{LO}$ given by (\ref{P_integral}) by means of
inserting the leading order term of the chiral expansion of the pion
transition formfactor (\ref{IR_asymptotics}) into (\ref{P_integral}). This
insertion however modifies significantly the high energy region of the loop
integration starting at the onset of the resonances where the chiral
expansion fails to converge. Such a modification of the loop has to be
compensated by local counterterm contribution in such a way that the $%
O(\alpha ^{2}p)$ term of the chiral expansion of the $P^{LO}$ is exactly
reproduced. This is the general idea of the matching of the coupling
constant $\chi ^{r}(\mu )$. In the absence of the first principle
determination of $F_{\pi ^{0}\gamma ^{\ast }\gamma ^{\ast }}$ this matching
procedure is however model dependent. We use here the value
\begin{equation}
\chi ^{r}(\mu =770\mathrm{MeV})=2.2\pm 0.9 \label{chirLMD}
\end{equation}%
which has been obtained in \cite{Knecht:1999gb} by means of using a large $%
N_{C}$ inspired Lowest Meson Dominance (LMD) ansatz for $F_{\pi ^{0}\gamma
^{\ast }\gamma ^{\ast }}$.

As a result we get\footnote{%
At this order we can put $F_{0}=F_{\pi }$. In fact, such a replacement
corresponds to partial re-summation of the higher order corrections, namely
the renormalization of the pion external leg.}
\begin{eqnarray}
P^{\chi LO}(m^{2},m^{2},s) &=&\left( \frac{\alpha }{2\pi }\right) ^{2}\frac{m%
}{F_{\pi }}\frac{1}{\beta (s)}\left[ \mathrm{Li}_{2}\left( x\right) -\mathrm{%
Li}_{2}\left( \frac{1}{x}\right) +\mathrm{i}\pi \ln \left( -x\right) \right]
\notag \\
&&+\left( \frac{\alpha }{2\pi }\right) ^{2}\frac{2m}{F_{\pi }}\left[ \frac{3%
}{2}\ln \left( \frac{m^{2}}{\mu ^{2}}\right) -\frac{5}{2}+\chi ^{r}(\mu )%
\right] .  \label{ChPT_LO}
\end{eqnarray}%
The structure of this expression can be easily understood. It corresponds to
the formula (\ref{P_LO_split}) where both $F_{\pi ^{0}\gamma ^{\ast }\gamma
^{\ast }}(0,0)$ and $\chi (s/\Lambda ^{2},m^{2}/\Lambda ^{2})$ have been
replaced with the leading terms of their chiral expansion, provided we
identify the renormalization scale $\mu $ with the intrinsic cut-off scale $%
\Lambda $ of the formfactor $F_{\pi ^{0}\gamma ^{\ast }\gamma ^{\ast }}$.

\begin{figure}[t]
\begin{center}
\epsfig{width=0.55\textwidth,file=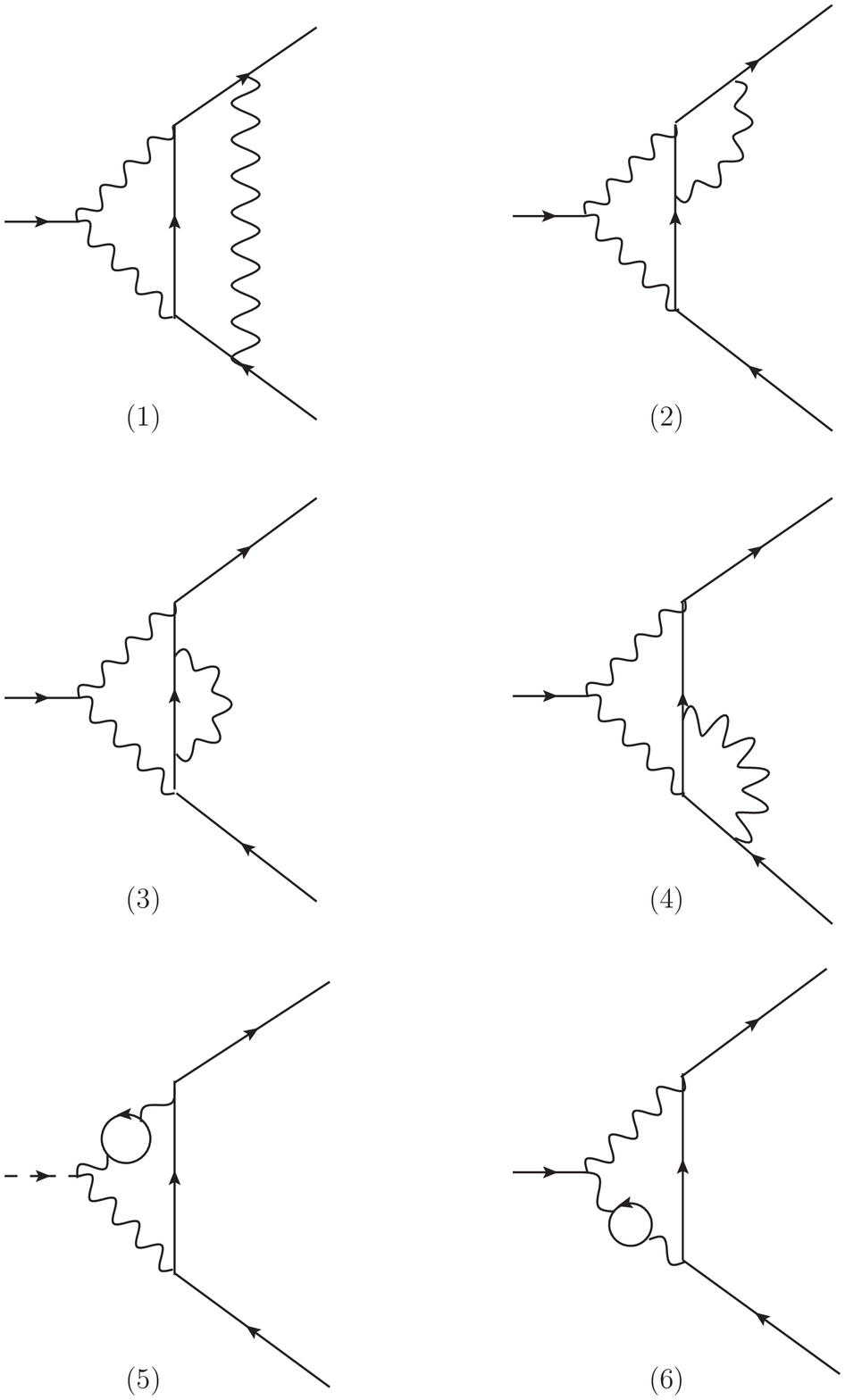}
\end{center}
\caption{The two-loop part of the next-to-leading order $O(\protect\alpha %
^{3}p^2)$ contributions to the amplitude.} \label{NLO_ChPT_2loop}
\end{figure}

\subsection{The \texorpdfstring{$O(\protect\alpha^{3}p^{2})$}{alpha3p2} part of the next-to-leading
order}

At the next-to-leading order the amplitude is generically $O(p^{8})$ in
terms of the simultaneous expansion according to the chiral order assignment (%
\ref{chiral_orders_assignment}). At this order there are two types of
contributions, which counts either as $O(\alpha ^{2}p^{4})$ or as $O(\alpha
^{3}p^{2})$ for the amplitude (or $O(\alpha ^{2}p^{3})$ and $O(\alpha ^{3}p)
$ for the form factor $P$). The contributions of the first type collect
two-loop and one-loop graphs with the same topology as depicted in the Fig. \ref{LO_QED}
(where now the blob represents either a one loop subgraph with pion internal
lines or \ $O(\alpha p^{4})$ order counterterm) and in addition tree graphs
with $O(\alpha ^{2}p^{4})~$counterterms\footnote{%
Strictly speaking there is also contribution from the pion external leg
renormalization, which we have however effectively added in the LO by means
of the replacement $F_{0}\rightarrow F_{\pi }$.}. Such contributions can be
understood as was mentioned above as the next-to-leading terms of the chiral expansion of (%
\ref{P_LO_split}).

\begin{figure}[t]
\begin{center}
\epsfig{width=0.80\textwidth,file=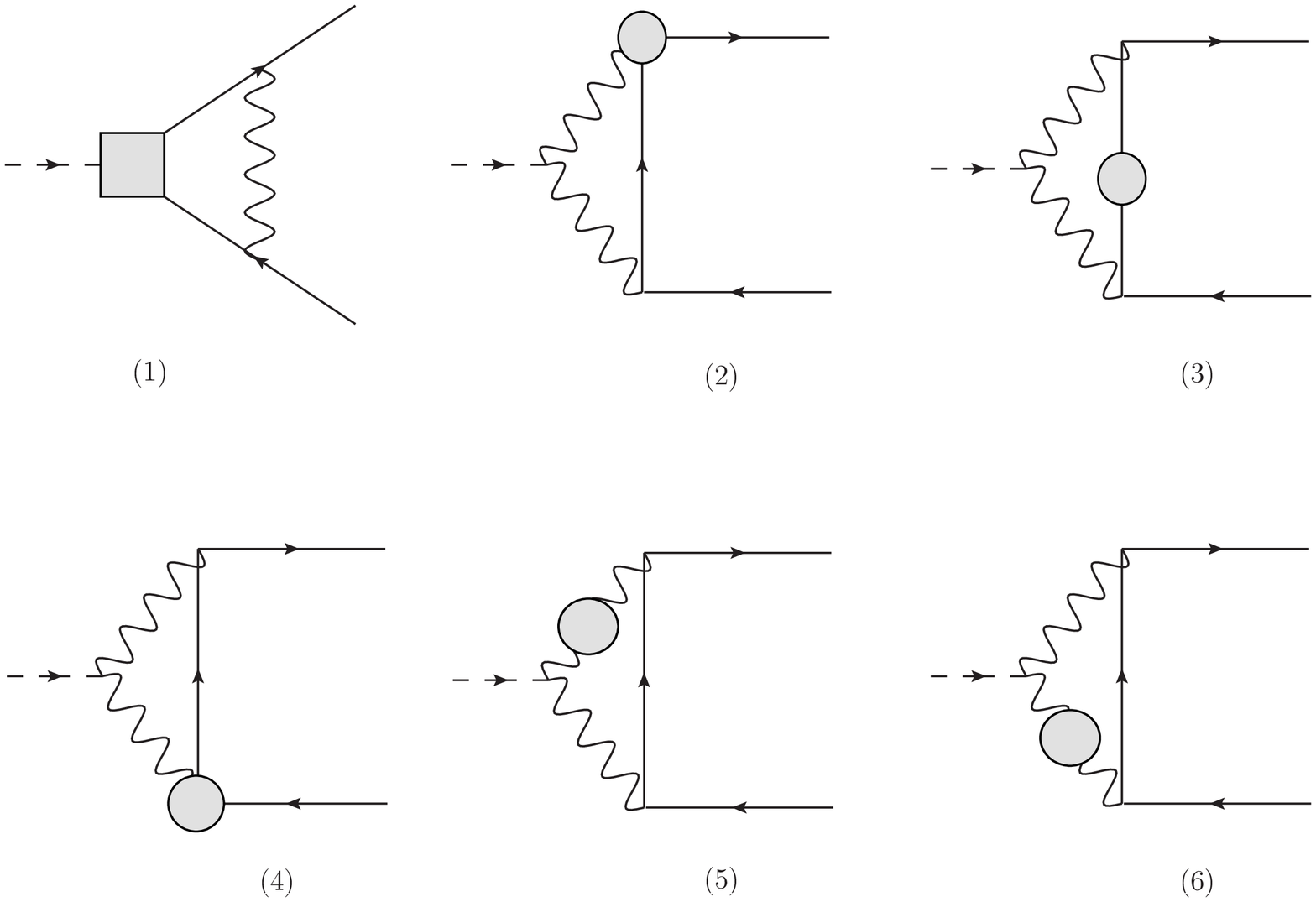}
\end{center}
\caption{The one-loop part of the next-to-leading order $O(\protect\alpha %
^{3}p^2)$ contributions to the amplitude. The numbers labeling
the graphs are in one-to-one correspondence with Fig.
\protect\ref{NLO_ChPT_2loop}. The shaded box is the
$O(\protect\alpha ^{2}p^{2})$ counterterm while the blobs represent
$O(\protect\alpha p^{2})$ ones.} \label{NLO_ChPT_1loop}
\end{figure}

In this paper we will concentrate on the contributions of the second type (%
\emph{i.e} $O(\alpha ^{3}p^{2})$) which represent the pure QED corrections.
In this case we get again three classes of graphs, namely six two-loops
graphs with virtual photons and electrons (see Fig. \ref{NLO_ChPT_2loop}),
six one-loop graphs with $O(\alpha p^{2})$ counterterms (see Fig. \ref%
{NLO_ChPT_1loop}, graphs (2)-(6)) or $O(\alpha ^{2}p^{2})$ counterterm (see
Fig. \ref{NLO_ChPT_1loop}, graph (1)) which renormalize the one-loop
subdivergences of the corresponding two-loop graphs and $O(\alpha ^{3}p^{2})$
tree-level graphs which are necessary to renormalize the remaining
superficial divergences. The latter are of the same order in $p$ as the
counterterm (\ref{CT_alpha^2p^2}) and therefore the relevant vertex from the
$O(\alpha ^{3}p^{2})$ Lagrangian can be summarily written in the form
\begin{equation}
\mathcal{L}_{\pi ee}^{\alpha ^{3}p^{2}}=-\mu ^{-4\varepsilon }\frac{1}{4}%
\left( \frac{\alpha }{\pi }\right) ^{3}\left[ \xi ^{r}(\mu )+O(\varepsilon
^{-2})+O(\varepsilon ^{-1})\right] \overline{e}\gamma ^{\mu }\gamma _{5}e%
\frac{\partial _{\mu }\pi ^{0}}{F_{0}},  \label{2loop_tree_CT_lagrangian}
\end{equation}%
where $\xi ^{r}(\mu )$ is the renormalized coupling and we have not written
the UV divergent part explicitly.

\section{Structure of the two-loop corrections and renormalization\label%
{renormalization_section}}

In this section we briefly discuss the general structure of the one and
two-loop contributions within the dimensional regularization and within
renormalization scheme suitable for power counting non-renormalizable
effective field theories \cite{Buchler:2003vw} (cf. also \cite%
{Bijnens:2010xg}). This discussion will be helpful for the organization of
the results of the explicit calculation and for consistency checks of the
results presented in the next sections.

Let us write the effective Lagrangian in the form
\begin{equation}
\mathcal{L}=\sum\limits_{n}\mu ^{-2\varepsilon n}\mathcal{L}_{n}
\end{equation}
where $\mathcal{L}_{n}$ accumulates the counterterms which are needed in
order to renormalize the superficial divergences of the $n-$loop graphs and $%
\mu $ is the dimensional regularization scale. Schematically
\begin{equation}
\mathcal{L}_{n}=\sum_{i}\left[ K_{i}^{r}(\mu )^{n-\mathrm{loop}%
}-\sum_{j=1}^{n}\frac{\gamma _{-j}^{(i),~n-\mathrm{loop}}}{\varepsilon ^{j}}%
\right] O_{n}^{(i)},
\end{equation}
where $K_{i}^{r}(\mu )^{n-\mathrm{loop}}$ are the renormalized counterterm
couplings (\emph{i.e.} the finite parts of the counterterms at the $n-$loop
level) and $O_{n}^{(i)}$ is a set of operators. In the above formula for $%
\mathcal{L}$ the factor $\mu ^{-2\varepsilon n}$ naturally appears when the
scale $\mu $ is artificially introduced into each loop integration writing $%
\mathrm{d}^{d}k=\mu ^{-2\varepsilon }(\mu ^{2\varepsilon }\mathrm{d}^{d}k)$
in order to restore the four-dimensional canonical dimension of the $d-$%
dimensional integration measure.

\subsection{Renormalization of the one-loop contributions}

For the total one-loop contributions to any one-particle irreducible vertex $%
\gamma $ entering the game we can write schematically
\begin{eqnarray}
\gamma ^{1-\mathrm{loop}} &=&\mu ^{-2\varepsilon }m^{D}\left[ \left( \frac{%
\mu }{m}\right) ^{2\varepsilon }\left( \frac{\gamma _{-1}^{1-\mathrm{loop}}}{%
\varepsilon }+\gamma _{0}^{1-\mathrm{loop}}+\varepsilon \gamma _{1}^{1-%
\mathrm{loop}}+O(\varepsilon ^{2})\right) \right.  \notag \\
&&\left. +\left( \chi _{\gamma }^{r}(\mu )-\frac{\gamma _{-1}^{1-\mathrm{loop%
}}}{\varepsilon }\right) \right]  \label{gamma_1loop}
\end{eqnarray}
where $D$ is the (four-dimensional) canonical dimension of $\gamma $.
In this formula\ the first line represents the result of the loops while the
second line accumulates the counterterm contributions. In this notation,
both $\gamma _{-1}^{1-\mathrm{loop}}$ and $\chi _{\gamma }^{r}(\mu )$ are
generally polynomials in the external momenta. Moreover $\chi _{\gamma
}^{r}(\mu )$ is linear in the renormalized counterterm couplings $%
K_{i}^{r}(\mu )^{1-\mathrm{loop}}$ introduced above. \ On the other hand,
the functions $\gamma _{i}^{1-\mathrm{loop}}$ for $i>-1$ have more
complicated analytical structure with branch points and cuts. In general, $%
\gamma _{i}^{1-\mathrm{loop}}$ depend nonlinearly on the tree-level
couplings $K_{i}^{r}(\mu )^{0-\mathrm{loop}}$. The chosen normalization
ensures that $\gamma _{i}^{1-\mathrm{loop}}$ are dimensionless. The above
structure (\ref{gamma_1loop}) is shared by the LO contribution to the
formfactor $P$ and also by the UV divergent one-loop subgraphs of the NLO
corrections to $P$, namely the off-shell one-loop $\pi e^{+}e^{-}$ and $%
\gamma e^{+}e^{-}$ vertices, the electron self-energy and the vacuum
polarization.

We use here the renormalization scheme suitable for power-counting
nonrenormalizable effective theories \cite{Buchler:2003vw} and require
renormalization scale independence order by order in the loop expansion. For
$\gamma ^{1-\mathrm{loop}}$ this means that the following finite quantity
\begin{eqnarray}
\overline{\chi }_{\gamma } &=&\left( \frac{\mu }{m}\right) ^{-2\varepsilon
}\left( \chi _{\gamma }^{r}(\mu )-\frac{\gamma _{-1}^{1-\mathrm{loop}}}{%
\varepsilon }\right) +\frac{\gamma _{-1}^{1-\mathrm{loop}}}{\varepsilon }
\notag \\
&=&\chi _{\gamma }^{r}(\mu )+\gamma _{-1}^{1-\mathrm{loop}}\ln \left( \frac{%
\mu ^{2}}{m^{2}}\right) +O(\varepsilon )  \label{Kbar}
\end{eqnarray}
has to be $\mu -$independent. This implies a running of $\chi _{\gamma
}^{r}(\mu )$ according to
\begin{equation}
\chi _{\gamma }^{r}(\mu )=\overline{\chi }_{\gamma }-\gamma _{-1}^{1-\mathrm{%
loop}}\ln \left( \frac{\mu ^{2}}{m^{2}}\right) +\varepsilon \left[ \overline{%
\chi }_{\gamma }\ln \left( \frac{\mu ^{2}}{m^{2}}\right) -\frac{1}{2}\gamma
_{-1}^{1-\mathrm{loop}}\ln ^{2}\left( \frac{\mu ^{2}}{m^{2}}\right) \right]
+O(\varepsilon ^{2})  \label{chi_run}
\end{equation}
and corresponding running of the (linear combinations of ) counterterm
couplings $K_{i}^{r}(\mu )^{1-\mathrm{loop}}$. As a result, (\ref%
{gamma_1loop}) can be re-organized in the following simple manifestly RG
invariant form
\begin{equation}
\gamma ^{1-\mathrm{loop}}=m^{D}\left( \gamma _{0}^{1-\mathrm{loop}}+%
\overline{\chi }_{\gamma }+O(\varepsilon )\right) .  \label{1loop_short}
\end{equation}

\subsection{Renormalization of the two-loop contributions}

For simplicity let us first assume that there are no infrared (IR)
divergences. The generalization with the presence of IR divergences will be
discussed in the next subsection.

At the two loop level, we have three types of contributions, namely
\begin{equation}
\gamma ^{2-\mathrm{loop}}=\gamma _{L}^{2-\mathrm{loop}}+\gamma _{CT}^{1-%
\mathrm{loop}}+\gamma _{CT}^{\mathrm{tree}}.
\end{equation}%
The first one corresponds the genuine two-loop contribution which can be
schematically written in the form
\begin{equation}
\gamma _{L}^{2-\mathrm{loop}}=\mu ^{-4\varepsilon }m^{D}\left( \frac{\mu }{m}%
\right) ^{4\varepsilon }\left( \frac{\gamma _{-2}^{2-\mathrm{loop}}}{%
\varepsilon ^{2}}+\frac{\gamma _{-1}^{2-\mathrm{loop}}}{\varepsilon }+\gamma
_{0}^{2-\mathrm{loop}}+O(\varepsilon )\right) .  \label{two_loop_general}
\end{equation}%
The second one represents a sum of one-loop graphs with one-loop level
counterterms which are necessary to renormalize the subdivergences of the
two-loop part
\begin{eqnarray}
\gamma _{CT}^{1-\mathrm{loop}} &=&\mu ^{-4\varepsilon }m^{D}\left( \frac{\mu
}{m}\right) ^{2\varepsilon }\sum_{i}\left( x_{i}^{r}(\mu )^{1-\mathrm{loop}}-%
\frac{\gamma _{-1}^{(i),~1-\mathrm{loop}}}{\varepsilon }\right)  \notag \\
&&\times \left( \frac{C_{-1}^{(i),~1-\mathrm{loop}}}{\varepsilon }%
+C_{0}^{(i),~1-\mathrm{loop}}+\varepsilon C_{1}^{(i),~1-\mathrm{loop}%
}+O(\varepsilon ^{2})\right)  \label{CT_1loop}
\end{eqnarray}%
where we have explicitly pulled out the dependence on the renormalized
one-loop counterterm $\ $couplings \ $x_{i}^{r}(\mu )^{1-\mathrm{loop}}$ and
coefficient of the corresponding infinite parts $\gamma _{-1}^{(i),~1-%
\mathrm{loop}}$. Finally we have also a tree level contribution of the
counterterms necessary to renormalize the remaining superficial divergence
of $\gamma ^{2-\mathrm{loop}}+\gamma _{CT}^{1-\mathrm{loop}}$, naively

\begin{eqnarray}
\gamma _{CT}^{\mathrm{tree}} &=&\mu ^{-4\varepsilon }m^{D}\left[ \xi
_{\gamma }^{r}(\mu )-\frac{\gamma _{-2}^{2-\mathrm{loop}}-\sum_{i}\gamma
_{-1}^{(i),~1-\mathrm{loop}}C_{-1}^{(i),~1-\mathrm{loop}}}{\varepsilon ^{2}}%
\right.  \notag \\
&&\left. -\frac{\gamma _{-1}^{2-\mathrm{loop}}-\sum_{i}\left( \gamma
_{-1}^{(i),~1-\mathrm{loop}}C_{0}^{(i),~1-\mathrm{loop}}-x_{i}^{r}(\mu )^{1-%
\mathrm{loop}}C_{-1}^{(i),~1-\mathrm{loop}}\right) }{\varepsilon }\right.
\notag \\
&&\left. -\frac{2\gamma _{-2}^{2-\mathrm{loop}}-\sum_{i}\gamma _{-1}^{(i),~1-%
\mathrm{loop}}C_{-1}^{(i),~1-\mathrm{loop}}}{\varepsilon }\ln \left( \frac{%
\mu ^{2}}{m^{2}}\right) \right]
\end{eqnarray}%
(here $\xi _{\gamma }^{r}(\mu )$ is polynomial in external momenta and
linear combination of the couplings $x_{i}^{r}(\mu )^{2-\mathrm{loop}}$ from
$\mathcal{L}_{2}$). However, in the absence of the \ IR divergences, in the
sum $\gamma _{L}^{2-\mathrm{loop}}+\gamma _{CT}^{1-\mathrm{loop}}$only the
local \ (\emph{i.e.} polynomial in the masses and external momenta) UV
divergences survive. This fact implies nontrivial relations between various $%
\gamma _{j}^{2-\mathrm{loop}}$, $C_{j}^{(i),~1-\mathrm{loop}}$ and $\gamma
_{-1}^{(i),~1-\mathrm{loop}}$ which can be used either in order to simplify
the above contributions or as a nontrivial check of the explicit
calculations. Cancellation of the explicitly $\mu -$independent nonlocal $%
O(\varepsilon ^{-1})$ terms in the sum $\gamma _{L}^{2-\mathrm{loop}}+\gamma
_{CT}^{1-\mathrm{loop}}$needs (note that $C_{-1}^{(i),~1-\mathrm{loop}}$ has
to be local because it corresponds to the UV divergence of the one-loop
graph)
\begin{equation}
\gamma _{-1}^{2-\mathrm{loop}}-\sum_{i}C_{0}^{(i),~1-\mathrm{loop}}\gamma
_{-1}^{(i),~1-\mathrm{loop}}=\left( \gamma _{-1}^{2-\mathrm{loop}}\right)
_{l}  \label{nonloc1}
\end{equation}%
where $\left( \gamma _{-1}^{2-\mathrm{loop}}\right) _{l}$ is
local. Similarly, cancellation of the nonlocal $O(\varepsilon
^{-1})$ terms proportional to $\ln \left( \mu ^{2}/m^{2}\right) $
requires
\begin{equation}
2\gamma _{-2}^{2-\mathrm{loop}}-\sum_{i}\gamma _{-1}^{(i),~1-\mathrm{loop}%
}C_{-1}^{(i),~1-\mathrm{loop}}=0.  \label{nonloc2}
\end{equation}%
In fact these relations are valid also graph by graph. \ Introducing RG
invariant counterterm couplings $\overline{x}_{i}$ according to (\ref{Kbar}%
), namely
\begin{equation}
\overline{x}_{i}^{1-\mathrm{loop}}=\left( \frac{\mu }{m}\right)
^{-2\varepsilon }\left( x_{i}^{r}(\mu )^{1-\mathrm{loop}}-\frac{\gamma
_{-1}^{(i),~1-\mathrm{loop}}}{\varepsilon }\right) +\frac{\gamma
_{-1}^{(i),~1-\mathrm{loop}}}{\varepsilon }
\end{equation}%
we can write using (\ref{nonloc1}) and (\ref{nonloc2})
\begin{eqnarray}
\gamma _{L}^{2-\mathrm{loop}}+\gamma _{CT}^{1-\mathrm{loop}} &=&\mu
^{-4\varepsilon }m^{D}\left( \frac{\mu }{m}\right) ^{4\varepsilon }\left[ -%
\frac{\gamma _{-2}^{2-\mathrm{loop}}}{\varepsilon ^{2}}+\frac{%
\sum_{i}C_{-1}^{(i),~1-\mathrm{loop}}\overline{x}_{i}^{1-\mathrm{loop}%
}+\left( \gamma _{-1}^{2-\mathrm{loop}}\right) _{l}}{\varepsilon }\right.
\notag \\
&&\left. +\gamma _{0}^{2-\mathrm{loop}}+\sum_{i}\left( C_{0}^{(i),~1-\mathrm{%
loop}}\overline{x}_{i}^{1-\mathrm{loop}}-C_{1}^{(i),~1-\mathrm{loop}}\gamma
_{-1}^{(i),~1-\mathrm{loop}}\right) \right.  \notag \\
&&\left. \phantom{\frac{1}{1}}+O(\varepsilon )\right]
\end{eqnarray}%
and the remaining tree-level contribution coming from the two-loop
counterterm can be then simplified as
\begin{equation}
\gamma _{CT}^{\mathrm{tree}}=\mu ^{-4\varepsilon }m^{D}\left( \frac{\mu }{m}%
\right) ^{4\varepsilon }\left[ \overline{\xi }_{\gamma }+\frac{\gamma
_{-2}^{2-\mathrm{loop}}}{\varepsilon ^{2}}-\frac{\sum_{i}C_{-1}^{(i),~1-%
\mathrm{loop}}\overline{x}_{i}^{1-\mathrm{loop}}+\left( \gamma _{-1}^{2-%
\mathrm{loop}}\right) _{l}}{\varepsilon }\right] .  \label{2loop_tree_CT}
\end{equation}%
Here the UV finite and RG invariant\footnote{%
As in the previous subsection we require RG scale invariance order by order
in the loop expansion. We also tacitly assume that all the relevant
contributions are included in $\gamma$ which represents a RG invariant
physical observable.} local quantity $\overline{\xi }_{\gamma } $ is a
two-loop analog of $\overline{\chi }_{\gamma }$%

\begin{eqnarray}
\overline{\xi }_{\gamma } &=&\left( \frac{\mu }{m}\right) ^{-4\varepsilon }%
\left[ \xi _{\gamma }^{r}(\mu )+\frac{\gamma _{-2}^{2-\mathrm{loop}}}{%
\varepsilon ^{2}}-\frac{\left( \gamma _{-1}^{2-\mathrm{loop}}\right)
_{l}+\sum_{i}x_{i}^{r}(\mu )^{1-\mathrm{loop}}C_{-1}^{(i),~1-\mathrm{loop}}}{%
\varepsilon }\right]  \notag \\
&&-\frac{\gamma _{-2}^{2-\mathrm{loop}}}{\varepsilon ^{2}}+\frac{\left(
\gamma _{-1}^{2-\mathrm{loop}}\right) _{l}+\sum_{i}\overline{x}_{i}^{1-%
\mathrm{loop}}C_{-1}^{(i),~1-\mathrm{loop}}}{\varepsilon }  \notag \\
&=&\xi _{\gamma }^{r}(\mu )+\ln \left( \frac{\mu ^{2}}{m^{2}}\right) \left[
2\left( \gamma _{-1}^{2-\mathrm{loop}}\right) _{l}+\sum_{i}x_{i}^{r}(\mu
)^{1-\mathrm{loop}}C_{-1}^{(i),~1-\mathrm{loop}}\right]  \notag \\
&&+\gamma _{-2}^{2-\mathrm{loop}}\ln ^{2}\left( \frac{\mu ^{2}}{m^{2}}%
\right) +O(\varepsilon ).
\end{eqnarray}%
This implies the following explicit dependence of $\xi _{\gamma }^{r}(\mu )$
on the renormalization scale $\mu $:
\begin{equation}\begin{split}
\xi _{\gamma }^{r}(\mu ) &=\overline{\xi }_{\gamma }-\left[ 2\left( \gamma
_{-1}^{2-\mathrm{loop}}\right) _{l}+\sum_{i}\overline{x}_{i}^{1-\mathrm{loop}%
}C_{-1}^{(i),~1-\mathrm{loop}}\right] \ln \left( \frac{\mu ^{2}}{m^{2}}%
\right)  \label{xi_run} \\
&+\gamma _{-2}^{2-\mathrm{loop}}\ln ^{2}\left( \frac{\mu ^{2}}{m^{2}}%
\right) +O(\varepsilon ).
\end{split}\end{equation}%
As a result we get the following simple and manifestly RG invariant form of
the total two-loop contribution $\gamma ^{2-\mathrm{loop}}$ to $\gamma $:
\begin{equation}
\gamma ^{2-\mathrm{loop}}=m^{D}\left[ \gamma _{0}^{2-\mathrm{loop}%
}+\sum_{i}\left( C_{0}^{(i),~1-\mathrm{loop}}\overline{x}_{i}^{1-\mathrm{loop%
}}-C_{1}^{(i),~1-\mathrm{loop}}\gamma _{-1}^{(i),~1-\mathrm{loop}}\right) +%
\overline{\xi }_{\gamma }+O(\varepsilon )\right] .
\end{equation}

\subsection{Treatment of IR divergences\label{IR_subsection}}

The presence of the IR divergences complicates the above simple picture a
bit. After the UV divergences are subtracted, additional divergent terms
survive, generally both in $\gamma _{L}^{2-\mathrm{loop}}$and $\gamma
_{CT}^{1-\mathrm{loop}}$, namely
\begin{equation}
\gamma _{L,IR}^{2-\mathrm{loop}}=\mu ^{-4\varepsilon }m^{D}\left( \frac{\mu
}{m}\right) ^{4\varepsilon }\left( \frac{\gamma _{-2,IR}^{2-\mathrm{loop}}}{%
\varepsilon ^{2}}+\frac{\gamma _{-1,IR}^{2-\mathrm{loop}}}{\varepsilon }%
\right)
\end{equation}%
and
\begin{eqnarray}
\gamma _{CT,IR}^{1-\mathrm{loop}} &=&\mu ^{-4\varepsilon }m^{D}\left( \frac{%
\mu }{m}\right) ^{2\varepsilon }\sum_{i}\frac{C_{-1,IR}^{(i),~1-\mathrm{loop}%
}}{\varepsilon }\left( x_{i}^{r}(\mu )^{1-\mathrm{loop}}-\frac{\gamma
_{-1}^{(i),~1-\mathrm{loop}}}{\varepsilon }\right)  \notag \\
&=&\mu ^{-4\varepsilon }m^{D}\left( \frac{\mu }{m}\right) ^{4\varepsilon
}\sum_{i}\frac{C_{-1,IR}^{(i),~1-\mathrm{loop}}}{\varepsilon }\left(
\overline{x}_{i}^{1-\mathrm{loop}}-\frac{\gamma _{-1}^{(i),~1-\mathrm{loop}}%
}{\varepsilon }\right) .
\end{eqnarray}%
However, in the sum of these two contributions, the $O(\varepsilon ^{-2})$
part has to vanish. This gives another useful relation which can be used as
nontrivial check of the explicit calculation, namely
\begin{equation}
\gamma _{-2,IR}^{2-\mathrm{loop}}=\sum_{i}C_{-1,IR}^{(i),~1-\mathrm{loop}%
}\gamma _{-1}^{(i),~1-\mathrm{loop}}  \label{IR_relation1}
\end{equation}%
and we get for the remaining IR divergent part
\begin{equation}
\gamma _{IR}^{2-\mathrm{loop}}=\mu ^{-4\varepsilon }m^{D}\left( \frac{\mu }{m%
}\right) ^{4\varepsilon }\frac{\gamma _{-1,IR}^{2-\mathrm{loop}%
}+\sum_{i}C_{-1,IR}^{(i),~1-\mathrm{loop}}\overline{x}_{i}^{1-\mathrm{loop}}%
}{\varepsilon }.  \label{gamma_2loop_IR}
\end{equation}

In our case the only one one-loop counterterm graph with IR
divergence is that corresponding to the counterterm of the $\pi
e^{+}e^{-}$ vertex graph
(let us denote the corresponding $C_{-1,IR}^{(i),~1-\mathrm{loop}}$ as $%
C_{-1,IR}^{(1),~1-\mathrm{loop}}$ in what follows) and in (\ref%
{gamma_2loop_IR}) the only relevant $\overline{x}_{i}^{1-\mathrm{loop}}$ we
denote as $\overline{\chi }$. The IR divergences in $|\gamma |^{2}$ where
\begin{equation}
|\gamma |^{2}=|\gamma ^{1-\mathrm{loop}}|^{2}+\gamma ^{2-\mathrm{loop}%
}\left( \gamma ^{1-\mathrm{loop}}\right) ^{\ast }+\gamma ^{1-\mathrm{loop}%
}\left( \gamma ^{2-\mathrm{loop}}\right) ^{\ast }
\end{equation}%
can be cancelled to the given order including  the soft
bremsstrahlung contribution $\Delta _{BS}$ into the inclusive
decay rate, where
\begin{equation}
\Delta _{BS}=|\gamma ^{1-\mathrm{loop}}|^{2}I_{BS}
\end{equation}%
and where $I_{BS}$ is a dimensionally regularized phase space bremsstrahlung
integral (see Section \ref{BS_section}) with a general structure
\begin{equation}
I_{BS}=\mu ^{-2\varepsilon }\left( \frac{\mu }{m}\right) ^{2\varepsilon
}\left( \frac{I_{-1}}{\varepsilon }+I_{0}+O(\varepsilon )\right) .
\label{I_BS_general}
\end{equation}%
It holds schematically
\begin{equation}
\gamma _{IR}^{2-\mathrm{loop}}\left( \gamma ^{1-\mathrm{loop}}\right) ^{\ast
}+\gamma ^{1-\mathrm{loop}}\left( \gamma _{IR}^{2-\mathrm{loop}}\right)
^{\ast }+|\gamma ^{1-\mathrm{loop}}|^{2}I_{BS}=O(\varepsilon ^{0}).
\label{BS_formula}
\end{equation}%
Inserting (\ref{1loop_short}) and (\ref{gamma_2loop_IR}) into (\ref%
{BS_formula}) and collecting the coefficients at various powers of $%
\overline{\chi }$ in the $O(\varepsilon ^{-1})$ term gives the following
conditions
\begin{equation}
I_{-1}+2\mathrm{Re}(C_{-1,IR}^{(1),~1-\mathrm{loop}})=0  \label{IR_relation2}
\end{equation}%
and
\begin{eqnarray}
\mathrm{Re}(\gamma _{0}^{1-\mathrm{loop}})I_{-1}+\mathrm{Re}(\gamma _{0}^{1-%
\mathrm{loop}}C_{-1,IR}^{(1),~1-\mathrm{loop}\ast })+\mathrm{Re}(\gamma
_{-1,IR}^{2-\mathrm{loop}}) &=&0 \\
\notag \\
|\gamma _{0}^{1-\mathrm{loop}}|^{2}I_{-1}+2\mathrm{Re}(\gamma _{0}^{1-%
\mathrm{loop}}\gamma _{-1,IR}^{2-\mathrm{loop}\ast }) &=&0.
\end{eqnarray}%
The latter two relations can be rewritten with help of (\ref{IR_relation2})
as
\begin{equation}
\gamma _{-1,IR}^{2-\mathrm{loop}}=C_{-1,IR}^{(1),~1-\mathrm{loop}}\gamma
_{0}^{1-\mathrm{loop}}  \label{IR_relation3}
\end{equation}%
which represents another relation which can be used as a check of the
explicit results. Using these relations, we get finally
\begin{eqnarray}
|\gamma |^{2}+\Delta _{BS} &=&\left\vert \gamma _{0}^{1-\mathrm{loop}}+%
\overline{\chi }\right\vert ^{2}\left( 1+I_{0}\right) -2\mathrm{Re}\left[
\left( \gamma _{0}^{1-\mathrm{loop}}+\overline{\chi }\right) ^{\ast }\gamma
_{1}^{1-\mathrm{loop}}C_{-1,IR}^{(1),~1-\mathrm{loop}}\right]  \notag \\
&&+2\mathrm{Re}\Bigg\{ \Big( \gamma _{0}^{1-\mathrm{loop}}+\overline{\chi }%
\Big) ^{\ast }  \notag \\
&& \times \Big[ \gamma _{0}^{2-\mathrm{loop}}+\overline{\xi }%
+\sum_{i}\left( C_{0}^{(i),~1-\mathrm{loop}}\overline{x}_{i}^{1-\mathrm{loop}%
}-C_{1}^{(i),~1-\mathrm{loop}}\gamma _{-1}^{(i),~1-\mathrm{loop}}\right) %
\Big] \Bigg\}  \notag \\
&&  \label{general_form_P}
\end{eqnarray}%
which is manifestly independent on the RG scale $\mu $.

\section{The one-loop graphs}

In this section we summarize the results for the one-loop graphs which are
relevant for the full $O(\alpha ^{3}p^{2})$ two-loop calculation. First we
will present \ the one-loop contribution to the form factor $%
P(s,m^{2},m^{2}) $ in more detail including also the order $O(\varepsilon )$
which will be necessary for treating the IR divergences. We will also
discuss the one-loop UV divergent subgraphs of the two-loop graphs and
one-loop graphs with one counterterm vertex. We will explicitly point out
the individual orders in $\varepsilon $ according to the general structure
discussed in the previous section. In order to simplify the results and to
avoid some repeating multiplicative factors, we will present all the results
in terms of the re-scaled form factor $\gamma (z)$ defined as
\begin{equation}
\gamma (s,m^{2},m^{2})\equiv 2\frac{F_{\pi }}{m}\left( \frac{\pi }{\alpha }%
\right) ^{2}P(s,m^{2},m^{2}).  \label{gamma_definition}
\end{equation}

\subsection{The leading order amplitude revisited\label{LO_subsection}}

As we have seen from the general formula (\ref{general_form_P}), we need
more detailed information on $\varepsilon-$ expansion of the LO amplitude $%
\gamma ^{\mathrm{1-loop}}$ (namely the $O(\varepsilon )$ term). In order to
be consistent with the two-loop calculations it is also convenient to
rewrite $\gamma ^{\mathrm{1-loop}}$ in terms of the Harmonic Polylogarithms
of Remiddi and~Vermaseren \cite{Remiddi:1999ew}. Let us note that in the
physical region the kinematical variable $x$ is negative (cf. (\ref%
{x_definition})), while the two-loop integrals are originally calculated in
their analyticity region corresponding to $0<x<1$ and only then analytically
continued (cf. Appendix \ref{MI appendix} ). This continuation apart from
generating imaginary parts brings about also additional minus signs into the
arguments of Harmonic Polylogarithms and therefore the most convenient way
how to present the result in the physical region is to use as an argument of
these functions a new variable%
\begin{equation}
z=-x=\frac{1-\beta }{1+\beta }>0.
\end{equation}%
In order to further simplify the long expressions we use the notation%
\begin{equation}
\overline{\gamma }\equiv \gamma -\ln 4\pi
\end{equation}%
and rewrite the rational function multiplying the Harmonic Polylogarithms $%
H(a_{1},\ldots ,a_{n};z)$ in terms of the variable $\beta $.

As a result we get then for the one-loop contribution $\gamma ^{\mathrm{%
1-loop}}(z)$ (cf. Section \ref{renormalization_section} )
\begin{equation}
\gamma ^{\mathrm{1-loop}}(z)=\mu ^{-2\varepsilon }\left( \frac{\mu }{m}%
\right) ^{2\varepsilon }\left( \gamma _{0}^{\mathrm{1-loop}}(z)+\overline{%
\chi }+\varepsilon \gamma _{1}^{\mathrm{1-loop}}(z)+O(\varepsilon
^{2})\right) .  \label{P_1loop}
\end{equation}%
In the above formula (\ref{P_1loop}) we have for $0<z<1$
\begin{eqnarray}
\gamma _{0}^{\mathrm{1-loop}} &=&\frac{1}{2\beta }\left( H(0,0;z)+\mathrm{i}%
\pi H(0;z)-2H(-2;z)+\frac{\pi ^{2}}{6}\right) -\frac{5}{2}+\frac{3}{2}%
\overline{\gamma }
\end{eqnarray}
and the $O(\varepsilon)$ term is

\begin{eqnarray}
\gamma _{1}^{\mathrm{1-loop}} &=&\frac{1}{2}\left[ -\frac{1}{\beta
}\left(
\frac{\pi ^{2}}{3}H(1;z)+\frac{\pi ^{2}}{3}H(-1;z)+2H(-3;z)+\frac{2\pi ^{2}}{%
3}H(0;z)\right. \right.  \notag \\
&&\left. \left. -4H(1,-2;z)-4H(-1,-2;z)-4H(-2,-1;z)\right. \right.  \notag \\
&&\left. \left. +2H(-2,0;z)+2H(1,0,0;z)+2H(-1,0,0;z)-H(0,0,0;z)\right.
\right.  \notag \\
&&\left. \left. +\mathrm{i}\pi \left( 2H(-2;z)+2H(1,0;z)+2H(-1,0;z)-H(0,0;z)+%
\frac{\pi ^{2}}{3}\right) \right. \right.  \notag \\
&&\left. \left. -\overline{\gamma }\left( 2H(-2;z)-H(0,0;z)-\mathrm{i}\pi
H(0;z)-\frac{\pi ^{2}}{6}\right) -5\zeta (3)\right) \right.  \notag \\
&&\left. +5\overline{\gamma }-\frac{3}{2}\overline{\gamma }^{2}-\frac{\pi
^{2}}{4}-9\right]  \notag \\
&&
\end{eqnarray}
where
\begin{eqnarray}
 \overline{\chi } &=&\left( \frac{\mu }{m}\right)
^{-2\varepsilon }\left(
\chi ^{r}(\mu )+\frac{3}{2}\left( \frac{1}{\varepsilon }-\overline{\gamma }%
\right) \right) -\frac{3}{2}\frac{1}{\varepsilon }
=\chi ^{r}(\mu )-\frac{3}{2}\overline{\gamma }-\frac{3}{2}\ln \left( \frac{%
\mu ^{2}}{m^{2}}\right) +O(\varepsilon ).  \notag \\
&&
\end{eqnarray}%
For further convenience let us also mention explicitly the UV divergent part
\begin{equation}
\gamma _{-1}^{\mathrm{1-loop}}=-\frac{3}{2}.  \label{P_1loop_UVdiv}
\end{equation}%
The formula (\ref{ChPT_LO}) can be then easily reconstructed as
\begin{equation}
P^{\chi LO}(m^{2},m^{2},s)=\frac{1}{2}\left( \frac{\alpha }{\pi }\right) ^{2}%
\frac{m}{F_{\pi }}\lim_{\varepsilon \rightarrow 0}\gamma ^{\mathrm{1-loop}}
\end{equation}%
and we can interpret the renormalization scale independent constant $%
\overline{\chi }$ in terms of the renormalized constant $\chi ^{r}\left( \mu
\right) $ at scale $\mu =m$ as
\begin{equation}
\overline{\chi }=\chi ^{r}(m)-\frac{3}{2}\overline{\gamma },
\end{equation}%
numerically
\begin{equation}
\overline{\chi }=-16.8\pm 0.9.  \label{chibar_num}
\end{equation}

\subsection{The one loop counterterms\label{one_loop_subsection_ct}}

In this subsection we summarize the counterterms needed for renormalization
of the one-loop sub-divergences of the two-loop graphs. We can write the
relevant counterterm Lagrangian in the general form either in terms of the
renormalized couplings (finite parts of the counterterms)
\begin{equation}
\mathcal{L}=\mu ^{-2\varepsilon }\left( \frac{\alpha }{\pi }\right) \left[
\left( x_{6}^{r}(\mu )-\frac{1}{4\varepsilon }\right) \mathrm{i}\overline{e}%
\gamma ^{\mu }D_{\mu }e+\left( x_{7}^{r}(\mu )+\frac{1}{\varepsilon }\right)
m\overline{e}e+\left( x_{8}^{r}(\mu )+\frac{1}{3\varepsilon }\right) \frac{1%
}{4}F^{\mu \nu }F_{\mu \nu }\right]
\end{equation}%
(here $D=\partial +\mathrm{i}eA$) or in terms of the renormalization scale
invariant constants (cf. (\ref{Kbar}))
\begin{equation}
\mathcal{L}=\mu ^{-2\varepsilon }\left( \frac{\alpha }{\pi }\right) \left(
\frac{\mu }{m}\right) ^{2\varepsilon }\left[ \left( \overline{x}_{6}-\frac{1%
}{4\varepsilon }\right) \mathrm{i}\overline{e}\gamma ^{\mu }D_{\mu }e+\left(
\overline{x}_{7}+\frac{1}{\varepsilon }\right) m\overline{e}e+\left(
\overline{x}_{8}+\frac{1}{3\varepsilon }\right) \frac{1}{4}F^{\mu \nu
}F_{\mu \nu }\right] .
\end{equation}%
These counterterms contribute to the electron self-energy $\Sigma (p)$, the
vertex function $\Gamma (p,p^{^{\prime }})$ and vacuum polarization $\Pi (p)$
which are related to the physical electron mass and physical charge, namely
\begin{eqnarray}
m_{phys} &=&m+\Sigma (m_{phys}) \\
e_{phys} &=&e\left[ 1+\frac{1}{2}\Pi (0)\right] .
\end{eqnarray}%
Within our regularization scheme and at one-loop level we get
\begin{eqnarray}
\Sigma (m) &=&m\left( \frac{\alpha }{\pi }\right) \mu ^{-2\varepsilon
}\left( \frac{\mu }{m}\right) ^{2\varepsilon }\left[ -\frac{3}{4}\left(
\overline{\gamma }-\frac{5}{3}\right) -\overline{x}_{6}-\overline{x}_{7}%
\right] \\
\Pi (0) &=&\left( \frac{\alpha }{\pi }\right) \mu ^{-2\varepsilon }\left(
\frac{\mu }{m}\right) ^{2\varepsilon }\left[ \frac{1}{3}\overline{\gamma }+%
\overline{x}_{8}\right] ,
\end{eqnarray}%
therefore adjusting
\begin{eqnarray}
\overline{x}_{6}+\overline{x}_{7} &=&-\frac{3}{4}\left( \overline{\gamma }-%
\frac{5}{3}\right) \\
\overline{x}_{8} &=&-\frac{1}{3}\overline{\gamma }
\end{eqnarray}%
(\emph{i.e.} choosing the on mass shell renormalization scheme) we ensure,
that the original parameters $m$ and $e$ coincide with the physical mass and
charge. The parameter $\overline{x}_{6}$ is connected with the electron wave
function renormalization, namely
\begin{equation}
Z_{e}^{-1}=1-\frac{\partial \Sigma (p)}{\partial {\slashed{p}}}
\end{equation}%
where in our scheme at one loop (here the $\varepsilon $ pole corresponds to
the IR divergence)
\begin{equation}
\frac{\partial \Sigma (p)}{\partial \slashed{p}}=\left( \frac{\alpha }{\pi }%
\right) \mu ^{-2\varepsilon }\left( \frac{\mu }{m}\right) ^{2\varepsilon
}\left( -\frac{1}{2\varepsilon }+\frac{3}{4}\left( \overline{\gamma }-\frac{5%
}{3}\right) -\overline{x}_{6}\right) .  \label{Z_factor}
\end{equation}%
Therefore the parameter $\overline{x}_{6}$ is not physical and has to cancel
in the physical amplitudes. We can conveniently set
\begin{equation}
\overline{x}_{6}=\frac{3}{4}\left( \overline{\gamma }-\frac{5}{3}\right) ,
\end{equation}%
then the only effect of the electron wave function renormalization
is the IR pole which is cancelled by the analogous pole in the
diagonal part of the bremsstrahlung integrals (see Section
\ref{BS_section}). We can therefore forget the electron wave
function renormalization completely provided we simultaneously
throw away the IR divergent part of the diagonal bremsstrahlung
integrals.

\subsection{One-loop graphs with counterterms\label{one_loop_subsection}}

There are six types of the one-loop graphs with $O(\alpha p^{2})$ and $%
O(\alpha ^{2}p^{2})$ counterterms which are depicted as $(1)-(6)$ in Fig. %
\ref{NLO_ChPT_1loop}. Let us denote their individual contributions as $%
\gamma _{\mathrm{CT~}}^{(i),~\mathrm{1-loop}}$. As a consequence of the
symmetries of the graphs we have the relations
\begin{eqnarray}
\gamma _{\mathrm{CT~}}^{(2),~\mathrm{1-loop}} &=&\gamma _{\mathrm{CT~}%
}^{(4),~\mathrm{1-loop}}\equiv \gamma _{\mathrm{CT~}}^{(\Gamma ),~\mathrm{%
1-loop}} \\
\gamma _{\mathrm{CT~}}^{(5),~\mathrm{1-loop}} &=&\gamma _{\mathrm{CT~}%
}^{(6),~\mathrm{1-loop}}\equiv \gamma _{\mathrm{CT~}}^{(\Pi ),~\mathrm{1-loop%
}}
\end{eqnarray}%
and as a result of straightforward algebra we find that $\gamma _{\mathrm{CT~%
}}^{\mathrm{1-loop}(\Gamma )}$ and $\gamma _{\mathrm{CT~}}^{\mathrm{1-loop}%
(\Pi )}$ are simply related to $\gamma ^{\mathrm{1-loop}}$ (cf. (\ref%
{P_1loop}) and (\ref{P_1loop_UVdiv})), namely%
\begin{eqnarray}
\gamma _{\mathrm{CT~}}^{(\Gamma ),~\mathrm{1-loop}} &=&\mu ^{-4\varepsilon
}\left( \frac{\mu }{m}\right) ^{4\varepsilon }\left( \frac{\alpha }{\pi }%
\right) \left( \overline{x}_{6}-\frac{1}{4\varepsilon }\right) \left( \frac{%
\gamma _{-1}^{\mathrm{1-loop}}}{\varepsilon }+\gamma _{0}^{\mathrm{1-loop}%
}+\varepsilon \gamma _{1}^{\mathrm{1-loop}}+O(\varepsilon ^{2})\right) \notag \\
\gamma _{\mathrm{CT~}}^{(\Pi ),~\mathrm{1-loop}} &=&\mu ^{-4\varepsilon
}\left( \frac{\mu }{m}\right) ^{4\varepsilon }\left( \frac{\alpha }{\pi }%
\right) \left( \overline{x}_{8}+\frac{1}{3\varepsilon }\right) \left( \frac{%
\gamma _{-1}^{\mathrm{1-loop}}}{\varepsilon }+\gamma _{0}^{\mathrm{1-loop}%
}+\varepsilon \gamma _{1}^{\mathrm{1-loop}}+O(\varepsilon ^{2})\right).\notag \\
&&
\end{eqnarray}%
Also $\gamma _{\mathrm{CT~}}^{(3),~\mathrm{1-loop}}$ can be rewritten in the
form%
\begin{eqnarray}
\gamma _{\mathrm{CT~}}^{(3),~\mathrm{1-loop}} &=&-\mu ^{-4\varepsilon
}\left( \frac{\mu }{m}\right) ^{4\varepsilon }\left( \frac{\alpha }{\pi }%
\right) \left( \overline{x}_{6}-\frac{1}{4\varepsilon }\right) \left( \frac{%
\gamma _{-1}^{\mathrm{1-loop}}}{\varepsilon }+\gamma _{0}^{\mathrm{1-loop}%
}+\varepsilon \gamma _{1}^{\mathrm{1-loop}}+O(\varepsilon ^{2})\right)\notag \\
&&+\gamma _{\mathrm{CT~}}^{(m),~\mathrm{1-loop}}
\end{eqnarray}%
where $\gamma _{\mathrm{CT~}}^{(m),~\mathrm{1-loop}}$ formally corresponds
to the topology $(3)$ of Fig. \ref{NLO_ChPT_1loop}, but now with the
insertion only of the modified mass counterterm
\begin{equation}
\mathcal{L}_{\mathrm{mod}}^{(m)}=\mu ^{-2\varepsilon }\left( \frac{\alpha }{%
\pi }\right) \left( \frac{\mu }{m}\right) ^{4\varepsilon }\left( \overline{x}%
_{6}+\overline{x}_{7}+\frac{3}{4\varepsilon }\right) m\overline{e}e.
\end{equation}

The only nontrivial one-loop graphs with counterterms are therefore $\gamma
_{\mathrm{CT~}}^{(1),~\mathrm{1-loop}}$ and $\gamma _{\mathrm{CT~}}^{(m),~%
\mathrm{1-loop}}$. \ The first one can be written as%
\begin{equation}\begin{split}
\gamma _{\mathrm{CT~}}^{(1),~\mathrm{1-loop}} &=\mu ^{-4\varepsilon }\left(
\frac{\mu }{m}\right) ^{4\varepsilon }\left( \frac{\alpha }{\pi }\right)
\left( \overline{\chi }+\frac{3}{2}\frac{1}{\varepsilon }\right) \\
&\times \left( \frac{C_{-1,\mathrm{UV}}^{(1),~\mathrm{1-loop}}}{\varepsilon
}+\frac{C_{-1,\mathrm{IR}}^{(1),~\mathrm{1-loop}}}{\varepsilon }+C_{0}^{(1),~%
\mathrm{1-loop}}+\varepsilon C_{1}^{(1),~\mathrm{1-loop}}+O(\varepsilon
^{2})\right)
\end{split}\end{equation}%
where we have explicitly pointed out the IR divergent part. The individual
orders of the $\varepsilon $ expansion are then
\begin{eqnarray}
C_{-1,\mathrm{UV}}^{(1),~\mathrm{1-loop}} &=&\frac{1}{4}  \notag \\
&& \\
C_{-1,\mathrm{IR}}^{(1),~\mathrm{1-loop}} &=&-\frac{1}{4}\left( \beta +\frac{%
1}{\beta }\right) \left( H(0;z)+\mathrm{i}\pi \right)  \notag \\
&& \\
C_{0}^{(1),~\mathrm{1-loop}} &=&-\frac{1}{4}\left( \beta +\frac{1}{\beta }%
\right) \left[ H(0,0;z)+2H(1,0;z)-\frac{2}{3}\pi ^{2}-\overline{\gamma }%
H(0;z)\right.  \notag \\
&&\left. \phantom{\frac{1}{2}}+\mathrm{i}\pi \left( H(0;z)+2H(1;z)\right) -%
\mathrm{i}\pi \overline{\gamma }\right] -\frac{1}{4}(\overline{\gamma }-3) \\
&& \notag \\
C_{1}^{(1),~\mathrm{1-loop}} &=&-\frac{1}{2}\left( \beta +\frac{1}{\beta }%
\right) \left[ H(2,0;z)+\frac{1}{2}H(0,0,0;z)+H(1,0,0;z)+2H(1,1,0;z)\right.
\notag \\
&&\left. -\frac{1}{2}\overline{\gamma }(H(0,0;z)+2H(1,0;z))\right.  \notag \\
&&\left. +\frac{1}{4}\overline{\gamma }^{2}H(0;z)-\frac{7\pi ^{2}}{24}H(0;z)-%
\frac{2\pi ^{2}}{3}H(1;z)\right.  \notag \\
&&\left. +\mathrm{i}\pi \left( H(2;z)+\frac{1}{2}H(0,0;z)+H(1,0;z)+2H(1,1;z)%
\right) \right.  \notag \\
&&\left. -\frac{\mathrm{i}\pi }{2}\overline{\gamma }(H(0;z)+2H(1;z))\right.
\notag \\
&&\left. +\frac{\mathrm{i}\pi }{4}\overline{\gamma }^{2}-\frac{\mathrm{i}\pi
^{3}}{8}-\zeta (3)+\frac{\pi ^{2}}{3}\overline{\gamma }\right]  \notag \\
&&+\frac{1}{8}\overline{\gamma }^{2}-\frac{3}{4}\overline{\gamma }+\frac{\pi
^{2}}{48}+\frac{7}{4}.
\end{eqnarray}

For $\gamma _{\mathrm{CT~}}^{(m),~\mathrm{1-loop}}$ we get%
\begin{equation}\begin{split}
\gamma _{\mathrm{CT~}}^{(m),~\mathrm{1-loop}} &=-\mu ^{-4\varepsilon
}\left( \frac{\mu }{m}\right) ^{4\varepsilon }\left( \frac{\alpha }{\pi }%
\right) \left( \overline{x}_{6}+\overline{x}_{7}+\frac{3}{4\varepsilon }%
\right) \\
&\times \left( \frac{C_{-1}^{(m),~\mathrm{1-loop}}}{\varepsilon }%
+C_{0}^{(m),~\mathrm{1-loop}}+\varepsilon C_{1}^{(m),~\mathrm{1-loop}%
}+O(\varepsilon ^{2})\right)
\end{split}\end{equation}%
where%
\begin{eqnarray}
C_{-1}^{(m),~\mathrm{1-loop}} &=&0  \notag \\
&& \\
C_{0}^{(m),~\mathrm{1-loop}} &=&\frac{1}{2\beta }\left[ H(0,0;z)+\mathrm{i}%
\pi H(0;z)-2H(-2;z)+\frac{\pi ^{2}}{6}\right]  \notag \\
&&-4H(-1;z)+2H(0;z)+2\mathrm{i}\pi +2  \notag \\
&& \\
C_{1}^{(m),~\mathrm{1-loop}} &=&\frac{1}{2\beta }\left[
-2H(-3;z)+4H(-2,z)+4H(-2,-1;z)\phantom{\frac{1}{2}}\right.  \notag \\
&&\left. +4H(1,-2;z)+4H(-1,-2;z)\right.  \notag \\
&&\left. -2H(0,0;z)-2H(-1,0,0;z)+H(0,0,0;z)\right.  \notag \\
&&\left. -2H(1,0,0;z)-2H(-2,0;z)\right.  \notag \\
&&\left. -\mathrm{i}\pi
(2H(0;z)+2H(-1,0;z)-H(0,0;z)+2H(1,0;z)+2H(-2;z))\right.  \notag \\
&&\left. -\overline{\gamma }\left( H(0,0;z)-2H(-2;z)+\mathrm{i}\pi H(0;z)+%
\frac{\pi ^{2}}{6}\right) \right.  \notag \\
&&\left. -\frac{\pi ^{2}}{3}(H(1;z)+H(-1;z)+2H(0;z)+1)+5\zeta (3)-\frac{%
\mathrm{i}\pi ^{3}}{3}\phantom{\frac{1}{2}}\right]  \notag \\
&&+2\left[ 4H(-1,-1;z)-2H(-2;z)-2H(-1,0;z)+H(0,0;z)\phantom{\frac{1}{2}}%
\right.  \notag \\
&&\left. +(\overline{\gamma }-1)(2H(-1;z)-H(0;z)-1)-\frac{2\pi ^{2}}{3}%
+1\right.  \notag \\
&&\left. -\mathrm{i}\pi (2H(-1;z)-H(0;z)+\overline{\gamma }-1)%
\phantom{\frac{1}{2}}\right].
\end{eqnarray}

\section{The two-loop graphs}

This section is devoted to the six $O(\alpha ^{3}p^{2})$ two-loop
graphs depicted in Fig. \ref{NLO_ChPT_2loop}. Let us denote their
contributions to
the re-scaled form factor $\gamma $ defined by (\ref{gamma_definition}) as $%
\gamma ^{(i),~\mathrm{2-loop}}$. Due to the symmetries of the graphs, we get
the relations
\begin{eqnarray}
\gamma ^{(2),~\mathrm{2-loop}} &=&\gamma ^{(4),~\mathrm{2-loop}}\equiv
\gamma ^{(\Gamma ),~\mathrm{2-loop}}  \\
\gamma ^{(5),~\mathrm{2-loop}} &=&\gamma ^{(6),~\mathrm{2-loop}}\equiv
\gamma ^{(\Pi ),~\mathrm{2-loop}}.  \label{2loop_symmetry}
\end{eqnarray}%
We have therefore only four independent two-loop contributions to the form
factor $\gamma $. For their determination we use the standard procedure
based on the reduction to the scalar Master Integrals (MI) and on
calculation of the latter by means of the differential equations technique.

\subsection{Reduction to Master Integrals}

\begin{figure}[t]
\begin{center}
\epsfig{width=0.70\textwidth,file=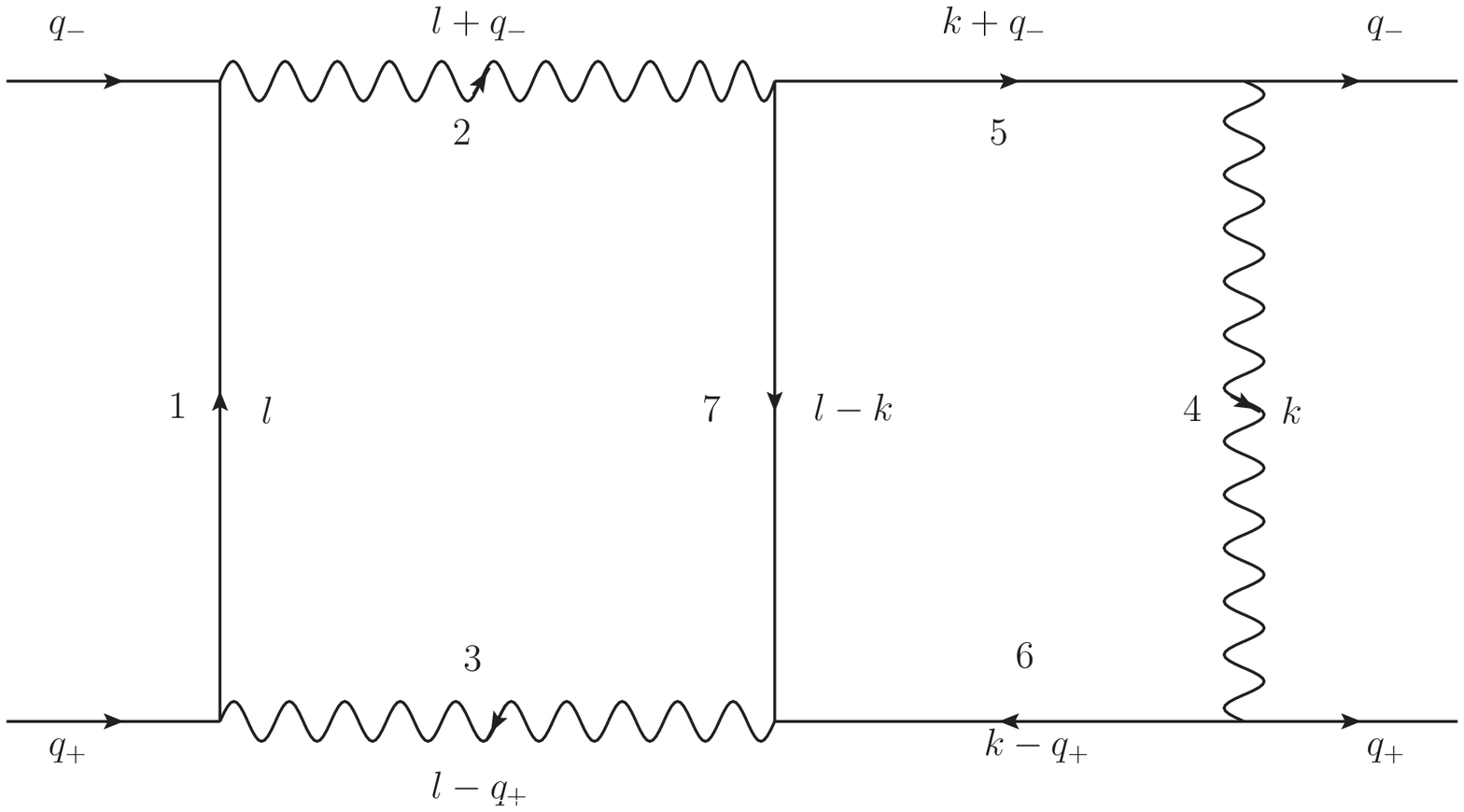}
\end{center}
\caption{The auxiliary diagram describing the momentum flow in the
scalar integrals $B(n_{1},\ldots ,n_{7})$. In this and the
following Figures \protect\ref{T2}.-\protect\ref{T5}., the full
internal lines correspond to the massive scalar propagators with
mass $m$ while wiggly lines stay for massless ones. The full
external lines carry the momenta on the electron/positron mass
shell and dashed external lines correspond to the momentum $Q$.}
\label{Aux}
\end{figure}

As a first step we define set of basic scalar integrals%
\begin{equation}
B(n_{1},\ldots ,n_{7})=\mu ^{4\varepsilon }\int \frac{\mathrm{d}^{d}k}{(2\pi
)^{d}}\frac{\mathrm{d}^{d}l}{(2\pi )^{d}}\prod_{i=1}^{7}\frac{1}{%
D_{i}(k,l)^{n_{i}}}
\end{equation}%
where\ $n_{i}$ are integers and $\{D_{i}(k,l)\}_{i=1}^{7}$ is a set of seven
independent propagator denominators, \ namely%
\begin{equation}\begin{split}
D_{1}(k,l) &=l^{2}-m^{2} \\
D_{2}(k,l) &=(l+q_{-})^{2} \\
D_{3}(k,l) &=(l-q_{+})^{2} \\
D_{4}(k,l) &=k^{2} \\
D_{5}(k,l) &=(k+q_{-})^{2}-m^{2} \\
D_{6}(k,l) &=(k-q_{+})^{2}-m^{2} \\
D_{7}(k,l) &=(l-k)^{2}-m^{2},
\end{split}\end{equation}%
which at the same time form the basis for the seven independent scalar
products build from the four-vectors $k$, $l$, $q_{+}$ and $q_{-}$. The
momentum flow in $B(n_{1},\ldots ,n_{7})$ corresponds to the auxiliary
diagram shown in Fig. \ref{Aux}. The integrals $B(n_{1},\ldots ,n_{7})$
depend besides the electron mass $m$ only on one independent scalar variable
$Q^{2}=(q_{+}+q_{-})^{2}$ and as a consequence of the symmetry properties of
the auxiliary diagram they satisfy the relation%
\begin{equation}
B(n_{1},n_{3},n_{2},n_{4},n_{6},n_{5},n_{7})=B(n_{1},n_{2},n_{3},n_{4},n_{5},n_{6},n_{7}).
\label{B_symmetry}
\end{equation}%
Note also, that some of $B(n_{1},\ldots ,n_{7})$ are identically equal to
zero. For instance whenever $n_{i}\leq 0$ for $i=1,2,7$ (or $i=5,6,7$)
simultaneously, we can factor out a massless tadpole and therefore $%
B(n_{1},\ldots ,n_{7})$ vanishes.

Using the projection (\ref{projection_P}) and expressing the
scalar products in the numerator of the integrands in terms of
$D_{i}(k,l)$ we get
schematically%
\begin{equation}
\gamma ^{(i),~\mathrm{2-loop}}=\mu ^{-4\varepsilon }\left( \frac{\alpha }{%
\pi }\right) \sum_{n_{1},\ldots ,n_{7}}c^{(i)}(n_{1},\ldots
,n_{7};y)B(n_{1},\ldots ,n_{7})  \label{2loop_reduction}
\end{equation}%
where $c^{(i)}(n_{1},\ldots ,n_{7};y)$ are known coefficients which depend
on the variable $y=Q^{2}/4m^{2}$ and electron mass $m$. The explicit form of
the reduction formulae (\ref{2loop_reduction}) can be found in the Appendix %
\ref{Reduction_appendix}. The relations (\ref{2loop_symmetry}) can be
explicitly verified for the right hand side of (\ref{2loop_reduction}) using
(\ref{B_symmetry}).

There are altogether 172 integrals $B(n_{1},\ldots ,n_{7})$
entering the sums in the reduction formulae
(\ref{2loop_reduction}), however, not all of
them are independent. In addition to the symmetry property (\ref{B_symmetry}%
) there are also additional relations based on the integration by parts
(IBP) \cite{Tkachov:1981wb,Chetyrkin:1981qh} and Lorentz invariance
identities (LI) \cite{Gehrmann:1999as}. The former relations are
consequences of the vanishing of the integral of the total divergence within
DR. In our case we get eight relations schematically written as
\begin{equation}
\int \frac{\mathrm{d}^{d}k}{(2\pi )^{d}}\frac{\mathrm{d}^{d}l}{(2\pi )^{d}}%
\left(
\begin{array}{c}
\frac{\partial }{\partial k^{\mu }} \\
\frac{\partial }{\partial l^{\mu }}%
\end{array}%
\right) \left(
\begin{array}{c}
k^{\mu } \\
l^{\mu } \\
q_{+}^{\mu } \\
q_{-}^{\mu }%
\end{array}%
\right) \left[ \prod_{i=1}^{7}\frac{1}{D_{i}(k,l)^{n_{i}}}\right] =0.
\label{IPB}
\end{equation}%
The remaining LI relations express the invariance of the scalar integrals $%
B(n_{1},\ldots ,n_{7})$ with respect to the Lorentz transformation of the
external momenta%
\begin{equation}
\left( q_{+}^{\mu }\frac{\partial }{\partial q_{+\nu }}-q_{+}^{\nu }\frac{%
\partial }{\partial q_{+\mu }}+q_{-}^{\mu }\frac{\partial }{\partial q_{-\nu
}}-q_{-}^{\nu }\frac{\partial }{\partial q_{-\mu }}\right) B(n_{1},\ldots
,n_{7})=0.  \label{LI}
\end{equation}%
The left hand sides of both (\ref{IPB}) and (\ref{LI}) (when contracted with
$q_{+}^{\mu }q_{-}^{\nu }$) can be expressed in terms of linear combinations
of $B(n_{1},\ldots ,n_{7})$ with various $n_{i}$ and with $y$ dependent
coefficients. The explicit form of the resulting relations is postponed to
Appendix \ref{IPB_appendix}, here we give only the LI identity as an
illustration:%
\begin{eqnarray}
&&\left[ \left( 1-2y\right) (n_{2}+n_{5})-n_{2}\mathbf{3}^{-}\mathbf{2}%
^{+}+4ym^{2}n_{2}\mathbf{2}^{+}+2yn_{2}\mathbf{1}^{-}\mathbf{2}^{+}-n_{5}%
\mathbf{6}^{-}\mathbf{5}^{+}+2yn_{5}\mathbf{4}^{-}\mathbf{5}^{+}\right.
\notag \\
&&\left. -(2\leftrightarrow 3,~5\leftrightarrow 6)\right] B(n_{1},\ldots
,n_{7})=0,  \label{LI1}
\end{eqnarray}%
where we have introduced the usual operators $\mathbf{j}^{\pm }$ ($%
j=1,\ldots ,7$) which act on $B(n_{1},\ldots ,n_{7})$ as%
\begin{equation}
\mathbf{j}^{\pm }B(n_{1},\ldots ,n_{j},\ldots n_{7})=B(n_{1},\ldots
,n_{j}\pm 1,\ldots n_{7}).  \label{jpm_operator}
\end{equation}

\begin{figure}[t]
\begin{center}
\epsfig{width=0.25\textwidth,file=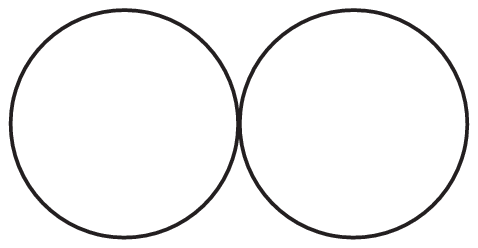}
\end{center}
\caption{The subset of MI with two propagators.} \label{T2}
\end{figure}
\begin{figure}[t]
\begin{center}
\epsfig{width=0.80\textwidth,file=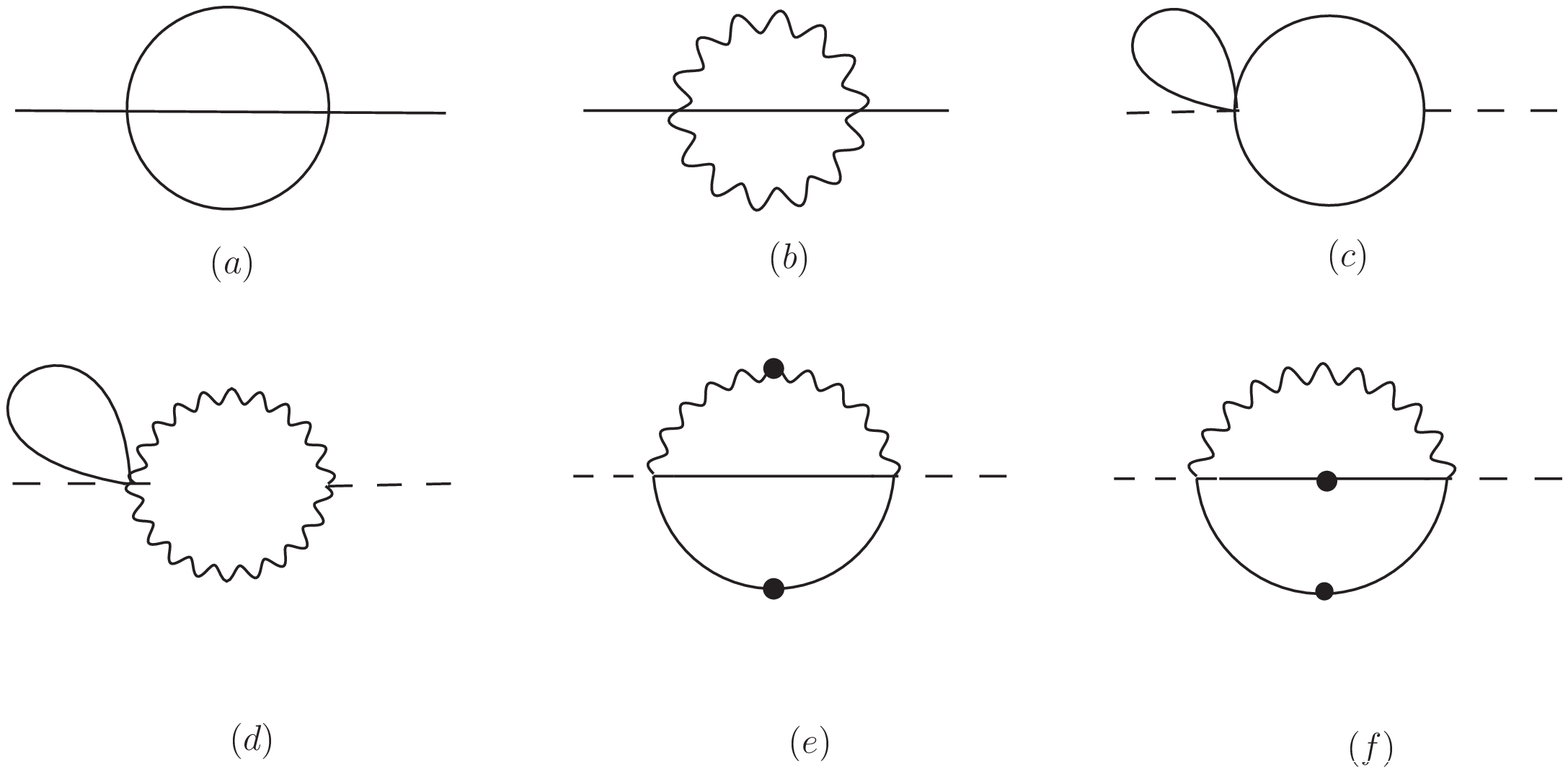}
\end{center}
\caption{The subset of MI with three propagators. There are five
types of different topologies, namely (a), (b), (c), (d) and (e,
f).} \label{T3}
\end{figure}

\begin{figure}[t]
\begin{center}
\epsfig{width=0.80\textwidth,file=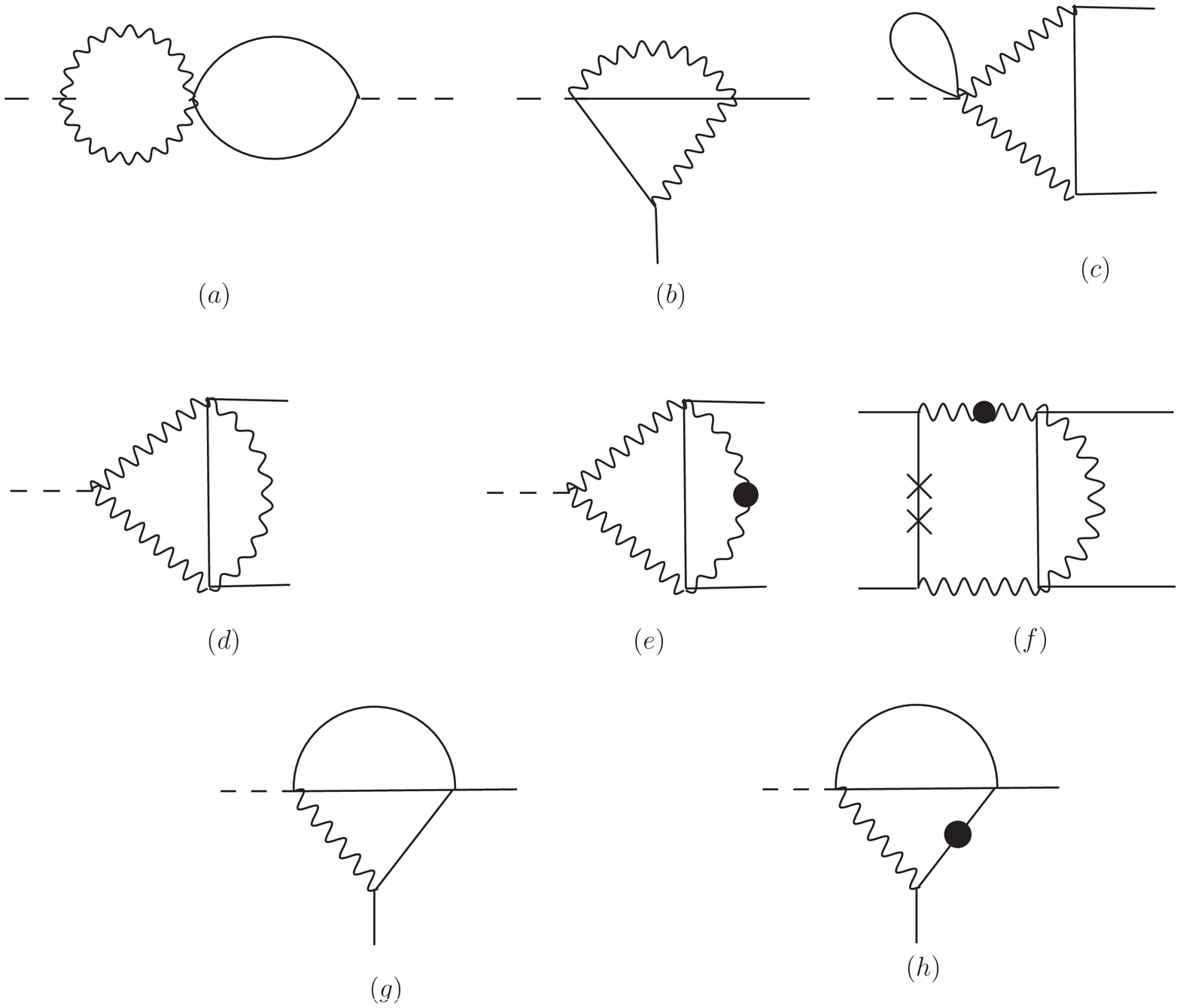}
\end{center}
\caption{The subset of MI with four propagators. There are five
types of different topologies, namely (a), (b), (c), (d, e, f) and
(g, h).} \label{T4}
\end{figure}

The above linear relations (\ref{IPB}) and (\ref{LI}) together with the
symmetry property (\ref{B_symmetry}) can be used as an input for the
Laporta-Remiddi algorithm \cite{Laporta:1996mq, Laporta:2001dd} which allows
further reduction of the number of the scalar integrals $B(n_{1},\ldots
,n_{7})$ expressing them in terms of smaller number of MI. We have used the
Maple implementation of the reduction procedure AIR \cite{Anastasiou:2004vj}
and found that all the necessary 172 integrals $B(n_{1},\ldots ,n_{7})$ can
be expressed in terms of 18 MI which belong to 12 different types of
topologies, namely one two propagator MI (Fig. \ref{T2}, one topology), six
three propagator MI (Fig. \ref{T3}, five topologies), eight four propagator
MI (Fig. \ref{T4}, five topologies) and three five propagator MI (Fig. \ref%
{T5}, two topologies). In the above figures, the dots added to the internal
propagator line mean that corresponding $n_{i}>1$ (number of dots is $n_{i}-1
$) while crossed line indicates that $n_{i}<0$, \emph{i.e.} the $i-$th
propagator is missing and the integral is a tensor one (number of crosses
corresponds to $|n_{i}|$). The set of MI is summarized in Tab. \ref{Tab_MI}.
\begin{figure}[t]
\begin{center}
\epsfig{width=0.80\textwidth,file=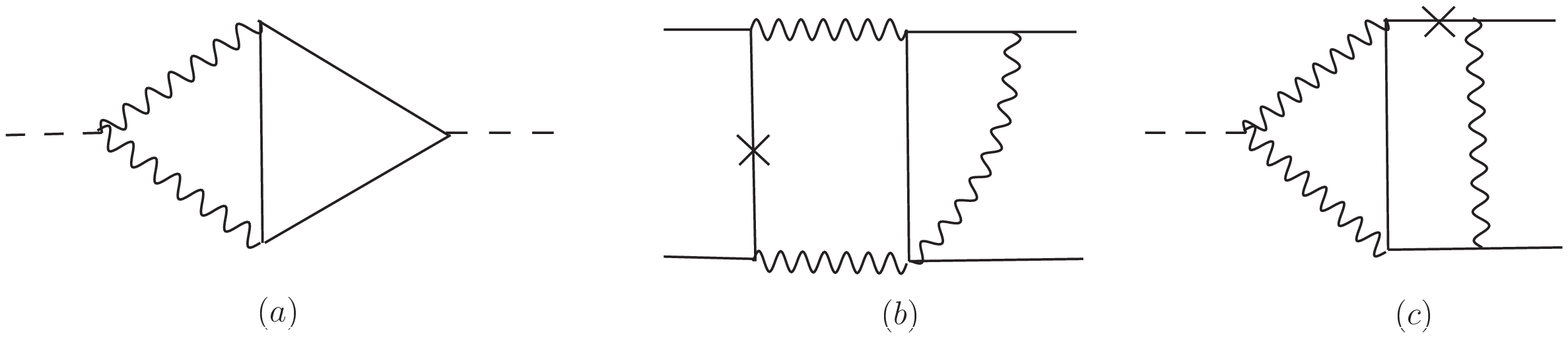}
\end{center}
\caption{The subset of MI with five propagators. There are two
topologies within this subset, namely (a) and (b,c) } \label{T5}
\end{figure}
\begin{table}[h]
\begin{center}
\begin{tabular}{|c|l|}
\hline
Number of propagators & MI \\ \hline\hline
2 & $B(0,0,0,0,0,1,1)^{*}$ \\ \hline
3 & $B(1,0,0,0,0,1,1),\;B(0,0,1,1,0,0,1),\;B(0,0,0,0,1,1,1)^{*},$ \\
& $B(0,1,1,0,0,0,1)^{*},\;\{B(0,2,0,0,0,2,1),\;B(0,1,0,0,0,2,2)\} $ \\ \hline
4 & $B(0,1,1,0,1,1,0)^{*},\;B(0,1,0,1,0,1,1),\;B(1,1,1,0,0,0,1)^{*},$ \\
& $\{B(1,1,0,0,0,1,1),\;B(2,1,0,0,0,1,1)\},$ \\
& $\{B(-2,2,1,1,0,0,1),\;B(0,1,1,2,0,0,1),\;B(0, 1, 1, 1, 0, 0, 1)\}$ \\
\hline
5 & $B(0, 1, 1, 0, 1, 1, 1),\{\;B(-1, 1, 1, 1, 0, 1, 1),\;B(0, 1,1, 1,-1, 1,
1)\} $ \\ \hline
\end{tabular}%
\end{center}
\caption{The MI for different number of propagators ordered according to the
figures \protect\ref{T2}-\protect\ref{T5}. The simple products of one-loop
integrals are denoted by star. The MI belonging to the same topology class
are placed in curly brackets.}
\label{Tab_MI}
\end{table}
As a final result of the reduction procedure we get the individual two-loop
contributions in the form of linear combinations of the MI with $y$
dependent coefficients%
\begin{eqnarray}
\gamma ^{(1),~\mathrm{2-loop}} &=&(2\pi )^{4}\left( \frac{2\alpha }{\pi }%
\right) \mu ^{-4\varepsilon }\left\{ \frac{2y-1}{y-1}\left( \frac{1}{%
4(\epsilon -1)}+\frac{1}{\epsilon }-\frac{3}{4(3\epsilon -1)}\right) \frac{%
B(-2,2,1,1,0,0,1)}{m^{2}}\right.  \notag \\
&&+\frac{2(2y-1)}{y-1}\left( \frac{1}{\epsilon }-2\right) B(-1,1,1,1,0,1,1)
\notag \\
&&+\left( \frac{2y(2y-1)}{\epsilon (y-1)}-\frac{12y^{3}-12y^{2}+7y-3}{3y(y-1)%
}+\frac{3y+2}{y(2\epsilon -1)}\right.  \notag \\
&&\left. -\frac{7y+3}{3y(3\epsilon -2)}-\frac{4(4y-3)}{3(y-1)(3\epsilon -1)}%
\right) \frac{B(0,0,0,0,0,1,1)}{m^{4}}  \notag \\
&&+\frac{2y-1}{y-1}\left( 4-\frac{3}{\epsilon }\right) \frac{B(0,0,0,0,1,1,1)%
}{m^{2}}  \notag \\
&&+\frac{2y-1}{y-1}\left( -\frac{5y-6}{2y}+\frac{1}{3\epsilon -1}-\frac{5}{%
2(4\epsilon -1)}-\frac{\left( 4y-1\right) (y-1)}{2y(\epsilon -1)(2y-1)}%
\right) \frac{B(0,0,1,1,0,0,1)}{m^{2}}  \notag \\
&&+\left( \frac{6(2y+1)(4y-3)}{(y-1)(3\epsilon -1)}+\frac{4(y-1)(4y-1)}{%
\epsilon -1}-\frac{2(2y-1)\left( 4y^{2}-6y+1\right) }{(y-1)\epsilon }\right.
\notag \\
&&\left. -\frac{16\left( 5y^{2}-2y-2\right) }{(y-1)(2\epsilon -1)}+\frac{%
12(7y+2)}{3\epsilon -2}\right) m^{2}B(0,1,0,0,0,2,2)  \notag \\
&&+\left( -\frac{2(2y-1)^{2}}{(y-1)\epsilon }+\frac{4\left(
12y^{3}-16y^{2}+6y-1\right) }{(y-1)y}\right.  \notag \\
&&\left. +\frac{(4y-1)^{2}}{y(\epsilon -1)}+\frac{1}{y(3\epsilon -2)}\right)
B(0,1,0,1,0,1,1)  \notag \\
&&+\frac{1}{y-1}\left( \frac{2y-1}{\epsilon }+\frac{5(2y-1)}{6(3\epsilon -1)}%
-\frac{4y+1}{3}-\frac{2(y-1)}{2\epsilon -1}\right) \frac{B(0,1,1,0,0,0,1)}{%
m^{2}}  \notag \\
&&+\frac{2y-1}{y-1}\left( \frac{4y}{\epsilon }-8y\right)
m^{2}B(0,1,1,0,1,1,1)  \notag \\
&&+\frac{2y-1}{y-1}\left( \frac{2(2y-1)}{\epsilon }-\frac{(4y-1)}{%
3(3\epsilon -1)}-\frac{2\left( 16y^{2}-12y-1\right) }{3(2y-1)}\right)
B(0,1,1,1,0,0,1)  \notag \\
&&-\frac{4y(2y-1)}{(y-1)(3\epsilon -1)}m^{2}B(0,1,1,2,0,0,1)  \notag \\
&&\left. +8\left( -\frac{12(y-1)}{3\epsilon -2}+\frac{2(8y-7)}{2\epsilon -1}-%
\frac{3(4y-3)}{3\epsilon -1}\right) m^{2}B(0,2,0,0,0,2,1)\right\}
\label{gamma_box_red}
\end{eqnarray}

\begin{eqnarray}
\gamma ^{(\Gamma ),~\mathrm{2-loop}} &=&(2\pi )^{4}\left( \frac{2\alpha }{%
\pi }\right) \mu ^{-4\varepsilon }\left\{ 2B(-1,1,1,1,0,1,1)%
\phantom{\frac{B(-2,2,1,1,0,0,1)}{m^{2}}}\right.  \notag \\
&&+\left( -\frac{1}{2\epsilon -1}+\frac{9}{8(3\epsilon -1)}+\frac{1}{%
8(\epsilon -1)}+\frac{1}{4(\epsilon -1)^{2}}\right) \frac{B(-2,2,1,1,0,0,1)}{%
m^{2}}  \notag \\
&&+\left( \frac{16y+39}{36y(3\epsilon -2)}+\frac{360y^{2}-26y+195}{144y}+%
\frac{1}{8}(4y+3)\epsilon -\frac{32y^{2}-36y+13}{8y\epsilon }\right.  \notag
\\
&&\left. +\frac{20y-97}{16(2\epsilon -1)}-\frac{16}{9(3\epsilon -1)}-\frac{%
4y^{2}-29y+13}{16y(2\epsilon -1)^{2}}\right) \frac{B(0,0,0,0,0,1,1)}{m^{4}}
\notag \\
&&+\frac{1}{y}\left( \frac{6y-1}{4(2\epsilon -1)}+\frac{24y-5}{4}+\frac{16y-3%
}{12(\epsilon -1)}\right.  \notag \\
&&\left. -\frac{10y-1}{2\epsilon }-\frac{3y}{2(3\epsilon -1)}+\frac{25y}{%
6(4\epsilon -1)}\right) \frac{B(0,0,1,1,0,0,1)}{m^{2}}  \notag \\
&&+\frac{1}{y}\left( \frac{2(y-1)(4y-1)(2y+1)}{\epsilon }+\frac{8(2y+1)y}{%
3\epsilon -1}-\frac{2(y-1)\left( 6y^{2}-1\right) }{(\epsilon -1)}\right.
\notag \\
&&\left. -\frac{4\left( 3y^{3}-3y-2\right) }{(2\epsilon -1)}-\frac{4\left(
4y^{2}+2y+3\right) }{(3\epsilon -2)}+\frac{4\left( y^{3}+2y^{2}+1\right) }{%
(2\epsilon -1)^{2}}\right) m^{2}B(0,1,0,0,0,2,2)  \notag \\
&&+\frac{1}{y}\left( -\frac{(4y-1)^{2}}{2(\epsilon -1)}-\frac{72y^{2}-24y+5}{%
3}+\frac{8y^{2}+1}{2\epsilon }-\frac{1}{3(3\epsilon -2)}\right)
B(0,1,0,1,0,1,1)  \notag \\
&&+\frac{1}{2}\left( \frac{5}{2\epsilon -1}-\frac{5}{2(3\epsilon -1)}-6-%
\frac{1}{2(\epsilon -1)}\right) \frac{B(0,1,1,0,0,0,1)}{m^{2}}  \notag \\
&&+\left( 1+\frac{1}{2\epsilon -1}\right) B(0,1,1,1,-1,1,1)+2\left( \frac{y}{\epsilon -1}+\frac{3y}{3\epsilon -1}\right)
m^{2}B(0,1,1,2,0,0,1)  \notag \\
&&+\frac{1}{2}\left( -\frac{4y-3}{\epsilon -1}+\frac{4y-1}{3\epsilon -1}%
+10\right) B(0,1,1,1,0,0,1)  \notag \\
&&+\frac{(y-1)^{2}}{y}\left( \frac{4(2y-1)}{\epsilon -1}-\frac{4(4y-1)}{%
\epsilon }+\frac{8(3y+2)}{2\epsilon -1}+\frac{8(y+3)}{(y-1)(3\epsilon -2)}%
\right.  \notag \\
&&\left. -\frac{32y}{(y-1)(3\epsilon -1)}-\frac{8\left( y^{2}+1\right) }{%
(2\epsilon -1)^{2}(y-1)}\right) m^{2}B(0,2,0,0,0,2,1)  \notag \\
&&+\frac{1}{y}\left( \frac{8y+1}{4\epsilon }+\frac{y-1}{8(2\epsilon -1)}-%
\frac{2y+1}{8(\epsilon -1)}-\frac{29y+5}{8}-\frac{9y\epsilon }{4}\right)
\frac{B(1,0,0,0,0,1,1)}{m^{2}}  \notag \\
&&+\frac{1}{y}\left( -\frac{8y^{2}+1}{2\epsilon }+\frac{8y^{2}-8y+3}{%
2(\epsilon -1)}-\frac{8y-5}{3(3\epsilon -2)}\right.  \notag \\
&&\left. -\frac{(y-1)^{2}}{(2\epsilon -1)}+\frac{39y^{2}-28y+10}{3}%
+6y^{2}\epsilon \right) B(1,1,0,0,0,1,1)  \notag \\
&&+6y\left( \frac{1}{2\epsilon -1}-1-\frac{2}{3}\epsilon \right)
B(1,1,1,0,0,0,1)  \notag \\
&&\left. +2\frac{y-1}{y}\left( \frac{4y-1}{\epsilon }-\frac{2(y-1)}{%
2\epsilon -1}+\frac{2y-1}{\epsilon -1}-\frac{2}{3\epsilon -2}+2y\right)
m^{2}B(2,1,0,0,0,1,1)\right\} \notag \\
&& \label{gamma_vertex_red}
\end{eqnarray}

\begin{eqnarray}
\gamma ^{(3),~\mathrm{2-loop}} &=&(2\pi )^{4}\left( \frac{2\alpha }{\pi }%
\right) \mu ^{-4\varepsilon }\left\{ 4y\left( 4\epsilon -\frac{1}{2\epsilon
-1}-1\right) B(1,1,1,0,0,0,1)\right.  \notag \\
&&\frac{1}{4}\left( \frac{3}{3\epsilon -1}-\frac{5}{\epsilon -1}-\frac{2}{%
(\epsilon -1)^{2}}\right) \frac{B(-2,2,1,1,0,0,1)}{m^{2}}  \notag \\
&&+\left( \frac{2}{2\epsilon -1}-\frac{1}{(2\epsilon -1)^{2}}+6-8\epsilon
\right) \frac{B(0,0,0,0,0,1,1)}{m^{4}}  \notag \\
&&+\left( \frac{7}{2}-\frac{1}{2\epsilon -1}-\frac{1}{3\epsilon -1}-\frac{5}{%
6(4\epsilon -1)}-\frac{2}{3(\epsilon -1)}\right) \frac{B(0,0,1,1,0,0,1)}{%
m^{2}}  \notag \\
&&+\left( \frac{1}{2(\epsilon -1)}-\frac{5}{6(3\epsilon -1)}-\frac{13}{3}%
+8\epsilon \right) \frac{B(0,1,1,0,0,0,1)}{m^{2}}  \notag \\
&&+\left( \frac{4y-3}{\epsilon -1}+\frac{2}{3}(8y-17)+\frac{4y-1}{%
3(3\epsilon -1)}\right) B(0,1,1,1,0,0,1)  \notag \\
&&\left. +4y\left( \frac{1}{3\epsilon -1}-\frac{1}{\epsilon -1}-\frac{2}{%
2\epsilon -1}\right) m^{2}B(0,1,1,2,0,0,1)\right\}  \\
&&  \label{gamma_selfenergy_red}\notag \\
\gamma ^{(\Pi ),~\mathrm{2-loop}} &=&(2\pi )^{4}\left( \frac{2\alpha }{\pi }%
\right) \mu ^{-4\varepsilon }\left\{ \frac{8}{3}\left( 2y-(y-3)\epsilon
-(y+3)\epsilon ^{2}\right) B(1,1,1,0,0,0,1)\right.  \notag \\
&&+\frac{1}{3y}\left( -\frac{2(y+2)}{\epsilon }+\frac{\left(
4y^{2}+19y+6\right) \epsilon ^{2}}{4}+\frac{\left( 4y^{2}-13y-39\right)
\epsilon }{8}-\frac{44y^{2}+9y-117}{16}\right.  \notag \\
&&\left. -\frac{16y-1}{20(2\epsilon -3)}+\frac{20y^{2}-9y-27}{16(2\epsilon
-1)}-\frac{4y-9}{5(3\epsilon -2)}\right) \frac{B(0,0,0,0,0,1,1)}{m^{4}}
\notag \\
&&+4\frac{y^{2}-1}{y}\left( \frac{12}{5(3\epsilon -2)}-\frac{4}{15(2\epsilon
-3)}-\frac{5y+3}{3(2\epsilon -1)}-\frac{y+2}{3}\right) m^{2}B(0,1,0,0,0,2,2)
\notag \\
&&+8\frac{(y-1)^{2}}{y}\left( \frac{4}{15(2\epsilon -3)}+\frac{5y+3}{%
3(2\epsilon -1)}-\frac{12}{5(3\epsilon -2)}+\frac{y+2}{3}\right)
m^{2}B(0,2,0,0,0,2,1)  \notag \\
&&+\frac{1}{y}\left( \frac{4(y+2)}{9\epsilon }-\frac{8y+7}{9(2\epsilon -3)}+%
\frac{4y-25}{6}\right.  \notag \\
&&\left. -\frac{3(4y-7)\epsilon }{4}-\frac{3(y+2)\epsilon ^{2}}{2}\right)
\frac{B(1,0,0,0,0,1,1)}{m^{2}}  \notag \\
&&+\left( 4(y+2)\epsilon ^{2}+\frac{4}{3}(5y-8)\epsilon -\frac{8y^{2}-5}{3y}%
\right.  \notag \\
&&\left. +\frac{16}{5(2\epsilon -3)}-\frac{4(8y-5)}{15y(3\epsilon -2)}%
\right) B(1,1,0,0,0,1,1)  \notag \\
&&\left. +4\frac{y-1}{y}\left( \frac{4y-3}{3}+\frac{2(y+2)\epsilon }{3}+%
\frac{8}{15(2\epsilon -3)}-\frac{4}{5(3\epsilon -2)}\right)
m^{2}B(2,1,0,0,0,1,1)\right\}.  \notag \\
&&  \label{gamma_vac_pol_red}
\end{eqnarray}

According to the above formulae we shall need the MI up to (and including)
the order $O(\varepsilon)$. However, for the calculation of MI as described
in the next subsection we need some of them to the order $O(\varepsilon^2)$.

\subsection{Calculation of the Master Integrals}

Some of the MI are not genuine two-loop integrals and can be calculated
simply as a product of one-loop integrals. This concerns four MI denoted by
star in Tab. \ref{Tab_MI}. The remaining fourteen MI can be obtained in a
standard way as a solution of the appropriate closed system of ordinary
linear differential equations \cite%
{Kotikov:1990kg,Kotikov:1991hm,Kotikov:1991pm,Remiddi:1997ny,Caffo:1998yd}.
The latter can be obtained by means of differentiating the integrals $%
B(n_{1},\ldots ,n_{7})$ with respect to $q_{\pm }^{\mu }$. On one hand, the
result of such a differentiation contracted with external momenta can be
related to the derivative of $B(n_{1},\ldots ,n_{7})$ with respect to the
variable $y$ as%
\begin{eqnarray}
y\frac{\partial }{\partial y}B(n_{1},\ldots ,n_{7}) &=&\frac{1}{y-1}\left[
\left( y-\frac{1}{2}\right) q_{+}^{\mu }\frac{\partial }{\partial q_{+}^{\mu
}}-\frac{1}{2}q_{-}^{\mu }\frac{\partial }{\partial q_{+}^{\mu }}\right]
B(n_{1},\ldots ,n_{7})  \notag \\
&=&\frac{1}{y-1}\left[ \left( y-\frac{1}{2}\right) q_{-}^{\mu }\frac{%
\partial }{\partial q_{-}^{\mu }}-\frac{1}{2}q_{+}^{\mu }\frac{\partial }{%
\partial q_{-}^{\mu }}\right] B(n_{1},\ldots ,n_{7})  \label{Dif_eq}
\end{eqnarray}%
(here the second identity is a consequence of the LI). On the other hand,
the right hand side of (\ref{Dif_eq}) can be expressed as a linear
combination (with $y$ dependent coefficients) of the integrals $%
B(n_{1},\ldots ,n_{7})$ belonging either to the same topology class or to
the topologies with less propagators. Finally we get (in terms of the
operators (\ref{jpm_operator}))\footnote{%
It is easy to see that the difference of both possible right hand sides of (%
\ref{Dif_eq1}) just corresponds to the LI identity (\ref{LI1}).}%
\begin{eqnarray}
y\frac{\partial }{\partial y}B(n_{1},\ldots ,n_{7}) &=&\frac{1}{y-1}\left[
\left( \frac{1}{2}-y\right) (n_{2}+n_{5})-\frac{1}{2}n_{2}\mathbf{3}^{-}%
\mathbf{2}^{+}+2ym^{2}n_{2}\mathbf{2}^{+}+yn_{2}\mathbf{1}^{-}\mathbf{2}%
^{+}\right.  \notag \\
&&\left. -\frac{1}{2}n_{5}\mathbf{6}^{-}\mathbf{5}^{+}+yn_{5}\mathbf{4}^{-}%
\mathbf{5}^{+}\right] B(n_{1},\ldots ,n_{7})  \notag \\
&=&\frac{1}{y-1}\left[ \left( \frac{1}{2}-y\right) (n_{3}+n_{6})-\frac{1}{2}%
n_{3}\mathbf{2}^{-}\mathbf{3}^{+}+2ym^{2}n_{3}\mathbf{3}^{+}+yn_{3}\mathbf{1}%
^{-}\mathbf{3}^{+}\right.  \notag \\
&&\left. -\frac{1}{2}n_{6}\mathbf{5}^{-}\mathbf{6}^{+}+yn_{6}\mathbf{4}^{-}%
\mathbf{6}^{+}\right] B(n_{1},\ldots ,n_{7}).  \label{Dif_eq1}
\end{eqnarray}%
Writing these equations for the MI we get at the right hand side along with
the original MI also other $B(n_{1},\ldots ,n_{7})$'s which can be
subsequently expressed in terms of the MI to close the system. From the form
of (\ref{Dif_eq1}) it can be seen that the resulting set of differential
equations has \textquotedblleft triangular structure\textquotedblright\ in
the sense that it connects the derivative of given MI either with MI with
the same topology or with topologies with smaller number of propagators.

For further convenience we introduce dimensionless quantities $%
b(n_{1},\ldots ,n_{7})$ defined as%
\begin{equation}
B(n_{1},\ldots ,n_{7})=\left( \mathrm{i}\Gamma (1+\varepsilon )(4\pi
)^{\varepsilon -2}\right) ^{2}\left( \frac{\mu }{m}\right) ^{4\varepsilon
}m^{2\left( 4-\sum_{i}n_{i}\right) }b(n_{1},\ldots ,n_{7})  \label{b}
\end{equation}%
and rewrite the differential equation for $b(n_{1},\ldots ,n_{7})$ in terms
of the variable $x$ (cf. (\ref{x_definition})). The system (\ref{Dif_eq1})
for $b(n_{1},\ldots ,n_{7})$ is then solved order by order in the $%
\varepsilon =2-d/2$ expansion writing%
\begin{equation}
b(n_{1},\ldots ,n_{7})=\sum_{i\geq -2}b_{i}(n_{1},\ldots ,n_{7})\varepsilon
^{i}
\end{equation}%
and expanding the right hand side of (\ref{Dif_eq1}) to the given power of $%
\varepsilon $. At each order, we solve the corresponding equations in the
unphysical region $0<x<1$ where the MI are analytic. The solution is then
fixed uniquely up to the integration constants which can be determined by
the requirement of the absence of singularities in some appropriately chosen
points of this analyticity region. For the calculation we have used the
Mathematica package HPL \cite{Maitre:2005uu, Maitre:2007kp}. The results are
summarized in Appendix \ref{MI appendix} where also comparison of our
independent calculations with those existing in the literature is
given.\bigskip

\subsection{Two-loop contributions\label{two_loop_subsection}}

Inserting the results of Appendix \ref{MI appendix} in formulae (\ref%
{gamma_box_red}-\ref{gamma_vac_pol_red}) we obtain the final form of the
two-loop contributions. Writing them in the form (cf. (\ref{two_loop_general}%
))%
\begin{equation}
\gamma ^{(i),~2-\mathrm{loop}}=\mu ^{-4\varepsilon }\left( \frac{\mu }{m}%
\right) ^{4\varepsilon }\left( \frac{\alpha }{\pi }\right) \left( \frac{%
\gamma _{-2}^{(i),~2-\mathrm{loop}}}{\varepsilon ^{2}}+\frac{\gamma
_{-1}^{(i),~2-\mathrm{loop}}}{\varepsilon }+\gamma _{0}^{(i),~2-\mathrm{loop}%
}+O(\varepsilon )\right)
\end{equation}%
we get
\begin{eqnarray}
\gamma _{-2}^{(1),~\mathrm{2-loop}} &=&\frac{3}{8}\left( \beta +\frac{1}{%
\beta }\right) \left( H(0,z)+\mathrm{i}\pi \right) -\frac{3}{16} \\
&& \notag \\
\gamma _{-1}^{(1),~\mathrm{2-loop}} &=&\frac{1}{4}\left( 1+\frac{1}{\beta
^{2}}\right) \left[ 2H(-3,z)+H(-2,0,z)-\frac{3}{2}H(0,0,0,z)\right. \notag \\
&&\phantom{\frac{1}{4}\left( 1+\frac{1}{\beta ^{2}}\right)}\left. +\mathrm{i}%
\pi \left( H(-2,z)-\frac{3}{2}H(0,0,z)-\frac{\pi ^{2}}{2}\right) \right] \notag \\
&&+\frac{3}{8}\left( \beta +\frac{1}{\beta }\right) \left[
H(0,0,z)+2H(1,0,z)+\mathrm{i}\pi \left( H(0,z)+2H(1,z)\right) -\frac{2}{3}%
\pi ^{2}\right] \notag \\
&&+\frac{5}{8}\left[ \left( 1-\frac{6}{5}\bar{\gamma}\right) \left( \beta +%
\frac{1}{\beta }\right) +\frac{\pi ^{2}}{6}\left( 1+\frac{1}{\beta ^{2}}%
\right) \right] \left( H(0,z)+\mathrm{i}\pi \right) \notag \\
&&+\frac{3}{8}\bar{\gamma}-\frac{35}{32} \\
&& \notag \\
\gamma _{0}^{(1),~\mathrm{2-loop}} &=&H(-1,-1,z)-H(-1,z) \notag \\
&&+\left( 1+\frac{1}{\beta ^{2}}\right) \left[ \frac{9}{4}%
H(-4,z)-2H(-3,-1,z)-\frac{3}{2}H(-2,-2,z)-H(-1,-3,z)\right. \notag \\
&&\left. -\frac{1}{8}\left( H(0,0,0,0,z)+\mathrm{i}\pi H(0,0,0,z)\right) +%
\frac{3}{4}\left( H(-1,0,0,0,z)+\mathrm{i}\pi H(-1,0,0,z)\right) \right. \notag \\
&&\left. +\frac{5}{4}\left( H(-2,0,0,z)+\mathrm{i}\pi H(-2,0,z)\right) -%
\frac{1}{4}\left( H(2,0,0,z)+\mathrm{i}\pi H(2,0,z)\right) \right. \notag \\
&&\left. +\frac{7}{4}\left( H(-3,0,z)+\mathrm{i}\pi H(-3,z)\right) -\frac{1}{%
4}\left( H(3,0,z)+\mathrm{i}\pi H(3,z)\right) \right. \notag \\
&&\left. -\frac{1}{2}\left( H(-1,-2,0,z)+\mathrm{i}\pi H(-1,-2,z)\right)
\right. \notag \\
&&\left. +\frac{1}{2}\left( H(-2,1,0,z)+\mathrm{i}\pi H(-2,1,z)\right) -%
\frac{1}{2}\left( H(-2,-1,0,z)+\mathrm{i}\pi H(-2,-1,z)\right) \right. \notag \\
&&+\frac{1}{2}\left( H(2,-1,0,z)+\mathrm{i}\pi H(2,-1,z)\right) +\frac{7\pi
^{2}}{24}H(2,z) \notag \\
&&\left. +\frac{\mathrm{i}\pi ^{3}}{4}H(-1,z)-\frac{\mathrm{i}\pi ^{3}}{24}%
H(0,z)-\frac{17\pi ^{4}}{576}\right] \notag \\
&&+\left( \beta +\frac{1}{\beta }\right) \left[ \frac{5-6\bar{\gamma}}{4}%
\left( H(1,0,z)+\mathrm{i}\pi H(1,z)\right) +\frac{3}{4}\left( H(1,0,0,z)+%
\mathrm{i}\pi H(1,0,z)\right) \right. \notag \\
&&\phantom{\left( \beta +\frac{1}{\beta }\right)}\left. +\frac{3}{4}\left(
H(2,0,z)+\mathrm{i}\pi H(2,z)\right) +\frac{3}{2}\left( H(1,1,0,z)+\mathrm{i}%
\pi H(1,1,z)\right) -\frac{\pi ^{2}}{2}H(1,z)\right] \notag \\
&&+\left( -\frac{5\pi ^{2}}{24}\left( 1+\frac{1}{\beta ^{2}}\right) -\frac{1%
}{2}\right) \left( H(-1,0,z)+\mathrm{i}\pi H(-1,z)\right) \notag \\
&&-\left( \frac{1}{2}\left( 1+\frac{1}{\beta }\right) +\frac{\pi ^{2}}{3}%
\left( 1+\frac{1}{\beta ^{2}}\right) \right) H(-2,z)+\frac{1}{2}\left( \beta
-\frac{1}{\beta }\right) H(-2,-1,z) \notag \\
&&-\left( \frac{1}{4}\left( \beta -\frac{1}{\beta }\right) +\frac{\bar{\gamma%
}}{2}\left( 1+\frac{1}{\beta ^{2}}\right) \right) \left( H(-2,0,z)+\mathrm{i}%
\pi H(-2,z)\right) \notag \\
&&+\frac{1}{16}\left( 5\beta ^{3}+22\beta +12\bar{\gamma}\left( 1+\frac{1}{%
\beta ^{2}}\right) +\frac{33}{\beta }\right) \left( H(0,0,0,z)+\mathrm{i}\pi
H(0,0,z)\right) \notag \\
&&+\frac{1}{16}\left( -\beta ^{2}+2\left( 5-6\bar{\gamma}\right) \beta +11+2%
\frac{7-6\bar{\gamma}}{\beta }+\frac{10\pi ^{2}}{3}\left( 1+\frac{1}{\beta
^{2}}\right) \right) \notag \\
&&\times \left( H(0,0,z)+\mathrm{i}\pi H(0,z)\right) +\left( \frac{1}{4\beta
}-\frac{\beta }{4}-\bar{\gamma}\left( 1+\frac{1}{\beta ^{2}}\right) \right)
H(-3,z) \notag \\
&&+\frac{1}{4}\left( \frac{3}{2}\beta +2+\frac{9}{2\beta }+\bar{\gamma}%
\left( 3\bar{\gamma}-5\right) \left( \beta +\frac{1}{\beta }\right) -\frac{5%
}{2}\left( \zeta (3)+\frac{\pi ^{2}\bar{\gamma}}{3}\right) \left( 1+\frac{1}{%
\beta ^{2}}\right) \right) \notag \\
&&\times \left( H(0,z)+\mathrm{i}\pi \right) -\frac{\pi ^{2}}{48}\left(
5\beta ^{3}+24\beta +\frac{37}{\beta }\right) H(0,z) \notag \\
&&+\frac{13\pi ^{2}\beta ^{2}}{96}+\left( \frac{1}{12}\pi ^{2}\left( 6\bar{%
\gamma}-5\right) -\zeta (3)\right) \beta +\frac{\left( 35-6\bar{\gamma}%
\right) \bar{\gamma}}{16}-\frac{19\pi ^{2}}{48}-\frac{277}{64}\notag \\
&&+\frac{\pi ^{2}\left( 4\bar{\gamma}-3\right) -4\zeta (3)}{8\beta }+\frac{%
\mathrm{i}\pi ^{3}}{12}\left[ 3\bar{\gamma}\left( 1+\frac{1}{\beta ^{2}}%
\right) -\frac{1}{2}\beta -\frac{1}{\beta }\right]
\end{eqnarray}

\begin{eqnarray}
\gamma _{-2}^{(\Gamma ),~\mathrm{2-loop}} &=&-\frac{3}{16} \\
&& \notag \\
\gamma _{-1}^{(\Gamma ),~\mathrm{2-loop}} &=&\frac{1}{8\beta }\left(
H(0,0,z)+\mathrm{i}\pi H(0,z)-2H(-2,z)+\frac{\pi ^{2}}{6}\right) \notag \\
&&+\frac{3}{32}(4\bar{\gamma}-7) \\
&& \notag \\
\gamma _{0}^{(\Gamma ),~\mathrm{2-loop}} &=&\frac{1}{8}\left( 3\beta +2+%
\frac{1}{\beta }\right) H(-3,z) \notag \\
&&-\frac{1}{8}\left( \pi ^{2}\beta ^{2}+11\pi ^{2}-12-\frac{\pi ^{2}}{3\beta
}\right) H(-1,z) \notag \\
&&-\frac{\pi ^{2}}{24\beta }H(1,z)-\frac{1}{2\beta }H(-1,-2,z)+\frac{1}{%
2\beta }H(1,-2,z) \notag \\
&&+\frac{3}{8}\left( \beta -\frac{1}{\beta }\right) \left( H(-2,0,z)+\mathrm{%
i}\pi H(-2,z)-2H(-2,-1,z)\right) \notag \\
&&-\frac{3}{2}H(-1,-1,z)+\frac{3}{4}\left( H(-1,0,z)+\mathrm{i}\pi
H(-1,z)\right) \notag \\
&&+\frac{1}{8}\left( \beta ^{3}-2\beta +\frac{9}{\beta }\right) \left(
H(-2,0,0,z)-H(2,0,0,z)-H(0,0,0,0,z)\right. \notag \\
&&\phantom{\frac{1}{8}\left( \beta ^{3}-2\beta +\frac{9}{\beta }\right)}%
\left. +\mathrm{i}\pi \left( H(-2,0,z)-H(2,0,z)-H(0,0,0,z)\right) \right) \notag \\
&&+\frac{1}{4}\left( \beta ^{2}+11+\frac{1}{\beta }\right) \left(
H(-1,0,0,z)+\mathrm{i}\pi H(-1,0,z)\right) \notag \\
&&-\frac{1}{4}\left( \beta ^{2}+9+\frac{1}{\beta }\right) \left( H(1,0,0,z)+%
\mathrm{i}\pi H(1,0,z)\right) \notag \\
&&-\frac{1}{16}\left( 4\beta ^{2}-9\beta +42-\frac{7}{\beta }\right) \left(
H(0,0,0,z)+\mathrm{i}\pi H(0,0,z)\right) \notag \\
&&-\frac{1}{16}\left( \pi ^{2}\beta ^{3}-2\pi ^{2}\beta -12-\frac{8\bar{%
\gamma}-9\pi ^{2}+4}{\beta }\right) H(-2,z) \notag \\
&&+\frac{1}{32}\left( \pi ^{2}\beta ^{3}-4\beta ^{2}-2\pi ^{2}\beta -32+%
\frac{-8\bar{\gamma}+9\pi ^{2}-4}{\beta }\right) H(0,0,z) \notag \\
&&-\frac{1}{16}\left( \zeta (3)\beta ^{3}-\pi ^{2}\beta ^{2}-\frac{1}{2}%
\left( 4\zeta (3)-7\pi ^{2}\right) \beta \right. \notag \\
&&\left. -\frac{32\pi ^{2}}{3}+12+\frac{13\pi ^{2}+54\zeta (3)}{6\beta }%
\right) H(0,z) \notag \\
&&+\frac{11\pi ^{4}\beta ^{3}}{1920}+\frac{1}{16}\left( \pi ^{2}-2\zeta
(3)\right) \beta ^{2}+\frac{1}{8}\left( 3\zeta (3)-\frac{11\pi ^{4}}{120}%
\right) \beta \notag \\
&&+\frac{1}{192}\left( 36\left( 7-2\bar{\gamma}\right) \bar{\gamma}-72\zeta
(3)+2\pi ^{2}-429\right) \notag \\
&&+\frac{1}{2\beta }\left( \frac{\pi ^{2}\left( -80\bar{\gamma}+99\pi
^{2}-40\right) }{960}-\zeta (3)\right) \notag \\
&&+\mathrm{i}\pi \left[ -\frac{1}{8}\left( \frac{1}{12}\pi ^{2}\beta
^{3}+\beta ^{2}-\frac{\pi ^{2}\beta }{6}+8+\frac{\left( 8\bar{\gamma}+3\pi
^{2}+4\right) }{4\beta }\right) H(0,z)\right. \notag \\
&&-\frac{1}{16}\left( \zeta (3)\beta ^{3}+\frac{\pi ^{2}\beta ^{2}}{3}+\frac{%
1}{2}\left( \pi ^{2}-4\zeta (3)\right) \beta \right. \notag \\
&&\left. \left. +\frac{2}{3}\left( 18+5\pi ^{2}\right) -\frac{\left( \pi
^{2}-54\zeta (3)\right) }{6\beta }\right) \right]
\end{eqnarray}%
\begin{eqnarray}
\gamma _{-2}^{(3),~\mathrm{2-loop}} &=&\frac{3}{16} \\
&& \notag \\
\gamma _{-1}^{(3),~\mathrm{2-loop}} &=&\frac{1}{4\beta }\left( H(0,0,z)+%
\mathrm{i}\pi H(0,z)-2H(-2,z)\right) \notag \\
&&-3H(-1,z)+\frac{3}{2}\left( H(0,z)+\mathrm{i}\pi \right) +\frac{3}{32}%
\left( 25-4\bar{\gamma}\right) +\frac{\pi ^{2}}{24\beta } \\
&& \notag \\
\gamma _{0}^{(3),~\mathrm{2-loop}} &=&\frac{1}{4}\left( \beta +\frac{1}{%
\beta }\right) \left( H(0,0,0,z)+\mathrm{i}\pi H(0,0,z)-2H(-3,z)\right) \notag \\
&&-\frac{1}{2\beta }\left( 2H(-1,0,0,z)+H(1,0,0,z)+\mathrm{i}\pi \left(
2H(-1,0,z)+H(1,0,z)\right) \right) \notag \\
&&+\frac{1}{\beta }\left( H(1,-2,z)+2H(-1,-2,z)\right)\notag \\
&&+8H(-1,-1,z)-4\left( H(-1,0,z)+\mathrm{i}\pi H(-1,z)\right)\notag \\
&&-\frac{1}{2}\beta \left( H(-2,0,z)+\mathrm{i}\pi H(-2,z)-2H(-2,-1,z)\right)
\notag \\
&&+\left( 6\bar{\gamma}-\frac{\pi ^{2}}{6\beta }-10\right) H(-1,z) \notag \\
&&+\left( 2-\frac{\bar{\gamma}}{2\beta }\right) \left( H(0,0,z)+\mathrm{i}%
\pi H(0,z)\right) +H(-2,z)\left( \frac{\bar{\gamma}}{\beta }-4\right) \notag \\
&&+\left( -\frac{\pi ^{2}\beta }{24}-3\bar{\gamma}+5-\frac{\pi ^{2}}{4\beta }%
\right) H(0,z)-\frac{\pi ^{2}}{12\beta }H(1,z) \notag \\
&&-\frac{\zeta (3)\beta }{2}+\frac{1}{192}\left( 36\bar{\gamma}\left( 2\bar{%
\gamma}-25\right) -322\pi ^{2}+1791\right) +\frac{33\zeta (3)-\pi ^{2}\bar{%
\gamma}}{12\beta } \notag \\
&&+\mathrm{i}\pi \left[ \frac{\pi ^{2}\beta }{24}-3\bar{\gamma}+5-\frac{\pi
^{2}}{6\beta }\right]
\end{eqnarray}%
\begin{eqnarray}
\gamma _{-2}^{(\Pi ),~\mathrm{2-loop}} &=&\frac{1}{4} \\
&& \notag \\
\gamma _{-1}^{(\Pi ),~\mathrm{2-loop}} &=&\frac{1}{6\beta }\left(
2H(-2,z)-H(0,0,z)-\mathrm{i}\pi H(0,z)\right)\notag \\
&&+\frac{1}{6}\left( 7-3\bar{\gamma}\right) -\frac{\pi ^{2}}{36\beta } \\
&& \notag \\
\gamma _{0}^{(\Pi ),~\mathrm{2-loop}} &=&\frac{1}{3\beta }\left(
H(-2,0,z)+H(-1,0,0,z)+H(1,0,0,z)\right. \notag \\
&&\left. -2H(-1,-2,z)-2H(1,-2,z)+H(-3,z)-2H(-2,-1,z)\right) \notag \\
&&+\frac{1}{8}\left( \beta ^{3}-2\beta -\frac{3}{\beta }\right) \left(
H(0,0,0,z)+\mathrm{i}\pi H(0,0,z)\right)\notag \\
&&-\left( \frac{\beta }{2}+\frac{2\left( \bar{\gamma}-1\right) }{3\beta }%
\right) H(-2,z)\notag \\
&&-\left( \frac{23\beta ^{2}}{72}-\frac{\beta }{4}+\frac{5}{24}-\frac{\bar{%
\gamma}-1}{3\beta }\right) \left( H(0,0,z)+\mathrm{i}\pi H(0,z)\right)\notag \\
&&-\frac{\pi ^{2}}{24}\left( \beta ^{3}-2\beta -\frac{13}{3\beta }\right)
H(0,z)+\frac{\pi ^{2}}{18\beta }\left( H(1,z)+H(-1,z)\right) \notag \\
&&+\frac{29\pi ^{2}\beta ^{2}}{144}+\frac{\pi ^{2}\beta }{24}+\frac{1}{48}%
\left( 8\bar{\gamma}\left( 3\bar{\gamma}-14\right) -7\pi ^{2}+206\right)\notag \\
&&+\frac{1}{6\beta }\left( \frac{\pi ^{2}}{3}\left( \bar{\gamma}-1\right)
-5\zeta (3)\right)\notag \\
&&+\frac{\mathrm{i}\pi }{3\beta }\left[ \left(
H(-2,z)+H(-1,0,z)+H(1,0,z)\right) +\frac{\pi ^{2}}{6}\right]
\end{eqnarray}%
The graph $(1)$ has beside the UV also IR divergences, which can be
identified using the general identities from Subsection \ref{IR_subsection}.
We get for $j=-1$, $-2$%
\begin{equation}
\gamma _{j}^{(1),~\mathrm{2-loop}}=\gamma _{j,\mathrm{UV}}^{(1),~\mathrm{%
2-loop}}+\gamma _{j,\mathrm{IR}}^{(1),~\mathrm{2-loop}}
\end{equation}%
where%
\begin{eqnarray}
\gamma _{-2,~\mathrm{UV}}^{(1),~\mathrm{2-loop}} &=&-\frac{3}{16} \\
\gamma _{-2,~\mathrm{IR}}^{(1),~\mathrm{2-loop}} &=&\frac{3}{8}\left( \beta +%
\frac{1}{\beta }\right) \left( H(0;z)+\mathrm{i}\pi \right) \\
\gamma _{-1,~\mathrm{UV}}^{(1),~\mathrm{2-loop}} &=&\frac{3}{8}\left( \beta +%
\frac{1}{\beta }\right) \left[ H(0,0,z)+2H(1,0,z)+\mathrm{i}\pi \left(
H(0,z)+2H(1,z)\right) -\frac{2}{3}\pi ^{2}\right.\notag \\
&&\left. \phantom{\frac{1}{4}\left( 1+\frac{1}{\beta ^{2}}\right)}-\overline{%
\gamma }\left( H(0,z)+\mathrm{i}\pi \right) \right] +\frac{3}{8}\overline{%
\gamma }-\frac{35}{32} \\
\gamma _{-1,~\mathrm{IR}}^{(1),~\mathrm{2-loop}} &=&\frac{1}{4}\left( 1+%
\frac{1}{\beta ^{2}}\right) \left[ 2H(-3,z)+H(-2,0,z)-\frac{3}{2}%
H(0,0,0,z)\right.\notag \\
&&\phantom{\frac{1}{4}\left( 1+\frac{1}{\beta ^{2}}\right)}\left. +\mathrm{i}%
\pi \left( H(-2,z)-\frac{3}{2}H(0,0,z)-\frac{\pi ^{2}}{2}\right) \right]\notag \\
&&+\frac{5}{8}\left[ \left( 1-\frac{3}{5}\bar{\gamma}\right) \left( \beta +%
\frac{1}{\beta }\right) +\frac{\pi ^{2}}{6}\left( 1+\frac{1}{\beta ^{2}}%
\right) \right] \left( H(0,z)+\mathrm{i}\pi \right) .
\end{eqnarray}%
From the above formulae the general relations (\ref{nonloc2}), which are
valid for each graph separately, can be easily verified.

\bigskip

\subsection{Two-loop counterterm contribution}

For the construction of the two-loop counterterm \ (\ref%
{2loop_tree_CT_lagrangian}) we need further the local parts of the $%
O(\varepsilon ^{-1})$ UV divergences of the two-loop graphs (cf. (\ref%
{nonloc1})). Using the results of the previous (sub)sections we get%
\begin{eqnarray}
\left( \gamma _{-1,~\mathrm{UV}}^{(1),~\mathrm{2-loop}}\right) _{l}
&=&\gamma _{-1,~\mathrm{UV}}^{(1),~\mathrm{2-loop}}-C_{0}^{(1),~\mathrm{%
1-loop}}\gamma _{-1}^{\mathrm{1-loop}}=\frac{1}{32} \notag \\
\left( \gamma _{-1}^{(\Gamma ),~\mathrm{2-loop}}\right) _{l} &=&\gamma
_{-1}^{(\Gamma ),~\mathrm{2-loop}}-\gamma _{0}^{\mathrm{1-loop}}\gamma
_{-1}^{(\Gamma ),~\mathrm{1-loop}}=-\frac{1}{32} \notag \\
\left( \gamma _{-1,~\mathrm{UV}}^{(3),~\mathrm{2-loop}}\right) _{l}
&=&\gamma _{-1,~\mathrm{UV}}^{(3),~\mathrm{2-loop}}+\gamma _{0}^{\mathrm{%
1-loop}}\gamma _{-1}^{(3),~\mathrm{1-loop}}-C_{0}^{(m),~\mathrm{1-loop}%
}\gamma _{-1}^{(m),~\mathrm{1-loop}}=\frac{7}{32}\notag \\
\left( \gamma _{-1}^{(\Pi ),~\mathrm{2-loop}}\right) _{l} &=&\gamma
_{-1}^{(\Pi ),~\mathrm{2-loop}}-\gamma _{0}^{\mathrm{1-loop}}\gamma
_{-1}^{(\Pi ),~\mathrm{1-loop}}=\frac{1}{3}
\end{eqnarray}%
where $\gamma _{-1}^{(i),~\mathrm{1-loop}}$ are the coefficients of the UV
divergent parts of the one-loop (sub)graphs (see (\ref{gamma_1loop})).
According to general formula (\ref{2loop_tree_CT}) we get for the two-loop
counterterm contribution
\begin{eqnarray}
\gamma _{CT}^{~\mathrm{tree}} &=&\mu ^{-4\varepsilon }\left( \frac{\mu }{m}%
\right) ^{4\varepsilon }\left( \frac{\alpha }{\pi }\right) \left[ \overline{%
\xi }+\frac{1}{\varepsilon ^{2}}\left( \gamma _{-2,~\mathrm{UV}}^{(1),~%
\mathrm{2-loop}}+2\gamma _{-2}^{(\Gamma ),~\mathrm{2-loop}}+\gamma
_{-2}^{(3),~\mathrm{2-loop}}+2\gamma _{-2}^{(\Pi ),~\mathrm{2-loop}}\right)
\right. \notag \\
&&\left. -\frac{1}{\varepsilon }\left( C_{-1,~\mathrm{UV}}^{(1),~\mathrm{%
1-loop}}\overline{\chi }+2\gamma _{-1}^{\mathrm{1-loop}}\overline{x}%
_{6}-\gamma _{-1}^{\mathrm{1-loop}}\overline{x}_{6}+C_{-1}^{(m),~\mathrm{%
1-loop}}(\overline{x}_{6}+\overline{x}_{7})+2\gamma _{-1}^{\mathrm{1-loop}}%
\overline{x}_{8}\right) \right.\notag \\
&&\left. -\frac{1}{\varepsilon }\left( \left( \gamma _{-1,~\mathrm{UV}%
}^{(1),~\mathrm{2-loop}}\right) _{l}+2\left( \gamma _{-1}^{(\Gamma ),~%
\mathrm{2-loop}}\right) _{l}+\left( \gamma _{-1,~\mathrm{UV}}^{(3),~\mathrm{%
2-loop}}\right) _{l}+2\left( \gamma _{-1}^{(\Pi ),~\mathrm{2-loop}}\right)
_{l}\right) \right]
\end{eqnarray}%
and for our choice of the renormalization scheme
\begin{equation}
\gamma _{CT}^{~\mathrm{tree}}=\mu ^{-4\varepsilon }\left( \frac{\mu }{m}%
\right) ^{4\varepsilon }\left( \frac{\alpha }{\pi }\right) \left[ \overline{%
\xi }+\frac{1}{8\varepsilon ^{2}}-\frac{1}{\varepsilon }\left( \frac{1}{4}%
\overline{\chi }-\frac{1}{8}\overline{\gamma }+\frac{131}{48}\right) \right]
\end{equation}%
where%
\begin{equation}
\overline{\xi }=\xi ^{r}(\mu )+\frac{1}{8}\left( 2\overline{\chi }-\overline{%
\gamma }+\frac{86}{3}\right) \ln \left( \frac{\mu ^{2}}{m^{2}}\right) -\frac{%
1}{8}\ln ^{2}\left( \frac{\mu ^{2}}{m^{2}}\right) +O(\varepsilon ).
\label{xi_bar}
\end{equation}%
To get the numerical values of the two-loop radiative corrections, we need
to know the NLO counterterm coupling $\overline{\xi }$ or its scale
dependent renormalized value $\xi ^{r}(\mu )$. In principle it can be
obtained similarly as $\chi ^{r}(\mu )$ (cf. \cite{Knecht:1999gb}) by means
of matching of the (complete) NLO chiral expansion of the amplitude with the
sum of the same types of graphs as we calculated but now with appropriate
model of nonlocal off-shell pion transition form factor $F_{\pi ^{0}\gamma
^{\ast }\gamma ^{\ast }}$ in place of the local $\pi ^{0}\gamma \gamma $
vertex (\ref{WZW}). This is however beyond the scope of our paper. Instead
we make a simple estimate of the value of $\xi ^{r}(\mu )$ using its running
with the renormalization scale. From (\ref{xi_bar}) with central value of $%
\overline{\chi }=-16.8$ (see (\ref{chibar_num})) we get%
\begin{equation}
\Delta \xi ^{r}\equiv \left\vert \xi ^{r}(1\mathrm{GeV})-\xi ^{r}(0.5\mathrm{%
GeV})\right\vert =5.5
\end{equation}%
We therefore roughly estimate
\begin{equation}
\xi ^{r}(770\mathrm{MeV})=0\pm 5.5
\end{equation}%
and get finally%
\begin{equation}
\overline{\xi}=-32.3+3.7(\overline{\chi}+16.8)\pm5.5.  \label{xi_bar_num}
\end{equation}

\section{Soft photon bremsstrahlung\label{BS_section}}

Within the soft photon approximation, the amplitude of the process $\pi
^{0}\rightarrow e^{+}e^{-}\gamma $ factorizes
\begin{equation}
\mathcal{M}_{\pi ^{0}\rightarrow e^{+}e^{-}\gamma }=e\mathcal{M}_{\pi
^{0}\rightarrow e^{+}e^{-}}\left( \frac{(q_{-}\cdot \varepsilon ^{\ast
}(k,\lambda ))}{(q_{-}\cdot k)}-\frac{(q_{+}\cdot \varepsilon ^{\ast
}(k,\lambda ))}{(q_{+}\cdot k)}\right)
\end{equation}%
where $\varepsilon (k,\lambda )$ is the polarization vector of the emitted
photon with soft momentum $k$ and helicity $\lambda $. Taking the square of
the modulus, summing over helicities and integrating over the soft photon
region $|\mathbf{k}|<\omega $ defined by the experimental energy cut
\begin{equation}
\omega =\frac{1}{2}M_{\pi ^{0}}(1-x_{D}^{\mathrm{cut}})  \label{dalitz_x}
\end{equation}%
we get
\begin{equation}
\Gamma _{\pi ^{0}\rightarrow e^{+}e^{-}\gamma }=I_{BS}\Gamma _{\pi
^{0}\rightarrow e^{+}e^{-}}
\end{equation}%
where
\begin{equation}
I_{BS}=e^{2}\int_{|\mathbf{k}|<\omega }\frac{\mathrm{d}^{3}\mathbf{k}}{(2\pi
)^{3}2|\mathbf{k}|}\left[ \frac{2(q_{+}\cdot q_{-})}{(q_{-}\cdot
k)(q_{+}\cdot k)}-\frac{m^{2}}{(q_{-}\cdot k)^{2}}-\frac{m^{2}}{(q_{+}\cdot
k)^{2}}\right] .
\end{equation}%
The latter integral is IR divergent and can be regularized using DR. Let us
divide the resulting $I_{BS}$ into the diagonal part%
\begin{equation}
I_{BS}^{\mathrm{diag}}=-\mu ^{-2\varepsilon }\left( \frac{\mu }{m}\right)
^{2\varepsilon }\left( \frac{\alpha }{\pi }\right) \int_{|\mathbf{k}|<\omega
}\frac{\mathrm{d}^{3-2\varepsilon }\mathbf{k}}{(2\pi )^{1-2\varepsilon }2|%
\mathbf{k}|}\left[ \frac{m^{2+2\varepsilon }}{(q_{-}\cdot k)^{2}}+\frac{%
m^{2+2\varepsilon }}{(q_{+}\cdot k)^{2}}\right]
\end{equation}%
and non-diagonal part
\begin{equation}
I_{BS}^{\mathrm{non-diag}}=\mu ^{-2\varepsilon }\left( \frac{\mu }{m}\right)
^{2\varepsilon }\left( \frac{\alpha }{\pi }\right) \int_{|\mathbf{k}|<\omega
}\frac{\mathrm{d}^{3-2\varepsilon }\mathbf{k}}{(2\pi )^{1-2\varepsilon }2|%
\mathbf{k}|}\frac{2m^{2\varepsilon }(q_{+}\cdot q_{-})}{(q_{-}\cdot
k)(q_{+}\cdot k)}.
\end{equation}%
In the rest system of the decaying pion both integrals are elementary and we
get%
\begin{eqnarray}
I_{BS}^{\mathrm{diag}} &=&\mu ^{-2\varepsilon }\left( \frac{\mu }{m}\right)
^{2\varepsilon }\left( \frac{\alpha }{\pi }\right) \frac{(4\pi
)^{\varepsilon }}{\Gamma (1-\varepsilon )}\left( \frac{m}{\omega }\right)
^{2\varepsilon }\left[ \frac{1}{\varepsilon }-\frac{1}{\beta }H(0;z)-\ln
4+O(\varepsilon )\right] \notag \\
&=&\mu ^{-2\varepsilon }\left( \frac{\mu }{m}\right) ^{2\varepsilon }\left(
\frac{\alpha }{\pi }\right) \left[ \frac{1}{\varepsilon }-\frac{1}{\beta }%
H(0;z)-\ln 4+2\ln \left( \frac{m}{\omega }\right) -\overline{\gamma }%
+O(\varepsilon )\right]
\end{eqnarray}%
and%
\begin{eqnarray}
I_{BS}^{\mathrm{non-diag}} &=&\mu ^{-2\varepsilon }\left( \frac{\mu }{m}%
\right) ^{2\varepsilon }\left( \frac{\alpha }{\pi }\right) \frac{(4\pi
)^{\varepsilon }}{2\Gamma (1-\varepsilon )}\left( \frac{m}{\omega }\right)
^{2\varepsilon }\left( \beta +\frac{1}{\beta }\right)\notag \\
&&\times \left[ \frac{H(0;z)}{\varepsilon }-2\ln 2H(0;z)-2H(1,0;z)-H(0,0;z)-%
\frac{\pi ^{2}}{3}+O(\varepsilon )\right]\notag \\
&=&\mu ^{-2\varepsilon }\left( \frac{\mu }{m}\right) ^{2\varepsilon }\left(
\frac{\alpha }{\pi }\right) \frac{1}{2}\left( \beta +\frac{1}{\beta }\right) %
\left[ \frac{H(0;z)}{\varepsilon }-2H(1,0;z)-H(0,0;z)-\frac{\pi ^{2}}{3}%
\right. \notag \\
&&\left. +H(0;z)\left( 2\ln \left( \frac{m}{\omega }\right) -\overline{%
\gamma }-\ln 4\right) +O(\varepsilon )\right] .
\end{eqnarray}%
Note that the IR divergent part of $I_{BS}^{\mathrm{diag}}$ coincides up to
a sign with that of the $2\partial \Sigma (p)/\partial \slashed{p}$ (see (%
\ref{Z_factor})). The latter factor is necessary for the renormalization of
the external fermion lines. Namely, on the level of the decay width
\begin{equation}
Z_{e}^{2}\Gamma _{\pi ^{0}\rightarrow e^{+}e^{-}}=\left( 1+2\frac{\partial
\Sigma (p)}{\partial \slashed{p}}+O(\alpha ^{2})\right) \Gamma _{\pi
^{0}\rightarrow e^{+}e^{-}}.
\end{equation}%
As we have mentioned in Subsection \ref{one_loop_subsection_ct}, within our
renormalization scheme
\begin{equation}
2\frac{\partial \Sigma (p)}{\partial \slashed{p}}=-\mu ^{-2\varepsilon
}\left( \frac{\mu }{m}\right) ^{2\varepsilon }\left( \frac{\alpha }{\pi }%
\right) \frac{1}{\varepsilon }.
\end{equation}%
Therefore up to the assumed accuracy we can effectively make the following
replacement $\ $\
\begin{equation}
Z_{e}^{2}\Gamma _{\pi ^{0}\rightarrow e^{+}e^{-}}+I_{BS}\Gamma _{\pi
^{0}\rightarrow e^{+}e^{-}}\rightarrow I_{BS}\Gamma _{\pi ^{0}\rightarrow
e^{+}e^{-}}
\end{equation}%
provided we replace $I_{BS}^{\mathrm{diag}}$ with its finite part.

Taking this modification into account we can finally write the
bremsstrahlung integral $I_{BS}$ in the form (\ref{I_BS_general})
\begin{equation}
I_{BS}=\mu ^{-2\varepsilon }\left( \frac{\mu }{m}\right) ^{2\varepsilon
}\left( \frac{\alpha }{\pi }\right) \left[ \frac{I_{-1}}{\varepsilon }%
+I_{0}+O(\varepsilon )\right]
\end{equation}%
with
\begin{eqnarray}
I_{-1} &=&\frac{1}{2}\left( \beta +\frac{1}{\beta }\right) H(0;z) \\
I_{0} &=&-\frac{1}{\beta }H(0;z)+\ln \left( \frac{m}{2\omega }\right) ^{2}-%
\overline{\gamma }\notag \\
&&+\frac{1}{2}\left( \beta +\frac{1}{\beta }\right) \left[ H(0;z)\left( \ln
\left( \frac{m}{2\omega }\right) ^{2}-\overline{\gamma }\right)
-2H(1,0;z)-H(0,0;z)-\frac{\pi ^{2}}{3}\right]
\end{eqnarray}%
The relation (\ref{IR_relation2}) is now manifest. This completes the list
of all the necessary ingredients for the final calculation of the radiative
correction $\delta (x_{D}^{\mathrm{cut}})$.\bigskip

\section{The two-loop radiative correction}

\subsection{The exact two-loop result}

Putting the results of the previous sections together and using the general
formula (\ref{general_form_P}) we can finally write for the complete QED
two-loop correction (\ref{delta_definition}) schematically%
\begin{eqnarray}
\delta (x_{D}^{\mathrm{cut}}) &=&\left( \frac{\alpha }{\pi }\right) \left[
I_{0}+\mathrm{Re}\left( \frac{1}{\gamma _{LO}} \left( 2\overline{\xi }%
+\Delta ^{(1)}+2\Delta ^{(\Gamma )}+\Delta ^{(3)~}+2\Delta ^{(\Pi )}\right)\right) %
\right]  \notag \\
&=&\delta ^{BS}+\delta ^{(\overline{\xi })}+\delta ^{(1)}+\delta ^{(\Gamma
)}+\delta ^{(3)~}+\delta ^{(\Pi )}.  \label{delta_x_D}
\end{eqnarray}%
In this formula $\gamma _{LO}=\gamma _{0}^{\mathrm{1-loop}}+\overline{\chi }$
\ and the variable $x_{D}^{\mathrm{cut}}$ is connected to the maximal energy
$\omega $ of the soft photon included into the inclusive $\pi
^{0}\rightarrow e^{+}e^{-}\gamma (|\mathbf{k}|<\omega )$ decay rate by (\ref%
{dalitz_x}). \ In the above expression we have explicitly separated the
contribution coming from the bremsstrahlung, the RG invariant NLO coupling $%
\overline{\xi }$ and the individual graphs:
\begin{eqnarray}
\Delta ^{(1)} &=&2\gamma _{0}^{(1),~\mathrm{2-loop}}+2\overline{\chi }%
C_{0}^{(1),~\mathrm{1-loop}}+3C_{1}^{(1),~\mathrm{1-loop}}-2\gamma _{1}^{%
\mathrm{1-loop}}C_{-1,\mathrm{IR}}^{(1),~\mathrm{1-loop}} \notag \\
\Delta ^{(\Gamma )} &=&2\gamma _{0}^{(\Gamma ),~\mathrm{2-loop}}+2\overline{x%
}_{6}\gamma _{0}^{\mathrm{1-loop}}-\frac{1}{2}\gamma _{1}^{\mathrm{1-loop}}
\notag \\
&=&2\gamma _{0}^{(\Gamma ),~\mathrm{2-loop}}+\frac{3}{2}\left( \overline{%
\gamma }-\frac{5}{3}\right) \gamma _{0}^{\mathrm{1-loop}}-\frac{1}{2}\gamma
_{1}^{\mathrm{1-loop}} \notag \\
\Delta ^{(3)~} &=&2\gamma _{0}^{(3),~\mathrm{2-loop}}-2\overline{x}%
_{6}\gamma _{0}^{\mathrm{1-loop}}+2(\overline{x}_{6}+\overline{x}%
_{7})C_{0}^{(m),~\mathrm{1-loop}}+\frac{1}{2}\gamma _{1}^{\mathrm{1-loop}}+%
\frac{3}{2}C_{1}^{(m),~\mathrm{1-loop}} \notag \\
&=&2\gamma _{0}^{(3),~\mathrm{2-loop}}-\frac{3}{2}\left( \overline{\gamma }-%
\frac{5}{3}\right) \left( \gamma _{0}^{\mathrm{1-loop}}+C_{0}^{(m),~\mathrm{%
1-loop}}\right) +\frac{1}{2}\gamma _{1}^{\mathrm{1-loop}}+\frac{3}{2}%
C_{1}^{(m),~\mathrm{1-loop}} \notag \\
\Delta ^{(\Pi )} &=&2\gamma _{0}^{(\Pi ),~\mathrm{2-loop}}+2\overline{x}%
_{8}\gamma _{0}^{\mathrm{1-loop}}+\frac{2}{3}\gamma _{1}^{\mathrm{1-loop}} \notag \\
&=&2\gamma _{0}^{(\Pi ),~\mathrm{2-loop}}-\frac{2}{3}\overline{\gamma }%
\gamma _{0}^{\mathrm{1-loop}}+\frac{2}{3}\gamma _{1}^{\mathrm{1-loop}}.
\end{eqnarray}%
Here we have inserted for $\overline{x}_{i}$ the particular values
corresponding to our choice of the renormalization scheme (cf. Section \ref%
{one_loop_subsection_ct}) and $\gamma _{0}^{(i),~\mathrm{2-loop}}$, $%
C_{0}^{(i),~\mathrm{1-loop}}$, $C_{1}^{(i),~\mathrm{1-loop}}$, $C_{-1,%
\mathrm{IR}}^{(1),~\mathrm{1-loop}}$ and $\gamma _{1}^{\mathrm{1-loop}}$ are
explicitly given in Subsections \ref{two_loop_subsection}, \ref%
{one_loop_subsection} and \ref{LO_subsection} respectively.

The numerical results for various contributions are summarized in
the second
column of Tab. \ref{table_num}. Here we use for the constants $\overline{%
\chi }$ and $\overline{\xi }$ the values (\ref{chibar_num}) and (\ref%
{xi_bar_num}) \ and the following numerical entries: $M_{\pi ^{0}}=135%
\mathrm{MeV}$, $m=0.51\mathrm{MeV}$ and $\alpha =1/137$. For the sum of all
the contributions we get finally
\begin{equation}
\delta ^{(\overline{\xi })}+\delta ^{(1)}+\delta ^{(\Gamma )}+\delta
^{(3)~}+\delta ^{(\Pi )}=\left( -0.8\pm 0.2\right) \%,
\end{equation}%
where as the only source of errors we take the uncertainties of $\overline{%
\chi }$ and $\overline{\xi }$. We observe considerable
cancellation between the large contributions $\delta ^{(1)}$ and
$\delta ^{(\Gamma )}$ which makes the role of the relatively
smaller contributions of the other graphs numerically important.

Adding the bremsstrahlung we have
\begin{equation}
\delta (x_{D}^{\mathrm{cut}})=\left( 4.7\ln (1-x_{D}^{\mathrm{cut}})+8.3\pm
0.2\right) \%
\end{equation}%
and for $x_{D}^{\mathrm{cut}}=0.95$ which is the cut used by KTeV
we get
\begin{equation}
\delta (x_{D}^{\mathrm{cut}}=0.95)=\left( -5.8\pm 0.2\right) \%.
\label{delta_num_exact_0.95}
\end{equation}

\subsection{Large-logarithm approximations to the exact result}

The exact two-loop expressions for $\delta (x_{D}^{\mathrm{cut}})$ and for $%
\Delta ^{(i)}$ are rather long but they can be approximated with a very good
accuracy performing the large-logarithm (LL) expansion in terms of
\begin{equation}
L=-\ln z=-\ln \left( \frac{1-\beta }{1+\beta }\right) \sim \ln \left( \frac{%
M_{\pi ^{0}}^{2}}{m^{2}}\right)
\end{equation}%
and taking into account only the leading terms (up to the order $O(L^{-1})$%
). \ For the various contributions we get\footnote{%
In these and following formulae we omit also all the terms of the order $%
O(1-\beta )$.}
\begin{eqnarray}
\gamma _{LO} &=&\frac{L^{2}}{4}+\bar{\chi}+\frac{3}{2}\bar{\gamma}+\frac{\pi
^{2}}{12}-\frac{5}{2}-\frac{\mathrm{i}\pi }{2}L+O\left( L^{-1}\right)  \notag
\\
&& \\
I_{0} &=&\frac{L^{2}}{2}+(L-1)\left( 2\ln \left( 1-x_{D}^{\mathrm{cut}%
}\right) +\bar{\gamma}\right) -\frac{\pi ^{2}}{3}+O\left( L^{-1}\right)
\notag \\
&& \\
\Delta ^{(1)} &=&\frac{1}{16}L^{4}-L^{3}\left( \frac{\bar{\gamma}}{4}%
+1\right) -L^{2}\left( \frac{1}{2}\bar{\chi}+\frac{3}{4}\bar{\gamma}+\frac{%
\pi ^{2}}{6}-\frac{17}{8}\right)  \notag \\
&&-L\left[ \bar{\chi}\bar{\gamma}+\bar{\gamma}\left( \frac{3}{2}\bar{\gamma}-%
\frac{5}{12}\pi ^{2}-\frac{5}{2}\right) -2\pi ^{2}-\frac{1}{2}\right]  \notag
\\
&&+\frac{1}{2}\left( \frac{4}{3}\pi ^{2}-\bar{\gamma}+3\right) \bar{\chi}-%
\frac{3}{8}\bar{\gamma}^{2}+\left( \pi ^{2}+\frac{17}{8}\right) \bar{\gamma}
\notag \\
&&+\frac{7}{144}\pi ^{4}-\frac{49}{24}\pi ^{2}-\frac{109}{32}  \notag \\
&&+\mathrm{i}\pi \left[ -\frac{1}{4}L^{3}+\frac{3}{4}L^{2}\left( \bar{\gamma}%
+4\right) +L\left( \bar{\chi}+\frac{3}{2}\bar{\gamma}-\frac{\pi ^{2}}{6}-%
\frac{17}{4}\right) \right.  \notag \\
&&\left. +\bar{\chi}\bar{\gamma}+\bar{\gamma}\left( \frac{3}{2}\bar{\gamma}+%
\frac{\pi ^{2}}{12}-\frac{5}{2}\right) -\frac{1}{2}\right] +O\left(
L^{-1}\right)  \notag \\
&& \\
\Delta ^{(\Gamma )} &=&-\frac{1}{12}L^{4}+\frac{2}{3}L^{3}+\frac{1}{4}%
L^{2}\left( \bar{\gamma}+\pi ^{2}-\frac{15}{2}\right) +L\left( \zeta (3)-%
\frac{11}{12}\pi ^{2}+\frac{3}{2}\right)  \notag \\
&&+\frac{15}{8}\bar{\gamma}^{2}+\frac{\pi ^{2}}{12}\bar{\gamma}-\frac{49}{8}%
\bar{\gamma}-\frac{5}{2}\zeta (3)+\frac{11}{120}\pi ^{4}-\frac{\pi ^{2}}{24}+%
\frac{129}{32}  \notag \\
&&+\mathrm{i}\pi \left( \frac{1}{3}L^{3}-2L^{2}+\frac{1}{2}L\left( \frac{\pi
^{2}}{3}-\bar{\gamma}+\frac{15}{2}\right) -\zeta (3)-\frac{5}{12}\pi ^{2}-%
\frac{3}{2}\right) +O\left( L^{-1}\right)  \notag \\
&& \\
\Delta ^{(3)~} &=&-\frac{L^{3}}{3}+L^{2}\left( 4-\frac{7}{4}\bar{\gamma}%
\right) +L\left( 12\bar{\gamma}+\frac{5}{4}\pi ^{2}-18\right)  \notag \\
&&-\frac{15}{8}\bar{\gamma}^{2}-\left( \frac{7}{12}\pi ^{2}+\frac{53}{8}%
\right) \bar{\gamma}+\frac{19\zeta (3)}{2}-\frac{21\pi ^{2}}{4}+\frac{677}{32%
}  \notag \\
&&+\mathrm{i}\pi \left( L^{2}+L\left( \frac{7}{2}\bar{\gamma}-8\right) -%
\frac{7}{12}\pi ^{2}-12\bar{\gamma}+18\right) +O\left( L^{-1}\right)  \notag
\\
&& \\
\Delta ^{(\Pi )} &=&\frac{1}{9}L^{3}-\frac{11}{18}L^{2}-\frac{2\pi ^{2}}{9}L+%
\frac{67}{12}-\frac{1}{2}\bar{\gamma}\left( \bar{\gamma}+\frac{8}{3}\right)
\notag \\
&&-\frac{\pi \mathrm{i}}{3}\left( L^{2}-\frac{11}{3}L\right) +O\left(
L^{-1}\right) \label{LL_approx}
\end{eqnarray}%
and
\begin{eqnarray}
\Delta ^{(1)}+2\Delta ^{(\Gamma )}+\Delta ^{(3)~}+2\Delta ^{(\Pi )} &=&-%
\frac{5L^{4}}{48}+L^{3}\left( \frac{2}{9}-\frac{\bar{\gamma}}{4}\right)
-L^{2}\left( \frac{\bar{\chi}}{2}+2\bar{\gamma}-\frac{\pi ^{2}}{3}-\frac{83}{%
72}\right) \notag \\
&&-L\left[ \bar{\gamma}\bar{\chi}+\frac{3\bar{\gamma}^{2}}{2}-\left( \frac{29%
}{2}+\frac{5\pi ^{2}}{12}\right) \bar{\gamma}-2\zeta (3)-\frac{35\pi ^{2}}{36%
}+\frac{29}{2}\right] \notag \\
&&-\left( \frac{\bar{\gamma}}{2}-\frac{2}{3}\pi ^{2}-\frac{3}{2}\right) \bar{%
\chi}+\frac{\bar{\gamma}^{2}}{2}+\left( \frac{7}{12}\pi ^{2}-\frac{233}{12}%
\right) \bar{\gamma} \notag \\
&&+\frac{9\zeta (3)}{2}+\frac{167\pi ^{4}}{720}-\frac{59\pi ^{2}}{8}+\frac{%
1775}{48} \notag \\
&&+\mathrm{i}\pi \left[ \bar{\gamma}\bar{\chi}+\frac{3\bar{\gamma}^{2}}{2}%
+\left( \frac{\pi ^{2}}{12}-\frac{29}{2}\right) \bar{\gamma}+L^{2}\left(
\frac{3\bar{\gamma}}{4}-\frac{2}{3}\right) \right. \notag \\
&&\left. +L\left( 4\bar{\gamma}+\bar{\chi}+\frac{\pi ^{2}}{6}-\frac{83}{36}%
\right) +\frac{5L^{3}}{12}-2\zeta (3)-\frac{17\pi ^{2}}{12}+\frac{29}{2}%
\right] \notag \\
&&+O\left( L^{-1}\right) .
\end{eqnarray}%
\begin{table}[tbp]
\begin{center}
\begin{tabular}{|c|c|c|c|c|c|}
\hline $\delta ^{(i)}[\%]$ & exact & LL, rational & LL, polynomial
& LL, polynomial, RG & $\overline{\chi }\rightarrow \infty $ \\
\hline
$\delta ^{(\overline{\xi })}$ & $-0.36$ & $-0.36$ & $0$ & $-0.92$ & $0$ \\
$\delta ^{(1)}$ & $18.4$ & $18.4$ & $18.8$ & $19.5$ & $-7.29$ \\
$\delta ^{(\Gamma )}$ & $-22.2$ & $-22.2$ & $-24.6$ & $-22.8$ & $0$ \\
$\delta ^{(3)~}$ & $0.92$ & $0.92$ & $3.43$ & $1.58$ & $0$ \\
$\delta ^{(\Pi )}$ & $2.45$ & $2.45$ & $1.17$ & $2.41$ & $0$ \\
$\delta(x_D^{\mathrm{cut}}=0.95)$ & $-5.84$ & $-5.84$ & $-6.29$ & $-5.28$ & $%
-12.4$ \\ \hline
\end{tabular}%
\end{center}
\caption{Numerical values of the contributions of the individual
graphs to the total two-loop correction $\protect\delta$ within
various approximations described in the main text.}
\label{table_num}
\end{table}
Inserting the above expressions into (\ref%
{delta_x_D}) we get an approximation of $\delta
(x_{D}^{\mathrm{cut}})$ in terms of the rational function of the
variable $L$.
It would be tempting to expand further the factor $1/\gamma _{LO}$ in (\ref%
{delta_x_D}) and approximate the whole $\delta (x_{D}^{\mathrm{cut}})$ as a
second order polynomial of $L$ with the result%
\begin{equation}
\delta ^{\mathrm{LL~polynomial}}(x_{D}^{\mathrm{cut}})=\left( \frac{\alpha }{%
\pi }\right) \left[ \frac{L^{2}}{12}+\frac{8}{9}L+2(L-1)\ln (1-x_{D}^{%
\mathrm{cut}})-\frac{1}{3}\bar{\chi}-\frac{13}{2}\bar{\gamma}-\frac{19}{36}%
\pi ^{2}+\frac{4}{9}\right]
\end{equation}%
or numerically%
\begin{equation}
\delta ^{\mathrm{LL~polynomial}}(x_{D}^{\mathrm{cut}})=\left( 4.7\ln
(1-x_{D}^{\mathrm{cut}})+7.9\right) \%
\end{equation}%
and%
\begin{equation}
\delta ^{\mathrm{LL~polynomial}}(x_{D}^{\mathrm{cut}}=0.95)=-6.3\%.
\end{equation}%
We can also proceed apparently more carefully and include part of the large
RG logarithms into the expansion writing $\ln \left( \mu ^{2}/m^{2}\right)
=-L+\ln \left( \mu ^{2}/M_{\pi ^{0}}^{2}\right) $; such a modification makes
a difference in the $O(L^{0})$ terms of the expansion. We get in this case%
\footnote{%
In the final expression we expressed again $\chi ^{r}(\mu )$ and $\ln (\mu
^{2}/M_{\pi ^{0}}^{2})$ in terms of $\overline{\chi }$ and $L$.}

\begin{eqnarray}
\delta _{RG}^{\mathrm{LL~polynomial}}(x_{D}^{\mathrm{cut}}) &=&\left( \frac{%
\alpha }{\pi }\right) \left[ \frac{L^{2}}{12}+\frac{8}{9}L+2(L-1)\ln
(1-x_{D}^{\mathrm{cut}})-\frac{1}{3}\bar{\chi}-\frac{13}{2}\bar{\gamma}-%
\frac{19}{36}\pi ^{2}+\frac{43}{9}\right]  \notag \\
&& \notag \\
&=&\left( 4.7\ln (1-x_{D}^{\mathrm{cut}})+8.9\right) \%
\end{eqnarray}%
which gives%
\begin{equation}
\delta _{RG}^{\mathrm{LL~polynomial}}(x_{D}^{\mathrm{cut}}=0.95)=-5.3\%.
\end{equation}

In both cases the reason for not very good agreement with the exact result
is that the expansion of $1/\gamma _{LO}$ does not converge well\footnote{%
While $1/\gamma _{LO}=0.024+0.044\mathrm{i}$, the approximation up to the
order $O(L^{-6})$ gives $[1/\gamma _{LO}]^{(5)}=0.044+0.037\mathrm{i}$.
Including part of the large RG logarithms into the expansion as described in
the main text we get $[1/\gamma _{LO}]^{(5)}=0.047+0.053\mathrm{i.}$} and
therefore it is much safer to approximate $\delta (x_{D}^{\mathrm{cut}})$ in
terms of rational function of $L$. We have compared all the possibilities
(LL rational, LL polynomial and LL polynomial with RG logs included) in the
third, fourth and fifth column of Tab. \ref{table_num} respectively. The
rational LL approximation is in excellent agreement with the exact result.

The LL approximation was originally calculated by Dorokhov\emph{\ et al.} in
\cite{Dorokhov:2008qn} including only the graphs (1)-(4) in the Fig. \ref%
{NLO_ChPT_2loop} and taking into account only\ selected regions of
the two-loop integration space. As we know from (\ref{LL_approx}),
the omitted graphs contribute only in the next to leading order in
the LL expansion, however due to the accidental cancellation of
the leading order terms they are in fact numerically important.
The result of \cite{Dorokhov:2008qn} for the virtual and soft
photon corrections is
\begin{eqnarray}
\delta _{\mathrm{Dorokhov}}^{virt.+soft~\gamma }(x_{D}^{%
\mathrm{cut}}) &=&\left( \frac{\alpha }{\pi }\right) \left[ -\frac{1}{24}%
L^{2}+2(L-1)\ln \left( 1-x_{D}^{\mathrm{cut}}\right) +\frac{3}{4}L-\frac{\pi
^{2}}{6}+2\right]  \notag \\
&&  \label{delta_Dorokhov} \\
&=&\left( 4.7\ln (1-x_{D}^{\mathrm{cut}})+6.7\right) \%  \notag
\end{eqnarray}%
which numerically gives for $x_{D}^{\mathrm{cut}}=0.95$ a correction with
significantly larger absolute value in comparison with (\ref%
{delta_num_exact_0.95}),
\begin{equation}
\delta _{\mathrm{Dorokhov}}^{virt.+soft~\gamma }(x_{D}^{%
\mathrm{cut}}=0.95)=-13.3\%.  \label{delta_Dorokhov_num}
\end{equation}%
The formula (\ref{delta_Dorokhov}) could be compared with our polynomial LL
approximations after subtracting the contributions of the graphs (5) and (6)
in Fig. \ref{NLO_ChPT_2loop} (\emph{i.e.} those with vacuum polarization
insertion into the internal photon lines). In \ the two variants described
above we get however a result substantially different from $\delta _{\mathrm{Dorokhov}}^{virt.+soft~\gamma }(x_{D}^{\mathrm{cut}})$,
namely
\begin{equation}
\delta _{(1)-(4)}^{\mathrm{LL~polynomial}}(x_{D}^{\mathrm{cut}})=\left(
\frac{\alpha }{\pi }\right) \left[ \frac{L^{2}}{12}+2(L-1)\ln \left(
1-x_{D}^{\mathrm{cut}}\right) -\frac{\bar{\chi}}{3}-\frac{13}{2}\bar{\gamma}-%
\frac{19\pi ^{2}}{36}+\frac{16}{3}\right]
\end{equation}%
and
\begin{equation}
\delta _{RG,~(1)-(4)}^{\mathrm{LL~polynomial}}(x_{D}^{\mathrm{cut}})=\left(
\frac{\alpha }{\pi }\right) \left[ \frac{L^{2}}{12}+2(L-1)\ln \left(
1-x_{D}^{\mathrm{cut}}\right) -\frac{\bar{\chi}}{3}-\frac{13}{2}\bar{\gamma}-%
\frac{19\pi ^{2}}{36}+\frac{13}{3}\right]
\end{equation}%
and numerical values for $x_{D}^{\mathrm{cut}}=0.95$
\begin{eqnarray}
\delta _{(1)-(4)}^{\mathrm{LL~polynomial}}(x_{D}^{\mathrm{cut}}
&=&0.95)=-7.9\% \\
\delta _{RG,~(1)-(4)}^{\mathrm{LL~polynomial}}(x_{D}^{\mathrm{cut}}
&=&0.95)=-7.6\%.
\end{eqnarray}%
The reason of this discrepancy is difficult to trace out because a
completely different framework has been used for the calculations in \cite%
{Dorokhov:2008qn} and we therefore left this problem open for further study.

\subsection{Point-like \texorpdfstring{$\protect\pi ^{0}e^{+}e^{-}$}{pi0e+e-} vertex approximation \label{Bergstrom_subsection}} 

Let us briefly comment on another type of approximation which is connected
with the model calculation of the QED radiative corrections by Bergstr\"{o}m
\cite{Bergstrom:1982wk}. His result including virtual corrections and soft
photon bremsstrahlung reads%
\begin{equation}
\delta _{\mathrm{Bergstr\ddot{o}m}}^{virt.+soft~\gamma }(x_{D}^{\mathrm{cut}%
})=\left( \frac{\alpha }{\pi }\right) \left[ 2(L-1)\ln \left( 1-x_{D}^{%
\mathrm{cut}}\right) +\frac{\pi ^{2}}{3}-1+O(1-x_{D}^{\mathrm{cut}})\right] ,
\label{delta_bergstroem}
\end{equation}%
where the $O(1-x_{D}^{\mathrm{cut}})$ terms, which we do not write
explicitly, stem from the fact that the real photon radiation was calculated
exactly \emph{i.e.} beyond the soft photon approximation. Taking also these
terms into account we get numerically
\begin{equation}
\delta _{\mathrm{Bergstr\ddot{o}m}}^{virt.+soft~\gamma }(x_{D}^{\mathrm{cut}%
}=0.95)=-13.8\%,
\end{equation}%
which is remarkably close to (\ref{delta_Dorokhov_num}). However,
the calculations \cite{Bergstrom:1982wk} used a completely
different approximation from the LL one that was used in
\cite{Dorokhov:2008qn}.

In the paper \cite{Bergstrom:1982wk} \ the approximation was based on the
substitution for the nonlocal one-loop $\pi ^{0}e^{+}e^{-}$ (sub)graph with
a local effective vertex of the form%
\begin{equation}
\mathcal{L}_{\mathrm{eff}}=\mathrm{i}g_{_{\mathrm{eff}}}\overline{e}\gamma
_{5}e\pi ^{0}.
\end{equation}%
In appropriate renormalization scheme this vertex is essentially equivalent%
\footnote{%
We ignore here the axial anomaly which does not contribute to the relevant $%
\langle e^{+}e^{-}|\partial _{\mu }\overline{e}\gamma ^{\mu }\gamma
_{5}e|0\rangle $ matrix element.} to the vertex%
\begin{equation}
\widetilde{\mathcal{L}}_{\mathrm{eff}}=-\frac{g_{_{\mathrm{eff}}}}{2m}%
\overline{e}\gamma ^{\mu }\gamma _{5}e\partial _{\mu }\pi ^{0}\equiv -\frac{1%
}{4F_{0}}\left( \frac{\alpha }{\pi }\right) ^{2}\chi _{\mathrm{eff}}%
\overline{e}\gamma ^{\mu }\gamma _{5}e\partial _{\mu }\pi ^{0},
\end{equation}%
which takes properly into account the GB nature of pion and has the same
structure as the counterterm (\ref{CT_alpha^2p^2}). The Bergstr\"{o}m's
calculation can be therefore qualitatively understood as the leading order
term in the formal large $\overline{\chi }$ expansion of the full two-loop
result \footnote{%
Note that in the limit $\overline{\chi }\rightarrow \infty $ only the
contribution of the one-loop graph (1) of Fig. \ref{NLO_ChPT_1loop} and soft
bremsstrahlung are effectively taken into account.} (\emph{i.e. }%
corresponding to the assumption of small nonlocal part of one-loop $\pi
^{0}e^{+}e^{-}$ (sub)graph \ in reference to its local part represented by $%
\overline{\chi }$). Performing further the LL expansion of this leading term
we get for the Bergstr\"{o}m-like approximation of our result\footnote{%
The difference between (\ref{delta_bergstroem}) and (\ref{delta_chi_infinity}%
) corresponds to different regularization of the IR divergences and
different renormalization scheme.}%
\begin{equation}
\delta _{\overline{\chi }\rightarrow \infty }(x_{D}^{\mathrm{cut}})=\left(
\frac{\alpha }{\pi }\right) \left[ 2(L-1)\ln \left( 1-x_{D}^{\mathrm{cut}%
}\right) +\frac{\pi ^{2}}{3}+\frac{3}{2}\left( 1-\bar{\gamma}\right)
+O\left( \overline{\chi }^{-1},L^{-1}\right) \right] .
\label{delta_chi_infinity}
\end{equation}%
For illustration purposes we add the corresponding numerical values as the
sixth column of Tab. \ref{table_num}. However for physical value of $%
\overline{\chi }$ such an expansion does not converge (note that $|\gamma
_{0}^{\mathrm{1-loop}}/\overline{\chi }|\sim 2$) and therefore the Bergstr%
\"{o}m's result can not be taken as a serious approximation of the full
two-loop $\delta (x_{D}^{\mathrm{cut}})$.

\section{The phenomenological applications of the
results: first look}

In this section we discuss several issues connected with the
phenomenological aspects of the results obtained above. Up to now
we have fixed or estimated the free  parameters $\chi^r(\mu)$ and
$\xi^r({\mu})$ of the effective Lagrangian and concentrated on the
corresponding prediction of the two-loop QED corrections. In such
a way obtained $\delta(x_D^{{\mathrm{cut}}})$ represents a part of
the theoretical tools necessary for a precise extraction of the
$B^{{\mathrm{no-rad}}}(\pi^0\to e^+e^-)$ from experimental data.
Therefore a consistency check of our final result (which uses one
specific value of $\chi^r(\mu)$) with that using other estimates
of $\chi^r(\mu)$ which are available in the literature should be
desirable. Such a check is the topic of the first subsection.

We can also, however, reverse the point of view and investigate
the sensitivity of the theoretical prediction for the branching
ratio $B(\pi^0\to e^+e^-(\gamma),x_D>0.95)$ on  the free parameter
$\chi^r(\mu)$ and try to extract the information on its actual
value from the experimental data. Though we have not all the
necessary ingredients at hand, we can make a preliminary analysis
of this issue. This is done in the second subsection.

Last but not least, our result is closely related to the
calculation of two-loop QED corrections to the class of processes
of the type $P\to l^+ l^-$ where $P=\pi ^{0},~\eta ,~K_{L}$ and
$l=e,~\mu $; this relation is briefly discussed in the last
subsection.

\begin{table}[t]
\begin{center}
\begin{tabular}{|c|c|c|c|c|c|c|c|}
\hline
& \text{CLEO bound} & \text{CLEO+OPE} & \text{QCDsr} & \text{LMD+V} & \text{%
QM} & \text{N$\chi $QM} & \text{VM}  \\ \hline
$\overline{\chi}$ & $-17.8$ & $-16.5\pm 0.3$ & $-16.3\pm 0.1$ & $-16.5$ & $%
-18.0\pm 0.5$ & $-16.7\pm 0.5$ & $-19.1$   \\
$\chi ^{r}$ & $1.3 $ & $2.6\pm 0.3$ & $2.8\pm 0.1 $ & $2.5$ & $1.1\pm 0.5$ &
$2.4\pm 0.5 $ & $-0.05$   \\
\ $\delta[\%]$ & $-5.68$ & $-5.89$ & $-5.92$ & $-5.88$ & $-5.65$ & $-5.85$ & $%
-5.50$   \\ \hline
\end{tabular}%
\end{center}
\caption{Illustration of the sensitivity of the central value of the two
loop QED correction $\protect\delta$ on the various values of $\overline{%
\protect\chi}$ described in the main text. The renormalized $\protect\chi^r$
is taken at $\protect\mu=M_\protect\rho$.}
\label{table_chi}
\end{table}

\begin{figure}[t]
\centering\includegraphics[scale=1]{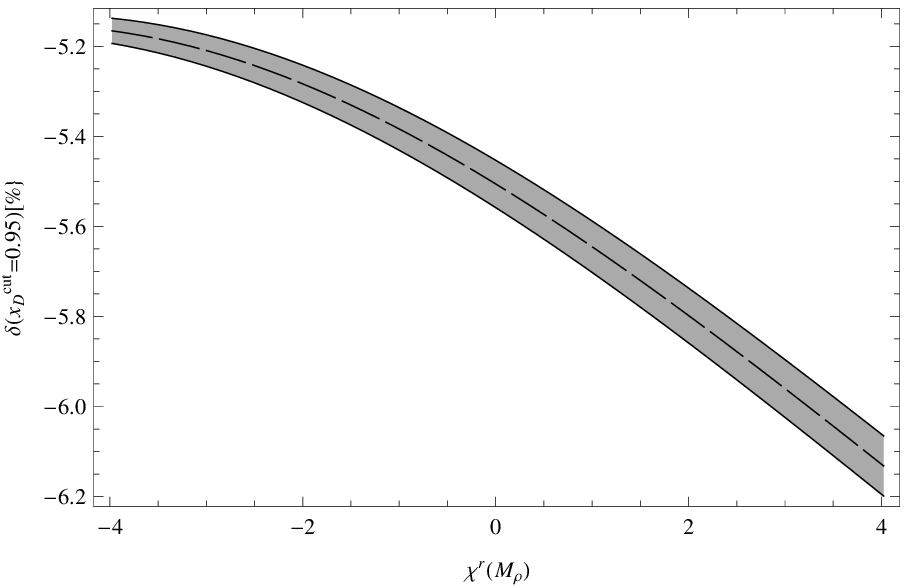}\newline
\caption{The dependence of $\protect\delta (x_{D}^{\mathrm{cut}}=0.95)$ on $%
\protect\chi^{r}(M_\protect\rho)$. The filled band corresponds to the
variation of $\protect\delta (x_{D}^{\mathrm{cut}}=0.95)$ with $\protect\xi%
^r(M_\protect\rho)$ inside its error bar.}
\label{chi_dependence}
\end{figure}

\subsection{Note on the dependence on \texorpdfstring{$\overline{\protect\chi }$}{chi}}

In the above numerical calculations we have fixed the value of the constant $%
\overline{\chi }$ according to the large $N_{C}$ inspired LMD estimate (\ref%
{chirLMD}) of the effective coupling $\chi ^{r}(\mu )$ entering the Lagrangian (\ref%
{CT_alpha^2p^2}). In the literature there exist, however, further model
dependent estimates of this constant based on various models or
phenomenological parameterizations of the pion transition form factor $%
F_{\pi ^{0}\gamma ^{\ast }\gamma ^{\ast }}$ (see \cite{Dorokhov:2007bd} for
comprehensive review). In Tab. \ref{table_chi} we summarize various values of
$\overline{\chi }$ and $\chi ^{r}(M_{\rho })$ which we take over\footnote{%
In \cite{Dorokhov:2007bd}, the values of ${\mathcal{A}}(0)=\overline{\chi }+%
\frac{3}{2}\overline{\gamma }-\frac{5}{2}$ are presented.} from \cite%
{Dorokhov:2007bd}. The first three columns (denoted as CLEO bound, CLEO+OPE
and QCDsr) correspond to various treatments of the parametrization of $F_{\pi
^{0}\gamma ^{\ast }\gamma ^{\ast }}(t,t)$ using the CLEO \cite%
{Gronberg:1997fj} data (see \cite{Dorokhov:2007bd} for details), next column
(LMD+V) is the improvement of the large $N_{C}$ estimate mentioned above
including two $1^{--}$ multiplets\footnote{%
This ansatz is denoted gVMD in \cite{Dorokhov:2007bd}} \cite{Knecht:1999gb}.
The column QM is based on the constituent quark model while the last two
columns (N$\chi $QM and VM) on two variants of the nonlocal chiral quark
model \cite{Dorokhov:2002iu} and \cite{Efimov:1981vh} respectively. In the
the last row we illustrate the sensitivity of our result for the central
value of $\delta (x_{D}^{\mathrm{cut}}=0.95)$ on $\overline{\chi }$. All but
the last resulting central values of $\delta (x_{D}^{\mathrm{cut}}=0.95)$
are compatible with LMD result (\ref{delta_num_exact_0.95}) within the
estimated error bar. The dependence of $\delta (x_{D}^{\mathrm{cut}}=0.95)$
on $\chi ^{r}(M_{\rho })$ in a wider range is plotted in Fig. \ref%
{chi_dependence} where also the variation with $\xi ^{r}(M_{\rho })$ inside
its estimated error bar is illustrated by the filled band. 
\footnote{Note that for fixed $\overline{\chi}$ the dependence of $\delta (x_{D}^{\mathrm{cut}}=0.95)$ 
on $\overline{\xi}$ is trivial (i.e. linear as can be seen from (\ref{delta_x_D})).}

\subsection{Note on the phenomenological determination of 
\texorpdfstring{$\chi^r(M_\rho)$ from $\pi^0\to e^+e^-$}{chir(Mrho) from pi0e+e-} decay}

Let us stress that the above analytical result for $\delta (x_{D}^{\mathrm{%
cut}})$ represents only a part of the problem of \ the complete QED
radiative corrections to the process under consideration, and that the
realistic analysis of the experimental data requires several additional
pieces of information. The first one corresponds to the issue of the hard
photon bremsstrahlung for which the soft photon approximation is not an
adequate framework and for which more appropriate calculations have to be
done. A closely related issue is the applicability of soft photon
approximation for the KTeV \ choice of the cut $x_{D}^{\mathrm{cut}}=0.95$.
\ Another missing information is connected with the \ Dalitz decay which
yields the same final state as the real photon bremsstrahlung in the $\pi
^{0}\rightarrow e^{+}e^{-}$ decay. Though this process is dominated by low $%
x_{D}$ and is therefore suppressed for $x_{D}>0.95$, its integrated
contribution is known to grow rapidly with decreasing $x_{D}^{\mathrm{cut}}$.

These additional issues have been addressed in the present context already
in the paper \cite{Bergstrom:1982wk}, (see also \cite{Dorokhov:2008qn}) and
in this form they have been used for the analysis of the experimental data
by the KTeV collaboration \cite{Abouzaid:2006kk}. However, the analysis performed in \cite%
{Bergstrom:1982wk} might be incomplete. As far as the hard photon
bremsstrahlung is concerned, the point-like $\pi ^{0}e^{+}e^{-}$
vertex approximation described in Subsection \ref{Bergstrom_subsection} has been used.
This approach is, however, well justified only for the soft photon
region where the details of the off-shell $\pi ^{0}e^{+}e^{-}$
vertex are inessential.

Also, the corrections due to the Dalitz decay calculated in \cite%
{Bergstrom:1982wk} (based on the radiative corrections to $\pi
^{0}\rightarrow e^{+}e^{-}\gamma (\gamma )$ obtained in \cite%
{Mikaelian:1972yg}) and used in \cite{Abouzaid:2006kk} should be
taken with some caution. As it
has been shown recently \cite{Kampf:2005tz}, the one-photon irreducible ($%
1\gamma IR$) contributions which were omitted in \cite{Mikaelian:1972yg} are
in fact important already for $x_{D}>0.6$ where they give a negative
contribution   $\delta ^{1\gamma IR}(x_{D})<-1\%$ of the leading order
differential decay rate $(\mathrm{d}\Gamma ^{\mathrm{Dalitz}}/\mathrm{d}%
x_{D})^{LO}$.

The importance of the detailed knowledge of $\mathrm{d}\Gamma
^{\mathrm{Dalitz}}/\mathrm{d}x_{D}$ is twofold. On one hand,
because the Dalitz decay has been used by KTeV-E799-II as a
normalization and because
the measured quantity was the ratio%
\begin{equation}
r=\frac{\Gamma (\pi ^{0}\rightarrow e^{+}e^{-},x_{D}>0.95)}{\Gamma (\pi
^{0}\rightarrow e^{+}e^{-}\gamma ,x_{D}>0.232)}=(1.685\pm 0.064\pm
0.027)\times 10^{-4},
\end{equation}%
it is necessary to extrapolate the Dalitz branching ratio to the full range
of $x_{D}$. As this extrapolation is concerned, the missing corrections to $%
\mathrm{d}\Gamma ^{\mathrm{Dalitz}}/\mathrm{d}x_{D}$ are in fact
inessential, since the contribution of $1\gamma IR$ corrections integrated
over the full phase space can be shown to be one order of magnitude smaller
then the experimental error of the $\Gamma (\pi ^{0}\rightarrow
e^{+}e^{-}\gamma )/\Gamma (\pi ^{0}\rightarrow \gamma \gamma )$ branching
ratio (see \cite{Kampf:2005tz} for details). On the other hand, $\mathrm{d}%
\Gamma ^{\mathrm{Dalitz}}/\mathrm{d}x_{D}$ has been used in order to
subtract the Dalitz decay background in the region $x_{D}>0.95$. Here the
effect of missing contributions might be more important.

Let us  give here only a preliminary illustration of the
interrelation of our partial result of the QED radiative
corrections and the precise branching ratio (\ref{KTeV inclusive})
obtained by KTeV-E799-II. Taking the above theoretical
uncertainties into account we can write our
prediction for the KTeV measured branching ratio as%
\begin{eqnarray}
B(\pi ^{0} &\rightarrow &e^{+}e^{-}(\gamma ),x_{D}>0.95)=B(\pi
^{0}\rightarrow \gamma \gamma ) \notag \\
&&\times \frac{\Gamma ^{LO}(\pi ^{0}\rightarrow e^{+}e^{-})~}{\Gamma (\pi
^{0}\rightarrow \gamma \gamma )}(1+\delta (0.95)+\Delta ^{BS}(0.95)+\Delta
^{1\gamma IR}(0.95)),
 \label{B_95_prediction}
\end{eqnarray}%
where the only experimental input is the precise branching ratio $B(\pi
^{0}\rightarrow \gamma \gamma )=$ $(98.823\pm 0.034)\%$. In the above
formula
\begin{equation}
\Delta ^{BS}(x_{D}^{\mathrm{cut}})\equiv \delta _{exact}^{BS}(x_{D}^{\mathrm{%
cut}})-\delta _{soft}^{BS}(x_{D}^{\mathrm{cut}})
\end{equation}%
is the difference between the soft photon and exact bremsstrahlung
correction and
\begin{equation}
\Delta ^{1\gamma IR}(x_{D}^{\mathrm{cut}})=\frac{1}{\Gamma ^{LO}(\pi
^{0}\rightarrow e^{+}e^{-})}\int_{x_{D}^{\mathrm{cut}}}^{1}\mathrm{d}%
x_{D}\left( \frac{\mathrm{d}\Gamma ^{\mathrm{Dalitz}}}{\mathrm{d}x_{D}}%
\right) ^{LO}\delta ^{1\gamma IR}(x_{D})
\end{equation}%
corresponds to the unsubtracted fraction of the Dalitz decay
background discussed above. Without detailed knowledge of the
exact bremsstrahlung we can only roughly estimate the error
$\Delta ^{BS}(x_{D}^{\mathrm{cut}})$ of the soft photon
approximation with help of the point-like $\pi ^{0}e^{+}e^{-}$
vertex approximation used in \cite{Bergstrom:1982wk}. We get
\begin{equation}
\Delta ^{BS}(x_{D}^{\mathrm{cut}})=-2\left( \frac{\alpha }{\pi }\right)
(L-1)(1-x_{D}^{\mathrm{cut}})+O\left( \frac{m^{2}}{M^{2}},(1-x_{D}^{\mathrm{%
cut}})^{2}\right)
\end{equation}%
which gives for $x_{D}^{\mathrm{cut}}=0.95$ a reasonable difference%
\begin{equation}
\Delta ^{BS}(0.95)\approx -0.25\%.
\end{equation}%
As far as the unsubtracted fraction  of the Dalitz decay background is
concerned, we can use the explicite formulae for the $1\gamma IR$
corrections taken from \cite{Kampf:2005tz} and arrive at\footnote{%
Here we neglect a very weak dependence on the constant $\chi ^{r}(M_{\rho })$
in the numerator.}%
\begin{equation}
\Delta ^{1\gamma IR}(0.95)=-\frac{1.75\times 10^{-15}}{[\Gamma ^{LO}(\pi
^{0}\rightarrow e^{+}e^{-})/\mathrm{MeV}]}
\end{equation}%
which gives for $\chi ^{r}(M_{\rho })=2.2$%
\begin{equation}
\Delta ^{1\gamma IR}|_{\chi ^{r}(M_{\rho })=2.2}=-0.35\%.
\end{equation}%
Note that both these additional contributions are larger then the variation
of $\delta (0.95)$ with $\xi ^{r}(M_{\rho })$ inside its error bar (which
yields $\Delta _{\xi }\delta (0.95)\lesssim 0.15\%$).

\begin{figure}[t]
\centering\includegraphics[scale=1]{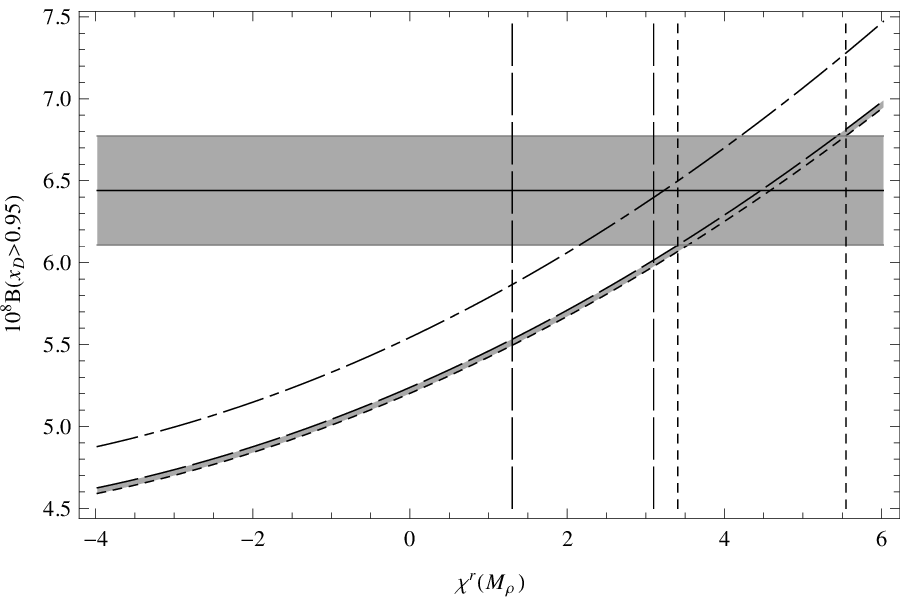} \newline
\caption{The dependence of predicted $B(\protect\pi ^{0}\rightarrow
e^{+}e^{-}(\protect\gamma ),x_{D}>0.95)$ on $\protect\chi^{r}(M_\protect\rho%
) $. The filled band corresponds to the variation of the additional
contributions $\Delta^{BS}+\Delta^{1\protect\gamma IR}$ from zero to its
maximal estimated value described in the main text (the dashed line here
represents $\Delta^{BS}+\Delta^{1\protect\gamma IR}=0$). The dash-dotted
line shows the leading order value. The horizontal band corresponds to the
KTeV-E799-II measurement. The dashed and dotted vertical lines delineate the
large $N_{C}$ inspired LMD estimate \protect\cite{Knecht:1999gb} and the
region compatible with experimental value respectively.}
\label{chi_experimental}
\end{figure}

In the Fig. \ref{chi_experimental} we have plotted the right hand side of (%
\ref{B_95_prediction}) as a function of $\chi ^{r}(M_{\rho })$
both with and without the estimated additional contributions
$\Delta ^{BS}$ and $\Delta ^{1\gamma IR}$ together with the
leading order branching ratio against the experimental value
(\ref{KTeV inclusive}). The vertical bands correspond to the large
$N_{C}$ inspired LMD estimate $\chi ^{r}(M_{\rho })=2.2\pm 0.9$
and to the range of values compatible with (\ref{KTeV inclusive}).
These bands do not overlap, the preferred region of $\chi
^{r}(M_{\rho })$ is shifted towards higher values\footnote{Quite
interestingly, even higher values of $\chi ^{r}(M_{\rho })$ have
been obtained from similar analysis of the related decays $K_L
\to \mu^+\mu^-$, namely $\chi ^{r}(M_{\rho })=8.07\pm 0.20$ or
$5.84\pm 0.20$ \cite{Cirigliano:2011ny}, and the bigger solution
obtained from $\eta \to \mu^+\mu^-$, namely $\chi ^{r}(M_{\rho
})=8.0\pm 0.9$ \cite{GomezDumm:1998gw}. Here the solution for
$\chi ^{r}(M_{\rho })$ shows two-fold ambiguity which is present
also in the $\pi^0\to e^+e^- $ case, however we have fixed it
keeping only the solution closer to the large $N_C$ prediction.},
namely $\chi ^{r}(M_{\rho })\approx 4.5\pm 1.1$ which corresponds
to $\delta (0.95,\chi ^{r}(M_{\rho })= 4.5 )=-6.2\%$.

\subsection{Generalization to the \texorpdfstring{$P\to l^+l^-$}{Pl+l-} decays}

\begin{figure}[t]
\center \scalebox{0.80}{\epsfig{figure=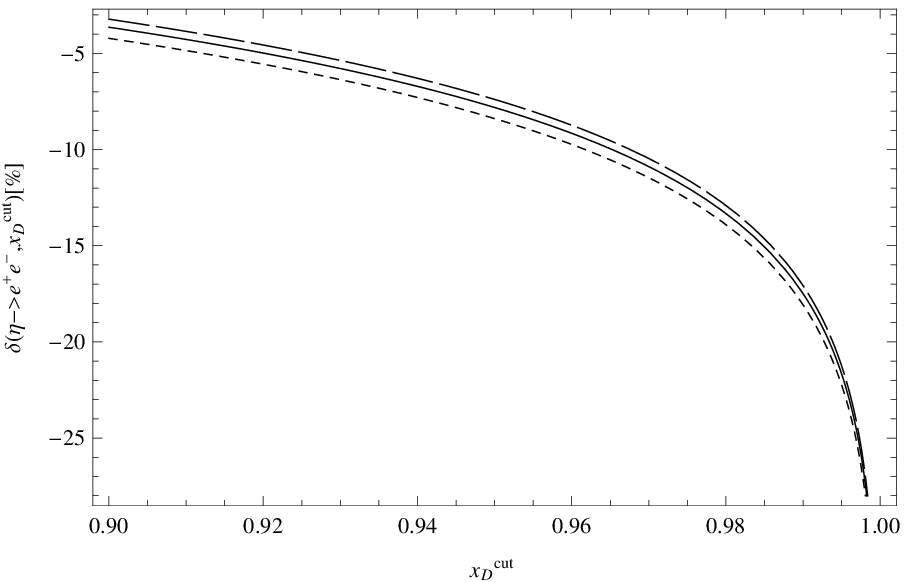}} \scalebox{0.80}{%
\epsfig{figure=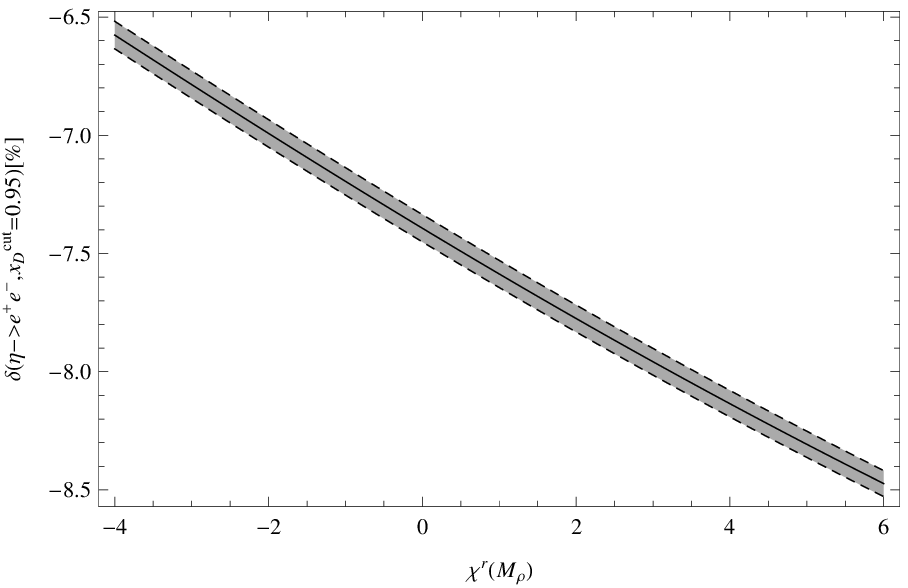}} \scalebox{0.80}{\epsfig{figure=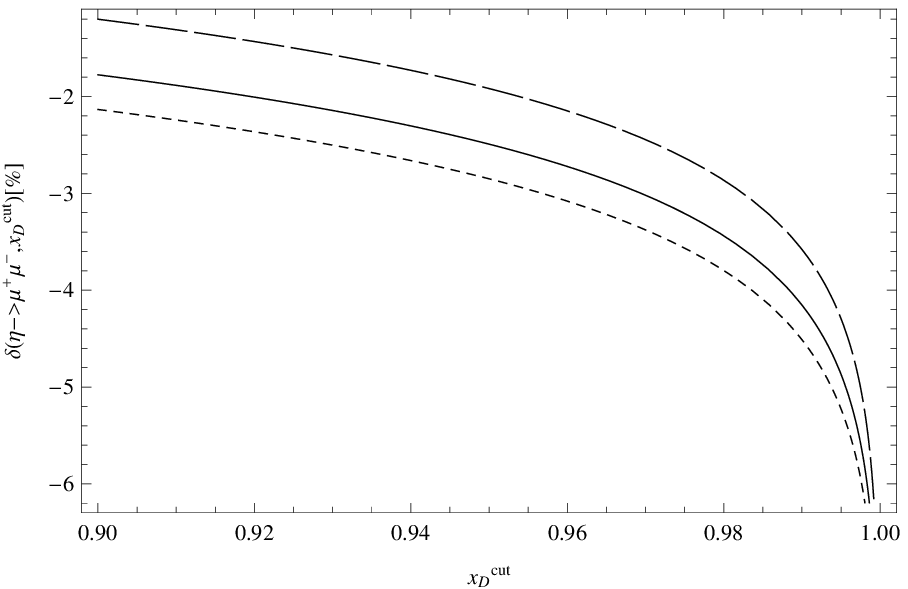}} %
\scalebox{0.80}{\epsfig{figure=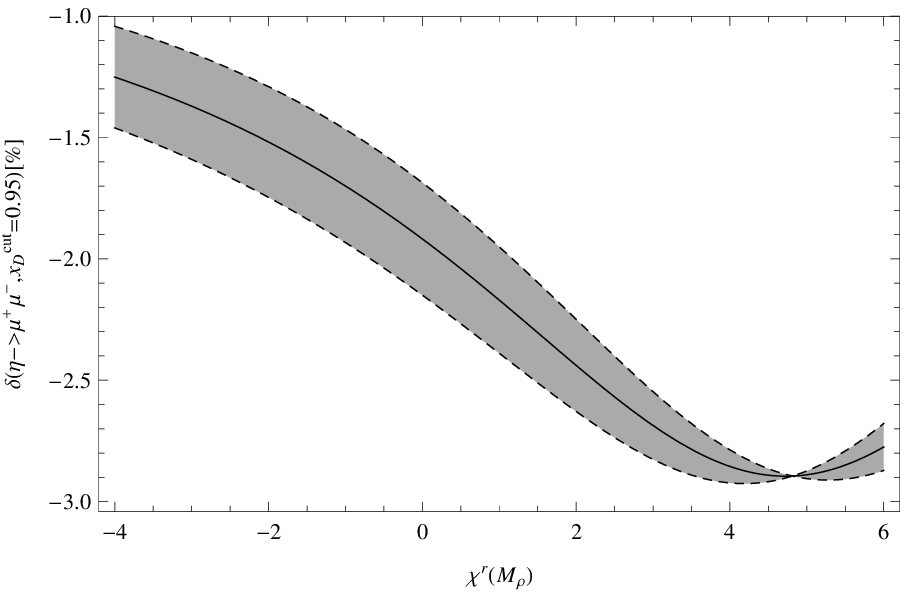}} \caption{On he left:
the $x_D^{{\mathrm{cut}}}$ dependence of the (partial) two-loop
QED corrections $\delta(x_D^{{\mathrm{cut}}})$  to $\eta\to
l^+l^-$ decays for $\chi^r(M_\rho)=2.2$ (solid line),
$\chi^r(M_\rho)=0$ (doted line) and $\chi^r(M_\rho)=5.5$ (dashed
line). On the right: the $\chi^r(M_\rho)$ dependence of
$\delta(x_D^{{\mathrm{cut}}}=0.95)$. The filled band corresponds
to variation of $\xi^r(M_\rho)$ inside its error bar.}
\label{plots_etall}
\end{figure}

\begin{figure}[t]
\center  \scalebox{0.80}{\epsfig{figure=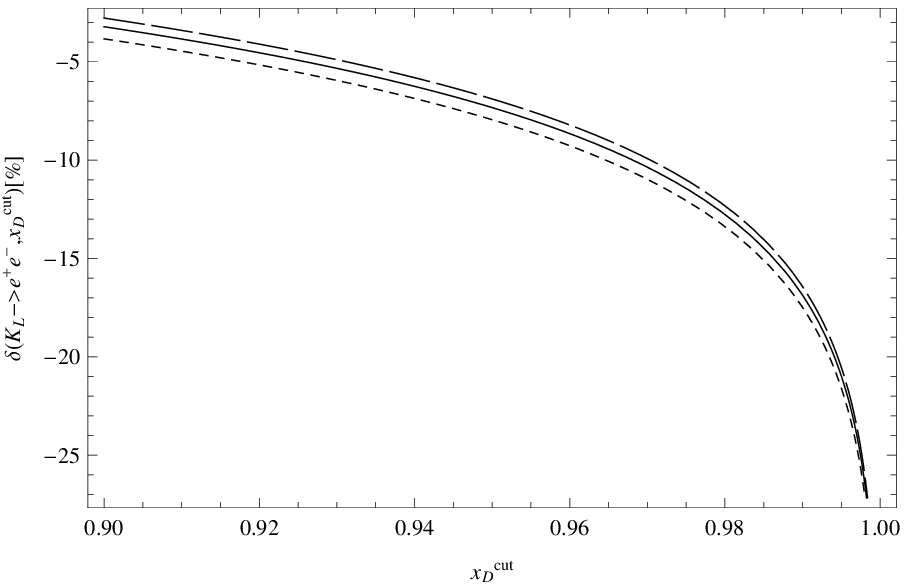}}
\scalebox{0.80}{\epsfig{figure=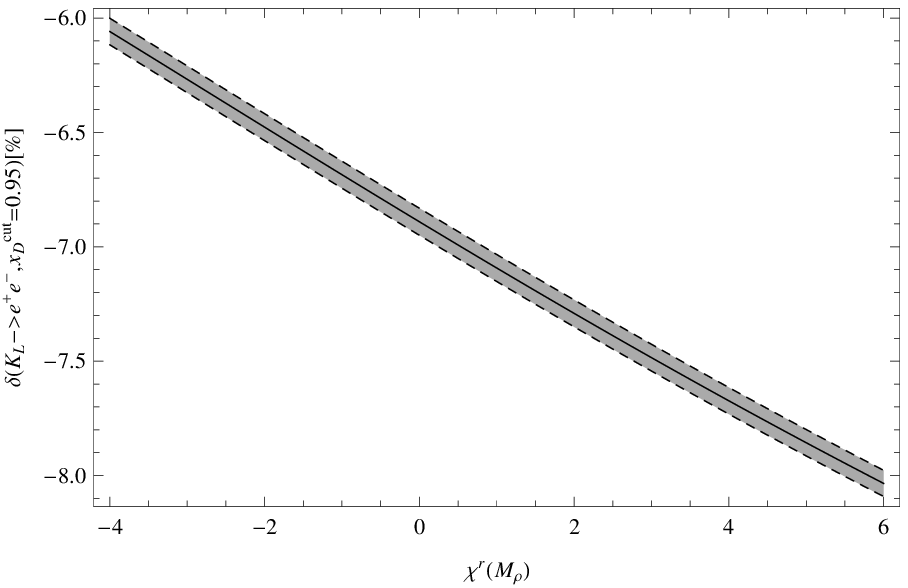}}
\scalebox{0.80}{\epsfig{figure=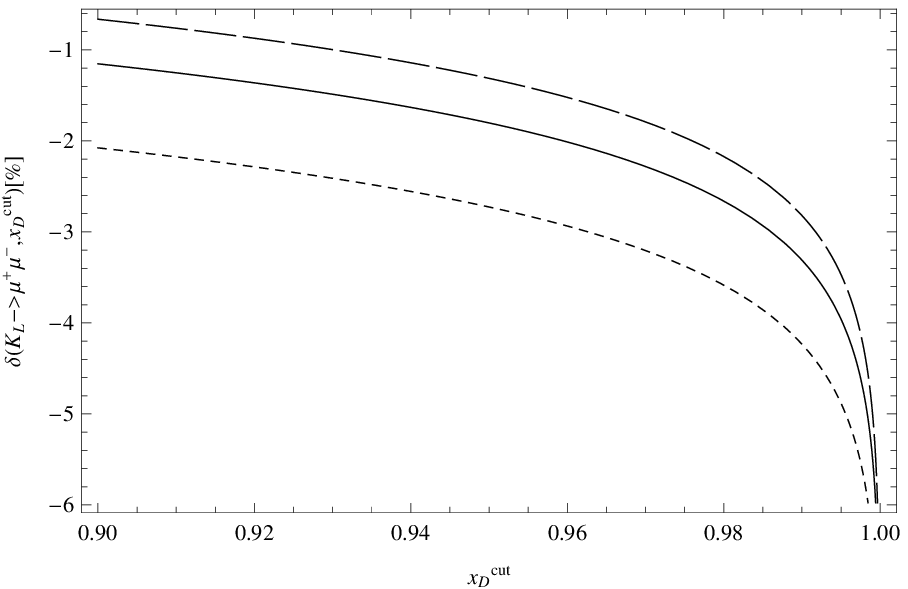}}
\scalebox{0.80}{\epsfig{figure=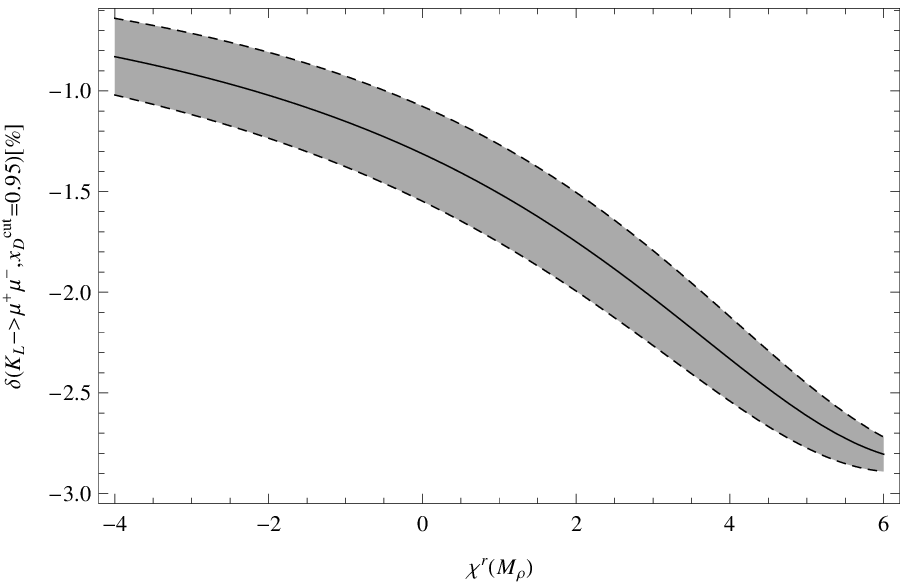}} \caption{The same plots
as in Fig.\ref{plots_etall} for the $K_L\to l^+l^-$ decays.}
\label{plots_KLll}
\end{figure}

From the point of view of $\chi PT$ the coupling constant $\chi
^{r}(\mu )$  is universal\footnote{The analogous statement is true
also for $\xi ^{r}(\mu )$.}, because it enters the $SU(3)$
generalization of the counterterm Lagrangian
${\cal{L}}^{(6)}_{{\mathrm{ct}}}$ given in Appendix
\ref{Lagrangian_appendix} and is therefore connected with other
processes of the type $P\rightarrow l^{+}l^{-}$ where $P=\pi
^{0},~\eta ,~K_{L}$ and $l=e,~\mu $. These decays are governed at
the leading order by the following effective long-distance
Lagrangian
\begin{equation}
\mathcal{L}_{\mathrm{eff},~Pl^{+}l^{-}}^{LD}=C_{P}\left( \frac{\alpha }{\pi }%
\right) \frac{1}{4F_{0}}\left\{ \frac{1}{2}P\varepsilon _{\mu \nu \alpha
\beta }F^{\mu \nu }F^{\alpha \beta }-\mu ^{-2\varepsilon }\left( \frac{%
\alpha }{\pi }\right) \left[ \chi _{P}^{r}(\mu )+\frac{3}{2}\left( \frac{1}{%
\varepsilon }-\overline{\gamma }\right)\right] \overline{l}\gamma ^{\mu }\gamma
^{5}l\partial _{\mu }P \right\}
\end{equation}%
where $C_{P}$ is a normalization factor and  $\chi _{P}^{r}(\mu )$ is the
effective long distance coupling which can be split into its universal $\chi
PT$ and specific short-distance components
\begin{equation}
\chi _{P}^{r}(\mu )=\chi ^{r}(\mu )+\chi _{P}^{sd}.
\end{equation}%
The latter is nontrivial only for the $P=K_{L}$ case where it is
well known \cite{Buchalla:1996fp} , \cite{Gorbahn:2006bm} (see
also \cite{GomezDumm:1998gw} and \cite{Isidori:2003ts})
\begin{equation}
\chi _{K_{L}}^{sd}=-1.82\pm 0.04.
\end{equation}%
\begin{table}[t]
\begin{center}
\begin{tabular}{|c|c|c|c|c|c|}
\hline
 $P\to l^+l^-$ & $\pi ^0\to e^+e^- $& $\eta \to e^+e^- $&
 $\eta \to \mu ^+\mu ^- $& $K_L\to \mu ^+\mu ^- $& $K_L\to e^+e^- $\\
\hline
 $\delta(\omega =3.37{\mathrm{MeV}})[\%]$ & $-5.8$ & $-16.3$ &
$-3.9$ & $-3.0$ & $-15.1$\\
\hline $\delta(x_D^{{\mathrm{cut}}} =0.95)[\%]$ & $-5.8$ & $-7.8$
&
$-2.5$ & $-1.8$ & $-7.3$\\
\hline
\end{tabular}%
\end{center}
\caption{Illustration of the typical values of the (partial)
two-loop QED corrections for processes $P\to l^+l^-$ with the same
cutoff $\omega$ for the soft photon energy and with the same
cutoff on $x_D$.} \label{table_Pll}
\end{table}
The Lagrangian $\mathcal{L}_{\mathrm{eff},~Pl^{+}l^{-}}^{LD}$ is up to the
normalization and shift in $\chi ^{r}(\mu )$ identical with that we have
used for the calculation of the $O(\alpha ^{3}p^{2})$ corrections to\ $\pi
^{0}\rightarrow e^{+}e^{-}$ decay which means that our general result can be
almost straightforwardly used for other $P\rightarrow l^{+}l^{-}$
processes by means of the substitution $M_{\pi ^{0}}\rightarrow M_{P}$ ,  $%
m\rightarrow m_{l}$ and  $\chi ^{r}\rightarrow \chi _{P}^{r}$.
More precisely, in such a way we obtain the contribution of the
graphs depicted in Fig. \ref{NLO_ChPT_2loop} where all the fermion
lines correspond to the final state lepton flavour, that means
that for the decays $P\rightarrow \mu ^{+}\mu ^{-}$ we miss the
vacuum polarization insertion graphs with electrons inside the
loop. The results of such a generalization are illustrated in
Tab. \ref{table_Pll} where
the corresponding corrections are listed (either for the same cutoff $\omega =3.37%
\mathrm{MeV}$ \ on the energy $E_\gamma$ of the soft photon or at the same cutoff  $x_{D}^{%
\mathrm{cut}}=0.95$ on $x_D=m^2_{l^+l^-}/M_P^2$) and compared with
the case of $\pi ^{0}\rightarrow e^{+}e^{-}$ decay. To obtain
these numbers we have fixed $\chi^r(M_\rho)=2.2$. As another
illustration of the typical values of the (partial in the sense of
the missing graphs) two-loop QED corrections we have plotted their
$x_{D}^{\mathrm{cut}}$ and $\chi ^{r}(M_{\rho })$ dependence in
Figs. \ref{plots_etall} and \ref{plots_KLll}.

\section{Summary and conclusion}

In this paper we have calculated the two-loop $O(\alpha ^{3}p^{2})$ QED
radiative corrections to the rare decay $\pi ^{0}\rightarrow e^{+}e^{-}$%
including all the relevant graphs. As a result we have obtained exact
analytical expression which takes into account the virtual photon
contributions without any approximation. The IR divergences has been treated
including real soft photon bremsstrahlung. The latter has been calculated
within the soft photon approximation and added to the inclusive decay rate $%
\pi ^{0}\rightarrow e^{+}e^{-}(\gamma )$. We have worked in the framework of
the $\chi PT$ with dynamical leptons and photons and parameterized the
missing information on the details of the pion transition form factor in
terms of two \emph{a priori} unknown RG invariant couplings $\overline{\chi }
$ and $\overline{\xi }$ which also incorporated the large RG logarithms. The
numerical value of the first of these couplings has been obtained using the
analysis of \cite{Knecht:1999gb} \ based on the large $N_{C}$ matching and
LMD ansatz, while the value of the second one (on the value of which our
result proved to be much less sensitive) has been estimated from the running
of the corresponding renormalized coupling $\xi ^{r}(\mu )$ with the RG
scale $\mu $.

The main motivation of our work was to test the validity of two
approximative calculations of the QED radiative corrections already existing
in the literature, namely the Bergstr\"{o}m's point-like $\pi ^{0}e^{+}e^{-}$
(sub)graph approximation \cite{Bergstrom:1982wk}, which has been used for
the analysis of the experimental data by the KTeV collaboration, and the
more sophisticated large-log approximation \cite{Dorokhov:2008qn} which
claimed to confirm the applicability of the previous one.

We have identified the Bergstr\"{o}m's calculation as a leading term in the
large $\overline{\chi }$ expansion of the exact two-loop result. We have
found that this expansion did not converge and therefore could not be
trusted without reservation. The numerical discrepancy between the exact
result and the Bergstr\"{o}m's one seems to confirm this conclusion.

Next we have discussed several variants of the LL approximations derived
from the exact two-loop result. The two polynomial ones differ from the
exact result by roughly twice the error estimated from the uncertainty of
the couplings $\overline{\chi }$ and $\overline{\xi }$ while the rational
one gives an excellent agreement. However we did not manage neither to
confirm the LL\ calculation of \cite{Dorokhov:2008qn} nor to reveal the
reason of  its large difference from our result. The approximation \cite%
{Dorokhov:2008qn} was obtained in completely different framework and we
therefore left the final resolution open to further studies.

Our final result numerically reads (see the definition (\ref{delta_definition}))
\begin{equation}
\delta (x_{D}^{\mathrm{cut}})=\left( 4.7\ln (1-x_{D}^{\mathrm{cut}})+8.3\pm
0.2\right) \%
\end{equation}%
and for the cut $x_{D}^{\mathrm{cut}}=0.95$ chosen by KTeV%
\begin{equation}
\delta (x_{D}^{\mathrm{cut}}=0.95)=\left( -5.8\pm 0.2\right) \%.
\end{equation}
Here the error stems from the uncertainty of $\overline{\chi }$ and $%
\overline{\xi }$. Our result significantly differs from the
previous approximative calculations
\cite{Bergstrom:1982wk,Dorokhov:2008qn}
\begin{equation}
\delta _{\mathrm{Bergstr\ddot{o}m}}^{virt.+soft~\gamma }(x_{D}^{\mathrm{cut}%
}=0.95)=-13.8\%,\;\;\;\delta _{\mathrm{Dorokhov}}^{virt.+soft~\gamma }(x_{D}^{%
\mathrm{cut}}=0.95)=-13.3\%.
\end{equation}%
and the change is in the right direction towards the agreement of
the experimental data with the SM prediction.

Let us note, that for the realistic analysis of the experimental
data it is necessary to discuss carefully two further topics,
which we have only partially  included in this work, namely the
real final state radiation beyond the soft photon approximation
and the incorporation of the Dalitz decay contribution. The former
is important because it is well known that the soft photon
approximation is not much reliable except of very limited region
in the phase space due to the small electron mass. The latter
process, which has been recently revisited in \cite{Kampf:2005tz},
is known to yield a non-negligible background
 of roughly $2-3\%$ \cite{Bergstrom:1982wk} near the cut $x_{D}^{\mathrm{cut}%
}=0.95$ chosen by KTeV. The more detailed analysis of these issues
is still in progress. In this work we have performed only
preliminary simplified analysis and found that the SM prediction
for the branching ratio $B(\pi^0\to e^+ e^-(\gamma),x_D>0.95)$
might be reconciled with the experimental value for $\chi
^{r}(M_{\rho })\approx 4.5\pm 1.1$, which is however off the
predictions for $\chi ^{r}(M_{\rho })$ based on the
phenomenological models of the pion transition form factor.

\section*{Acknowledgement}

We would like to thank Marc Knecht for initiating this project and for
valuable discussions at the beginning and Karol Kampf for valuable
discussions. This work is supported in part by Center for Particle Physics
(project no.\ LC 527) of the Ministry of Education of the Czech Republic.

\appendix

\section{ The \texorpdfstring{$\protect\chi PT$}{ChPT} Lagrangian with dynamical photons and
leptons \label{Lagrangian_appendix}}

In this appendix we summarize the relevant parts of the $\chi PT$
Lagrangian with dynamical photons and leptons. Following the
notation of \cite{Urech:1994hd} and using the $SU(2)_{L}\times
SU(2)_{R}$ variant of the theory (see also \cite{Kaiser:2000ck})
we need the following terms of the
complete Lagrangian%
\begin{equation}
\mathcal{L}=\mathcal{L}_{WZW}+\mathcal{L}_{e-\gamma }^{(2)}+\mathcal{L}%
_{e-\gamma }^{(4)}+\mathcal{L}_{\mathrm{ct}}^{(6)}
\end{equation}%
where%

\begin{equation}\begin{split}
\mathcal{L}_{WZW} &=\frac{N_{C}}{32\pi ^{2}}\varepsilon ^{\mu \nu \rho
\sigma }\left[ \langle U^{+}\widehat{r}_{\mu }U\widehat{l}_{\nu }-\widehat{r}%
_{\mu }\widehat{l}_{\nu }+\mathrm{i}\Sigma _{\mu }(U^{+}\widehat{r}_{\nu }U+%
\widehat{l}_{\nu })\rangle \langle v_{\rho \sigma }\rangle +\frac{2}{3}%
\langle \Sigma _{\mu }\Sigma _{\nu }\Sigma _{\rho }\rangle \langle v_{\sigma
}\rangle \right] \\
\mathcal{L}_{e-\gamma }^{(2)} &=-\frac{1}{4}F_{\mu \nu }F^{\mu \nu }+%
\overline{e}(\mathrm{i}\gamma^{\mu} D_{\mu}-m)e+\frac{1}{2}\left( \partial
\cdot A\right) ^{2} \\
\mathcal{L}_{e-\gamma }^{(4)} &=\left( \frac{\alpha }{\pi }\right) x_{6}%
\overline{e}\mathrm{i}\gamma ^{\mu }D_{\mu }e+\left( \frac{\alpha }{\pi }%
\right) x_{7}m\overline{e}e+\frac{1}{4}\left( \frac{\alpha }{\pi }\right)
x_{8}F_{\mu \nu }F^{\mu \nu } \\
\mathcal{L}_{\mathrm{ct}}^{(6)} &=\frac{3}{32}\mathrm{i}\left( \frac{\alpha
}{\pi }\right) ^{2}\overline{e}\gamma ^{\mu }\gamma ^{5}e\left[ \chi
_{1}\langle Q^{2}(D_{\mu }UU^{+}-D_{\mu }U^{+}U\rangle +\chi _{2}\langle
U^{+}QD_{\mu }UQ-UQD_{\mu }U^{+}Q\rangle \right]
\end{split}\end{equation}%
and where%
\begin{eqnarray}
 U &=&\exp \frac{\mathrm{i}}{F_{0}}\left(
\begin{array}{cc}
\pi ^{0} , & \sqrt{2}\pi ^{+}\\ \sqrt{2}\pi ^{-},& -\pi ^{0}
\end{array} \right), \qquad
Q =\mathrm{diag}\left(\frac{2}{3},-\frac{1}{3}\right)\notag \\
DU &=&\partial U-\mathrm{i}rU+\mathrm{i}Ul, \qquad
\Sigma =U^{+}\partial U \\
\widehat{r} &=&r-\frac{1}{2}\langle r\rangle, \qquad
\widehat{l} =l-\frac{1}{2}\langle l\rangle \notag\\
v_{\mu \nu } &=&\partial _{\mu }v_{\nu }-\partial _{\nu }v_{\mu
}-i[v_{\mu },v_{\nu }]\notag
\end{eqnarray}%
The coupling $\chi $ is connected with $\chi _{1,2}$ according to%
\begin{equation}
\chi =-\frac{1}{4}\left( \chi _{1}+\chi _{2}\right)
\end{equation}

\section{Reduction to scalar integrals\label{Reduction_appendix}}

In this appendix we give the reduction of the individual two-loop graphs to
172 scalar integrals.
\begin{eqnarray}
\gamma ^{(1),~\mathrm{2-loop}} &=&2\frac{\left( 8\pi \right) ^{4}}{m^{2}}%
\left( \frac{\alpha }{\pi }\right) \mu ^{-4\varepsilon } \notag \\
&&\times \left[ -64m^{6}y^{3}B(0,1,1,1,1,1,1)+32m^{6}y^{2}B(0,1,1,1,1,1,1)%
\right. \notag \\
&&+32m^{4}y^{2}B(0,0,1,1,1,1,1)+32m^{4}y^{2}B(0,1,0,1,1,1,1) \notag \\
&&-16m^{4}yB(0,0,1,1,1,1,1)-16m^{4}yB(0,1,0,1,1,1,1) \notag \\
&&-4m^{2}yB(0,-1,1,1,1,1,1)+8m^{2}yB(0,0,0,1,1,1,1) \notag \\
&&-4m^{2}yB(0,0,1,1,1,0,1)-4m^{2}yB(0,1,-1,1,1,1,1) \notag \\
&&-4m^{2}yB(0,1,0,1,0,1,1)+8m^{2}yB(0,1,1,1,0,0,1) \notag \\
&&-4m^{2}yB(0,1,1,1,0,1,0)-4m^{2}yB(0,1,1,1,1,0,0) \notag \\
&&+2m^{2}B(0,-1,1,1,1,1,1)-4m^{2}B(0,0,0,1,1,1,1) \notag \\
&&+2m^{2}B(0,1,-1,1,1,1,1)+B(0,-1,1,1,1,0,1) \notag \\
&&-B(0,0,0,1,0,1,1)-B(0,0,0,1,1,0,1) \notag \\
&&-2B(0,0,1,1,0,0,1)+B(0,0,1,1,0,1,0) \notag \\
&&+2B(0,0,1,1,1,-1,1)-B(0,0,1,1,1,0,0) \notag \\
&&+B(0,1,-1,1,0,1,1)+2B(0,1,0,1,-1,1,1) \notag \\
&&-2B(0,1,0,1,0,0,1)-B(0,1,0,1,0,1,0) \notag \\
&&\left. +B(0,1,0,1,1,0,0)\right]  \\
&& \notag \\
\gamma ^{(\Gamma ),~\mathrm{2-loop}} &=&\frac{\left( 8\pi \right) ^{4}}{m^{2}%
}\left( \frac{\alpha }{\pi }\right) \mu ^{-4\varepsilon }\left[
64m^{6}y^{2}B(1,1,1,1,1,0,1)+48m^{4}y^{2}B(0,1,1,1,1,0,1)\right. \notag \\
&&-32m^{4}y^{2}B(1,0,1,1,1,0,1)-48m^{4}y^{2}B(1,1,1,0,1,0,1) \notag \\
&&+32m^{4}y^{2}B(1,1,1,1,0,0,1)+16m^{4}y^{2}B(1,1,1,1,1,0,0) \notag \\
&&-32m^{4}yB(1,0,1,1,1,0,1)-32m^{4}yB(1,1,0,1,1,0,1) \notag \\
&&-12m^{2}yB(0,0,1,1,1,0,1)-12m^{2}yB(0,1,0,1,1,0,1) \notag \\
&&-12m^{2}yB(0,1,1,1,0,0,1)-12m^{2}yB(0,1,1,1,1,-1,1) \notag \\
&&+24m^{2}yB(0,1,1,1,1,0,0)+16m^{2}yB(1,-1,1,1,1,0,1) \notag \\
&&+16m^{2}yB(1,0,0,1,1,0,1)+24m^{2}yB(1,0,1,0,1,0,1) \notag \\
&&-16m^{2}yB(1,0,1,1,0,0,1)+4m^{2}yB(1,0,1,1,1,-1,1) \notag \\
&&-12m^{2}yB(1,0,1,1,1,0,0)+24m^{2}yB(1,1,0,0,1,0,1) \notag \\
&&-12m^{2}yB(1,1,0,1,0,0,1)-12m^{2}yB(1,1,0,1,1,0,0) \notag \\
&&-8m^{2}yB(1,1,1,1,0,-1,1)+8m^{2}yB(1,1,1,1,0,0,0) \notag \\
&&+8m^{2}yB(1,1,1,1,1,-1,0)-8m^{2}yB(1,1,1,1,1,0,-1) \notag \\
&&+4m^{2}B(1,-1,1,1,1,0,1)-8m^{2}B(1,0,0,1,1,0,1) \notag \\
&&+4m^{2}B(1,1,-1,1,1,0,1)+3B(0,0,1,1,0,0,1) \notag \\
&&-3B(0,0,1,1,1,-1,1)-3B(0,1,0,1,0,0,1)+3B(0,1,0,1,1,-1,1) \notag \\
&&-2B(1,-2,1,1,1,0,1)+4B(1,-1,0,1,1,0,1)-3B(1,-1,1,0,1,0,1) \notag \\
&&+2B(1,-1,1,1,0,0,1)-B(1,-1,1,1,1,-1,1)+2B(1,-1,1,1,1,0,0) \notag \\
&&-2B(1,0,-1,1,1,0,1)+6B(1,0,0,0,1,0,1)-3B(1,0,0,1,0,0,1) \notag \\
&&+B(1,0,0,1,1,-1,1)-4B(1,0,0,1,1,0,0)+2B(1,0,1,1,0,-1,1) \notag \\
&&-2B(1,0,1,1,0,0,0)-2B(1,0,1,1,1,-2,1)+2B(1,0,1,1,1,-1,0) \notag \\
&&-3B(1,1,-1,0,1,0,1)+B(1,1,-1,1,0,0,1)+2B(1,1,-1,1,1,0,0) \notag \\
&&-2B(1,1,0,1,-1,0,1)+2B(1,1,0,1,0,-1,1) \notag \\
&&\left. +2B(1,1,0,1,0,0,0)-2B(1,1,0,1,1,-1,0)\right]  \\
&& \notag \\
\gamma ^{(3),~\mathrm{2-loop}} &=&2\frac{\left( 8\pi \right) ^{4}}{m^{2}}%
\left( \frac{\alpha }{\pi }\right) \mu ^{-4\varepsilon } \notag \\
&&\times \left[ 32m^{6}y^{2}B(2,1,1,1,0,0,1)+16m^{4}y^{2}B(1,1,1,1,0,0,1)%
\right. \notag \\
&&+16m^{4}y^{2}B(2,1,1,0,0,0,1)-16m^{4}y^{2}B(2,1,1,1,0,0,0) \notag \\
&&-16m^{4}yB(2,0,1,1,0,0,1)-16m^{4}yB(2,1,0,1,0,0,1) \notag \\
&&+16m^{2}yB(0,1,1,1,0,0,1)-4m^{2}yB(1,0,1,1,0,0,1) \notag \\
&&-4m^{2}yB(1,1,0,1,0,0,1)+4m^{2}yB(1,1,1,1,-1,0,1) \notag \\
&&+4m^{2}yB(1,1,1,1,0,-1,1)-8m^{2}yB(1,1,1,1,0,0,0) \notag \\
&&-8m^{2}yB(2,0,1,0,0,0,1)+8m^{2}yB(2,0,1,1,0,0,0) \notag \\
&&-8m^{2}yB(2,1,0,0,0,0,1)+8m^{2}yB(2,1,0,1,0,0,0) \notag \\
&&+2m^{2}B(2,-1,1,1,0,0,1)-4m^{2}B(2,0,0,1,0,0,1) \notag \\
&&+2m^{2}B(2,1,-1,1,0,0,1)-B(1,0,1,1,-1,0,1) \notag \\
&&+B(1,0,1,1,0,-1,1)+B(1,1,0,1,-1,0,1) \notag \\
&&-B(1,1,0,1,0,-1,1)+B(2,-1,1,0,0,0,1) \notag \\
&&-B(2,-1,1,1,0,0,0)-2B(2,0,0,0,0,0,1) \notag \\
&&+2B(2,0,0,1,0,0,0)+B(2,1,-1,0,0,0,1) \notag \\
&&\left. -B(2,1,-1,1,0,0,0)\right]  \\
&& \notag \\
\gamma ^{(\Pi ),~\mathrm{2-loop}} &=&-2\frac{\left( 8\pi \right) ^{4}}{m^{2}}%
\left( \frac{\alpha }{\pi }\right) \mu ^{-4\varepsilon} \notag \\
&&\times \left[ 32m^{6}y^{2}B(1,2,1,0,1,0,1)+16m^{4}y^{2}B(1,1,1,0,1,0,1)%
\right. \notag \\
&&-16m^{4}y^{2}B(1,2,1,0,0,0,1)+16m^{4}y^{2}B(1,2,1,0,1,0,0) \notag \\
&&-16m^{4}yB(1,1,1,0,1,0,1)-16m^{4}yB(1,2,0,0,1,0,1) \notag \\
&&-8m^{2}yB(1,0,1,0,1,0,1)+8m^{2}yB(1,1,1,0,0,0,1) \notag \\
&&-8m^{2}yB(1,1,1,0,1,-1,1)+8m^{2}yB(1,2,1,0,0,-1,1) \notag \\
&&-8m^{2}yB(1,2,1,0,0,0,0)-8m^{2}yB(1,2,1,0,1,-1,0) \notag \\
&&+8m^{2}yB(1,2,1,0,1,0,-1)+2m^{2}B(1,0,1,0,1,0,1) \notag \\
&&-4m^{2}B(1,1,0,0,1,0,1)+2m^{2}B(1,2,-1,0,1,0,1) \notag \\
&&+B(1,-1,1,0,1,0,1)-2B(1,0,0,0,1,0,1) \notag \\
&&-B(1,0,1,0,0,0,1)+2B(1,0,1,0,1,-1,1) \notag \\
&&-B(1,0,1,0,1,0,0)+B(1,1,-1,0,1,0,1) \notag \\
&&-2B(1,1,0,0,1,-1,1)+2B(1,1,0,0,1,0,0) \notag \\
&&-2B(1,1,1,0,0,-1,1)+2B(1,1,1,0,0,0,0) \notag \\
&&+2B(1,1,1,0,1,-2,1)-2B(1,1,1,0,1,-1,0) \notag \\
&&+B(1,2,-1,0,0,0,1)-B(1,2,-1,0,1,0,0) \notag \\
&&+2B(1,2,0,0,-1,0,1)-2B(1,2,0,0,0,-1,1) \notag \\
&&\left. -2B(1,2,0,0,0,0,0)+2B(1,2,0,0,1,-1,0)\right]
\end{eqnarray}

\section{The IBP identities\label{IPB_appendix}}

Here we list the IBP identities%
\begin{equation}
\int \frac{\mathrm{d}^{d}k}{(2\pi )^{d}}\frac{\mathrm{d}^{d}l}{(2\pi )^{d}}%
\frac{\partial }{\partial p^{\mu }}q^{\mu }\left[ \prod_{i=1}^{7}\frac{1}{%
D_{i}(k,l)^{n_{i}}}\right] =0
\end{equation}%
where $p=k,l$ and $q=k$, $l$, $q_{\pm }$. We present these identities in the
form%
\begin{equation}
\mathbf{O}(p,q)B(n_{1},\ldots ,n_{7})=0
\end{equation}%
where the operators $\mathbf{O}(p,q)$ corresponding to the insertion of ${%
\partial }/{\partial p^{\mu }}q^{\mu }$ into the loop integral can be
rewritten in terms of the operators (\ref{jpm_operator}) as
\begin{eqnarray}
\mathbf{O}(k,q_{-}) &=&2m^{2}n_{5}\mathbf{5}^{+}+2m^{2}n_{7}\mathbf{7}%
^{+}-n_{4}+n_{5}-2m^{2}(2y-1)n_{6}\mathbf{6}^{+}+n_{7}\mathbf{7}^{+}\mathbf{1%
}^{-} \notag \\
&&-n_{7}\mathbf{7}^{+}\mathbf{2}^{-}-n_{5}\mathbf{5}^{+}\mathbf{4}^{-}-n_{6}%
\mathbf{6}^{+}\mathbf{4}^{-}-n_{7}\mathbf{7}^{+}\mathbf{4}^{-}+n_{4}\mathbf{4%
}^{+}\mathbf{5}^{-}+n_{6}\mathbf{6}^{+}\mathbf{5}^{-}+n_{7}\mathbf{7}^{+}%
\mathbf{5}^{-} \notag \\
&& \notag \\
\mathbf{O}(k,q_{+}) &=&2m^{2}n_{6}\mathbf{6}^{+}+2m^{2}n_{7}\mathbf{7}%
^{+}-n_{4}+n_{6}-2m^{2}(2y-1)n_{5}\mathbf{5}^{+}+n_{7}\mathbf{7}^{+}\mathbf{1%
}^{-} \notag \\
&&-n_{7}\mathbf{7}^{+}\mathbf{3}^{-}-n_{5}\mathbf{5}^{+}\mathbf{4}^{-}-n_{6}%
\mathbf{6}^{+}\mathbf{4}^{-}-n_{7}\mathbf{7}^{+}\mathbf{4}^{-}+n_{4}\mathbf{4%
}^{+}\mathbf{6}^{-}+n_{5}\mathbf{5}^{+}\mathbf{6}^{-}+n_{7}\mathbf{7}^{+}%
\mathbf{6}^{-} \notag \\
&& \notag \\
\mathbf{O}(l,q_{-}) &=&2m^{2}n_{1}\mathbf{1}^{+}+2m^{2}n_{3}\mathbf{3}%
^{+}+2m^{2}n_{7}\mathbf{7}^{+}+n_{1}-n_{2}+2m^{2}(2y-1)n_{3}\mathbf{3}%
^{+}+n_{2}\mathbf{2}^{+}\mathbf{1}^{-} \notag \\
&&+n_{3}\mathbf{3}^{+}\mathbf{1}^{-}+n_{7}\mathbf{7}^{+}\mathbf{1}^{-}-n_{1}%
\mathbf{1}^{+}\mathbf{2}^{-}-n_{3}\mathbf{3}^{+}\mathbf{2}^{-}-n_{7}\mathbf{7%
}^{+}\mathbf{2}^{-}-n_{7}\mathbf{7}^{+}\mathbf{4}^{-}+n_{7}\mathbf{7}^{+}%
\mathbf{5}^{-} \notag \\
&& \notag \\
\mathbf{O}(l,q_{+}) &=&2m^{2}n_{1}\mathbf{1}^{+}+2m^{2}n_{2}\mathbf{2}%
^{+}+2m^{2}n_{7}\mathbf{7}^{+}+n_{1}-n_{3}+2m^{2}(2y-1)n_{2}\mathbf{2}%
^{+}+n_{2}\mathbf{2}^{+}\mathbf{1}^{-} \notag \\
&&+n_{3}\mathbf{3}^{+}\mathbf{1}^{-}+n_{7}\mathbf{7}^{+}\mathbf{1}^{-}-n_{1}%
\mathbf{1}^{+}\mathbf{3}^{-}-n_{2}\mathbf{2}^{+}\mathbf{3}^{-}-n_{7}\mathbf{7%
}^{+}\mathbf{3}^{-}-n_{7}\mathbf{7}^{+}\mathbf{4}^{-}+n_{7}\mathbf{7}^{+}%
\mathbf{5}^{-} \notag \\
&& \notag \\
\mathbf{O}(k,l) &=&2m^{2}n_{5}\mathbf{5}^{+}+2m^{2}n_{6}\mathbf{6}%
^{+}+2m^{2}n_{7}\mathbf{7}^{+}+n_{7}-n_{4}-n_{4}\mathbf{4}^{+}\mathbf{1}%
^{-}+n_{7}\mathbf{7}^{+}\mathbf{1}^{-}-n_{5}\mathbf{5}^{+}\mathbf{2}^{-} \notag \\
&&-n_{6}\mathbf{6}^{+}\mathbf{3}^{-}-n_{5}\mathbf{5}^{+}\mathbf{4}^{-}-n_{6}%
\mathbf{6}^{+}\mathbf{4}^{-}-n_{7}\mathbf{7}^{+}\mathbf{4}^{-}+n_{4}\mathbf{4%
}^{+}\mathbf{7}^{-}+n_{5}\mathbf{5}^{+}\mathbf{7}^{-}+n_{6}\boldsymbol{6}^{+}%
\mathbf{7}^{-} \notag \\
&& \notag \\
\mathbf{O}(l,k) &=&-n_{1}\mathbf{1}^{+}\mathbf{4}^{-}+n_{1}\mathbf{1}^{+}%
\mathbf{7}^{-}-n_{1}+n_{7}-n_{2}\mathbf{2}^{+}\mathbf{1}^{-}-n_{3}\mathbf{3}%
^{+}\mathbf{1}^{-}-n_{7}\mathbf{7}^{+}\mathbf{1}^{-} \notag \\
&&+n_{7}\mathbf{7}^{+}\mathbf{4}^{-}-n_{2}\mathbf{2}^{+}\mathbf{5}^{-}-n_{3}%
\mathbf{3}^{+}\mathbf{6}^{-}+n_{2}\mathbf{2}^{+}\mathbf{7}^{-}+n_{3}\mathbf{3%
}^{+}\mathbf{7}^{-} \notag \\
&& \notag \\
\mathbf{O}(l,l) &=&d-2n_{1}-n_{2}-n_{3}-n_{7}-2m^{2}n_{1}\mathbf{1}%
^{+}-2m^{2}n_{7}\mathbf{7}^{+} \notag \\
&&-n_{2}\mathbf{2}^{+}\mathbf{1}^{-}-n_{3}\mathbf{3}^{+}\mathbf{1}^{-}-n_{7}%
\mathbf{7}^{+}\mathbf{1}^{-}+n_{7}\mathbf{7}^{+}\mathbf{4}^{-} \notag \\
&& \notag \\
\mathbf{O}(k,k) &=&d-2n_{4}-n_{5}-n_{6}-n_{7}+n_{7}\mathbf{7}^{+}\mathbf{1}%
^{-}-n_{5}\mathbf{5}^{+}\mathbf{4}^{-}-n_{6}\mathbf{6}^{+}\mathbf{4}%
^{-}-n_{7}\mathbf{7}^{+}\mathbf{4}^{-}.
\end{eqnarray}

\section{Results for Master Integrals \label{MI appendix}}
The results for MI are written in terms of the dimensionless quantities $b(n_{1},\ldots,n_{7})$ defined in (\ref{b}).
\subsection{Two propagator topology}

\begin{figure}[h]
\begin{center}
\includegraphics[scale=0.5]{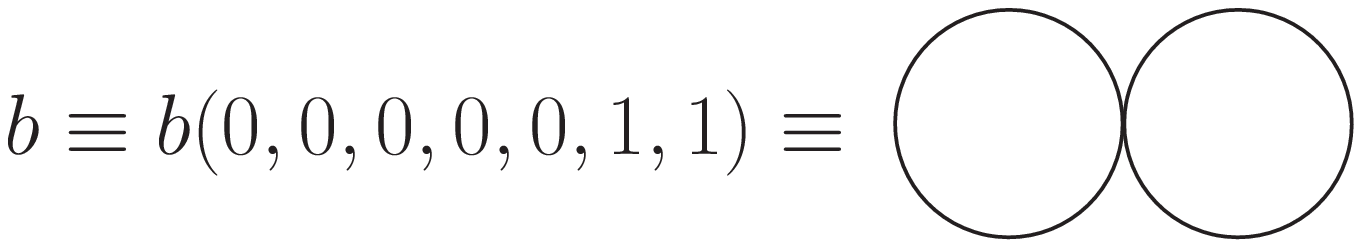}
\end{center}
\caption{The two propagator MI.}
\label{Fig T2_appendix}
\end{figure}

This topology contains only one MI depicted in Fig. \ref{Fig T2_appendix}.
It factorizes to one loop diagrams (i.e. a square of a tadpole) and hence
can be computed straightforwardly.
We get the result
\begin{equation}
b=\frac{\Gamma ^{2}(\epsilon -1)}{\Gamma ^{2}(\epsilon +1)},
\label{square tadpole}
\end{equation}
which has the following expansion
\begin{eqnarray*}
b^{(-2)} &=&1 \\
b^{(-1)} &=&2 \\
b^{(0)} &=&3 \\
b^{(1)} &=&4 \\
b^{(2)} &=&5 .
\end{eqnarray*}

\subsection{Three propagator topology, type \it{a}}

\begin{figure}[h]
\centering\includegraphics[scale=0.5]{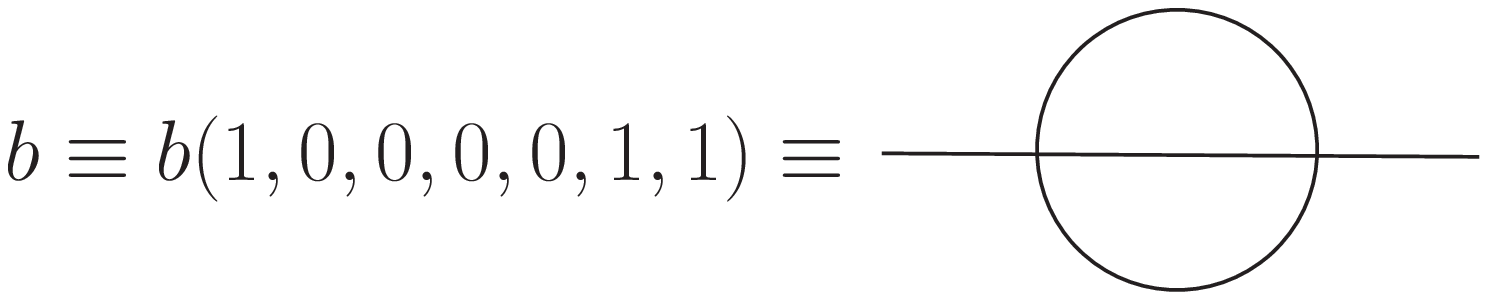}
\caption{The three propagator MI, type $a$.}
\label{Fig T3a_appendix}
\end{figure}

There is a single MI in this topology, see Fig. \ref{Fig T3a_appendix} . It
is $y$ independent thus can not be computed using the differential equations
technique as formulated in this paper. Moreover, it does not factorize so
the evaluation is more involved. However, the result can be found in the
literature. The pioneering calculation was done in \cite{Fleischer:1999hp} as far as we know.
It was further generalized in \cite{Argeri:2002wz} and both results agree.
The expansion in powers of $\epsilon$ reads \cite{Fleischer:1999hp,Argeri:2002wz}
\begin{eqnarray*}
b^{(-2)} &=&\frac{3}{2} \\
b^{(-1)} &=&\frac{17}{4} \\
b^{(0)} &=&\frac{59}{8} \\
b^{(1)} &=&32\left( \frac{65}{512}+\frac{\pi ^{2}}{24}\right) \\
b^{(2)} &=&-64\left( -\frac{7\zeta (3)}{16}+\frac{1117}{2048}-\frac{13\pi
^{2}}{96}+\frac{1}{8}\pi ^{2}\log (2)\right) .
\end{eqnarray*}

\subsection{Three propagator topology, type \it{b}}

\begin{figure}[h]
\centering\includegraphics[scale=0.5]{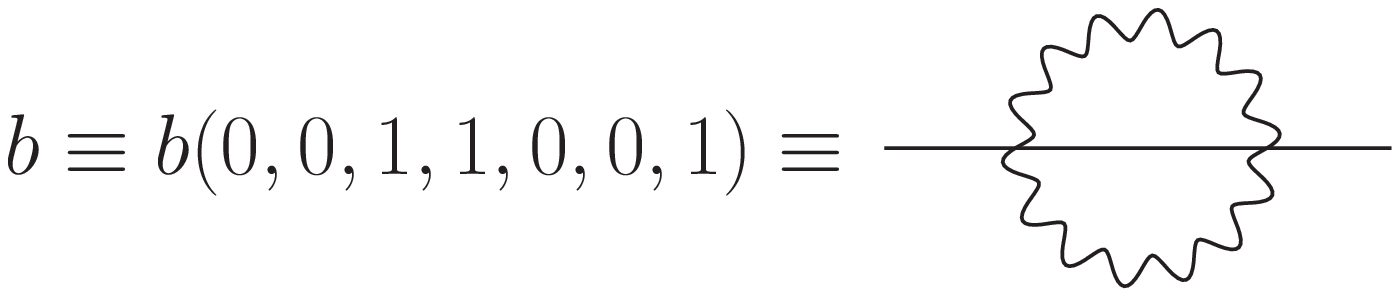}
\caption{The three propagator MI, type $b$.}
\label{Fig_T3b_appendix}
\end{figure}

This single Master integral depicted in Fig. \ref{Fig_T3b_appendix} does not
depend on $y$ and could be calculated directly using Feynman
parametrization.

\begin{equation}
b=-\frac{\Gamma (3-4\epsilon )\Gamma (1-\epsilon )^{2}\Gamma (\epsilon
)\Gamma (2\epsilon -1)}{\Gamma (3-3\epsilon )\Gamma (2-2\epsilon )\Gamma
(\epsilon +1)^{2}}.
\end{equation}%
Expanding the above formula gives%
\begin{eqnarray*}
b^{(-2)} &=&\frac{1}{2} \\
b^{(-1)} &=&\frac{5}{4} \\
b^{(0)} &=&\frac{1}{24}\left( 33+8\pi ^{2}\right) \\
b^{(1)} &=&\frac{1}{12}\left( 26+10\pi ^{2}+5\psi ^{(2)}(1)+8\psi
^{(2)}(2)-37\psi ^{(2)}(3)\right) \\
b^{(2)} &=&\frac{1}{720}\left( 660\pi ^{2}+256\pi ^{4}+15(-751+50\psi
^{(2)}(1)+80\psi ^{(2)}(2)-370\psi ^{(2)}(3))\right) .
\end{eqnarray*}

\subsection{Three propagator topology, type \it{c}}

\begin{figure}[h]
\centering\includegraphics[scale=0.5]{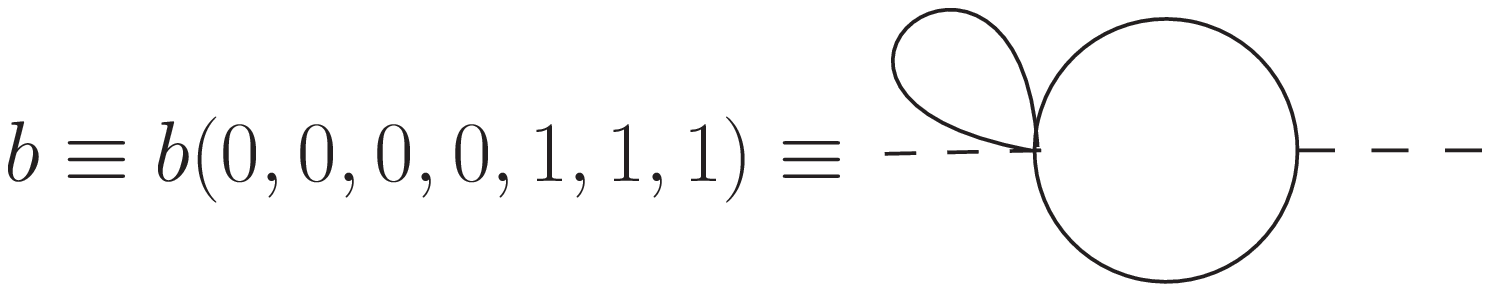}
\caption{Three propagator MI, type $c$.}
\label{Fig_T3c_appendix}
\end{figure}

This topology includes one MI (see Fig. \ref{Fig_T3c_appendix}) and corresponds to the product of one-loop integrals. The $\varepsilon$ expansion of the latter was studied in \cite{Davydychev:2000na}.
The convenient point for fixing the integration constants was in this case $%
x=1$ which is below the physical threshold. The solution is%
\begin{eqnarray*}
b^{(-2)} &=&1 \\
b^{(-1)} &=&-\frac{(x+1)H(0,x)-3x+3}{x-1} \\
b^{(0)} &=&-\frac{3(x+1)H(0,x)}{x-1}+\left( \frac{4}{x-1}+2\right) H(-1,0,x)+%
\frac{(x+1)H(0,0,x)}{1-x}+\frac{1}{6}\left( \frac{\pi ^{2}(x+1)}{x-1}%
+42\right) \\
b^{(1)} &=&-\frac{\pi ^{2}(x+1)H(-1,x)}{3(x-1)}+\frac{\left( \pi
^{2}-42\right) (x+1)H(0,x)}{6(x-1)}+\left( \frac{4}{x-1}+2\right) H(-2,0,x)
\\
&&+\frac{6(x+1)H(-1,0,x)}{x-1}-\frac{3(x+1)H(0,0,x)}{x-1}-\frac{%
4(x+1)H(-1,-1,0,x)}{x-1} \\
&&+\left( \frac{4}{x-1}+2\right) H(-1,0,0,x)+\frac{(x+1)H(0,0,0,x)}{1-x}+%
\frac{4(x+1)\zeta (3)+30(x-1)+\pi ^{2}(x+1)}{2(x-1)}.
\end{eqnarray*}
We checked this MI with \cite{Bonciani:2003te} where it is computed using the same formalism to $\mathcal{%
O}(\epsilon^{1})$ including. Comparing the results gives an agreement.

\subsection{Three propagator topology, type \it{d}}

\begin{figure}[h]
\centering\includegraphics[scale=0.5]{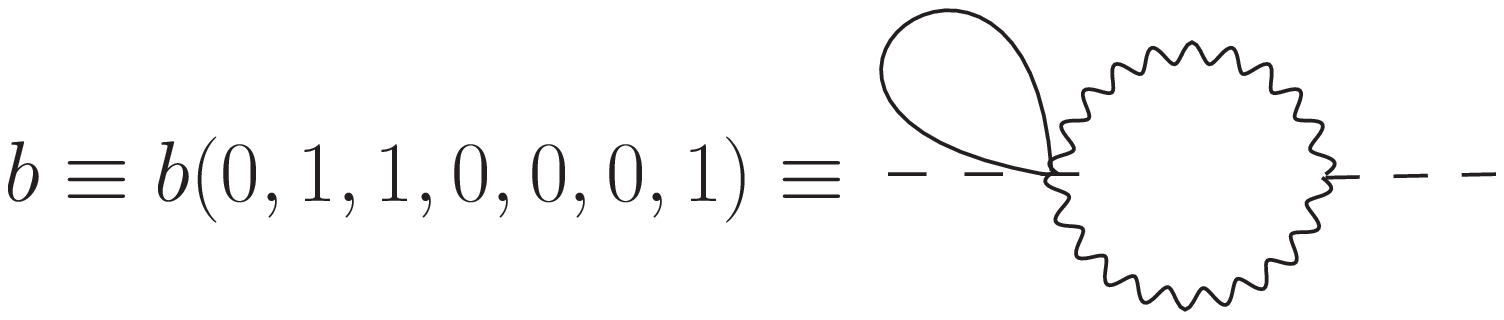}
\caption{Three propagator MI, type $d$.}
\label{Fig_T3d_appendix}
\end{figure}

This MI depicted in Fig. \ref{Fig_T3d_appendix} could be computed directly
using loop integration (it factorizes to an off-shell massless bubble and a
tadpole).
The result reads
\begin{equation}
b=-\frac{4^{-\epsilon }(-y)^{-\epsilon }\Gamma (1-\epsilon )^{2}\Gamma
(\epsilon -1)\Gamma (\epsilon )}{\Gamma (2-2\epsilon )\Gamma (\epsilon
+1)^{2}},
\end{equation}%
and has the following expansion%
\begin{eqnarray*}
b^{(-2)} &=&1 \\
b^{(-1)} &=&H(0,x)+2H(1,x)+3 \\
b^{(0)} &=&3H(0,x)+6H(1,x)+2H(2,x)+H(0,0,x)+2H(1,0,x)+4H(1,1,x)-\frac{\pi
^{2}}{6}+7 \\
b^{(1)} &=&\left( 7-\frac{\pi ^{2}}{6}\right) H(0,x)+\left( 14-\frac{\pi ^{2}%
}{3}\right) H(1,x)+6H(2,x)+2H(3,x)+3H(0,0,x)+6H(1,0,x) \\
&&+12H(1,1,x)+4H(1,2,x)+2H(2,0,x)+4H(2,1,x)+H(0,0,0,x)+2H(1,0,0,x) \\
&&+4H(1,1,0,x)+8H(1,1,1,x)-2\zeta (3)-\frac{\pi ^{2}}{2}+15 \\
b^{(2)} &=&\left( -2\zeta (3)+15-\frac{\pi ^{2}}{2}\right) H(0,x)+\left(
-4\zeta (3)+30-\pi ^{2}\right) H(1,x)+\left( 14-\frac{\pi ^{2}}{3}\right)
H(2,x) \\
&&+6H(3,x)+2H(4,x)+\left( 7-\frac{\pi ^{2}}{6}\right) H(0,0,x)+\left( 14-%
\frac{\pi ^{2}}{3}\right) H(1,0,x) \\
&&+\left( 28-\frac{2\pi ^{2}}{3}\right)
H(1,1,x)+12H(1,2,x)+4H(1,3,x)+6H(2,0,x)+12H(2,1,x) \\
&&+4H(2,2,x)+2H(3,0,x)+4H(3,1,x)+3H(0,0,0,x)+6H(1,0,0,x) \\
&&+12H(1,1,0,x)+24H(1,1,1,x)+8H(1,1,2,x)+4H(1,2,0,x)+8H(1,2,1,x) \\
&&+2H(2,0,0,x)+4H(2,1,0,x)+8H(2,1,1,x)+H(0,0,0,0,x)+2H(1,0,0,0,x) \\
&&+4H(1,1,0,0,x)+8H(1,1,1,0,x)+16H(1,1,1,1,x)-6\zeta (3)-\frac{\pi ^{4}}{40}-%
\frac{7\pi ^{2}}{6}+31.
\end{eqnarray*}
Also this MI was checked with \cite{Bonciani:2003hc} and a complete match
was reached.

\subsection{Three propagator topology, type \it{e}}

\begin{figure}[h]
\centering\includegraphics[scale=0.5]{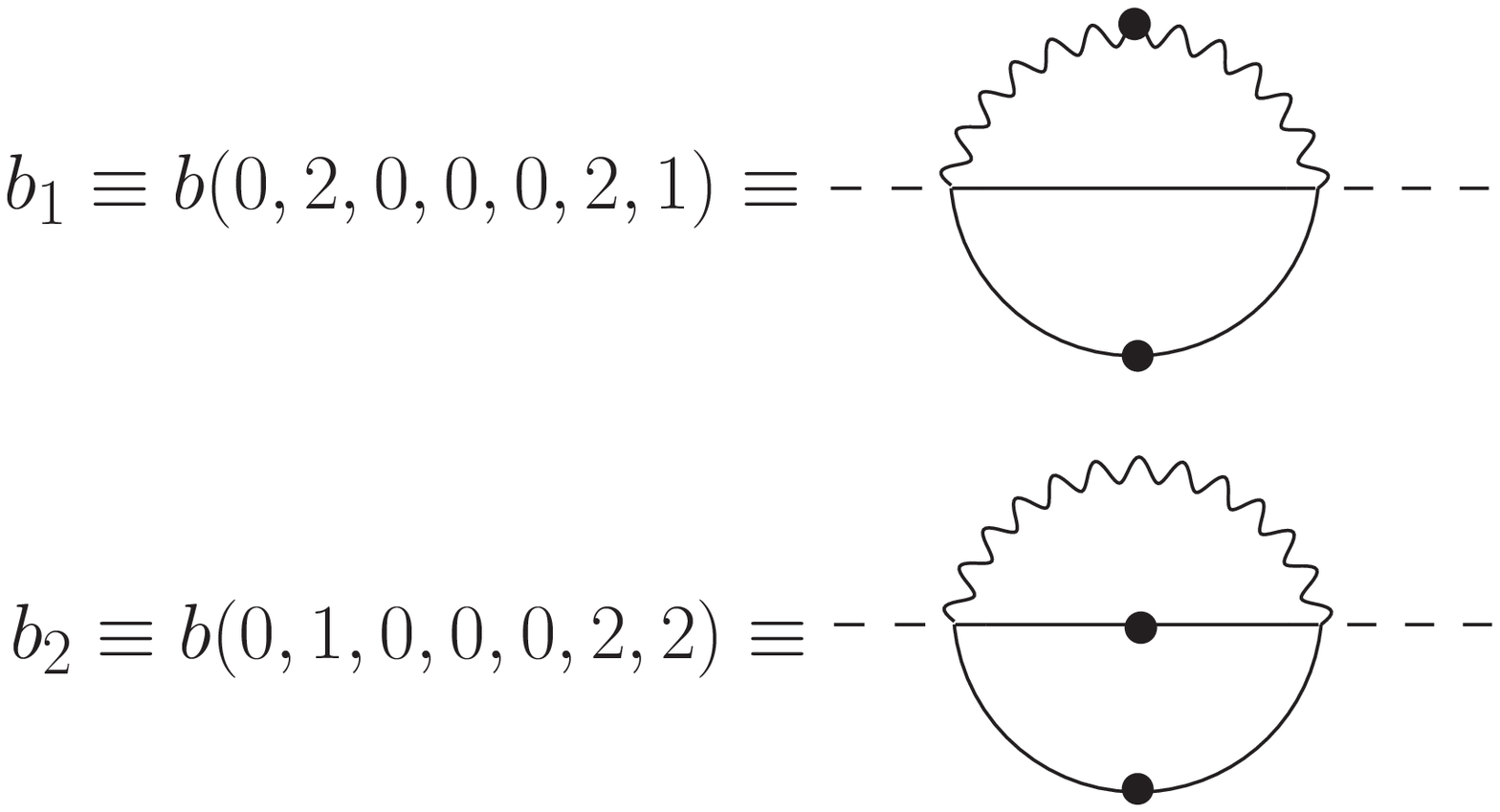}\newline \caption{The
two nested MIs with three propagators; type $e$.}
\label{T3e_appendix}
\end{figure}

There are two nested MIs (see Fig. \ref{T3e_appendix}).
The suitable point for fixing the integration constants was chosen as $x=1$.
Then the solution reads
\begin{eqnarray*}
b_{1}^{(-1)} &=&\frac{xH(0,x)}{x^{2}-1} \\
b_{2}^{(-1)} &=&0 \\
b_{1}^{(0)} &=&-\frac{6xH(-1,0,x)}{x^{2}-1}+\frac{2xH(1,0,x)}{x^{2}-1}+\frac{%
(5x-3)xH(0,0,x)}{(x-1)^{2}(x+1)}+\frac{\pi ^{2}x}{6-6x^{2}} \\
b_{2}^{(0)} &=&-\frac{2xH(0,0,x)}{(x-1)^{2}} \\
b_{1}^{(1)} &=&\frac{\pi ^{2}xH(-1,x)}{x^{2}-1}+\frac{\pi ^{2}xH(1,x)}{%
3-3x^{2}}+\frac{36xH(-1,-1,0,x)}{x^{2}-1}-\frac{24xH(-1,0,0,x)}{x^{2}-1} \\
&&-\frac{12xH(-1,1,0,x)}{x^{2}-1}-\frac{12xH(1,-1,0,x)}{x^{2}-1}+\frac{%
4xH(1,1,0,x)}{x^{2}-1} \\
&&+\frac{\pi ^{2}(3-5x)xH(0,x)}{6(x-1)^{2}(x+1)}+\frac{6(3-5x)xH(-2,0,x)}{%
(x-1)^{2}(x+1)}+\frac{2(5x-3)xH(2,0,x)}{(x-1)^{2}(x+1)} \\
&&+\frac{(13x-7)xH(0,0,0,x)}{(x-1)^{2}(x+1)}+\frac{2(3x-5)xH(1,0,0,x)}{%
(x-1)^{2}(x+1)}+\frac{2(4-7x)x\zeta (3)}{(x-1)^{2}(x+1)} \\
b_{2}^{(1)} &=&\frac{\pi ^{2}xH(0,x)}{3(x-1)^{2}}+\frac{12xH(-2,0,x)}{%
(x-1)^{2}}-\frac{4xH(2,0,x)}{(x-1)^{2}}-\frac{6xH(0,0,0,x)}{(x-1)^{2}} \\
&&+\frac{4xH(1,0,0,x)}{(x-1)^{2}}+\frac{6x\zeta (3)}{(x-1)^{2}} \\
b_{1}^{(2)} &=&\frac{\pi ^{2}(5x-3)H(-2,x)x}{(x-1)^{2}(x+1)}+\frac{\pi
^{2}(3-5x)H(2,x)x}{3(x-1)^{2}(x+1)}+\frac{6(7-13x)H(-3,0,x)x}{(x-1)^{2}(x+1)}
\\
&&-\frac{6\pi ^{2}H(-1,-1,x)x}{x^{2}-1}+\frac{4\pi ^{2}H(-1,0,x)x}{x^{2}-1}+%
\frac{2\pi ^{2}H(-1,1,x)x}{x^{2}-1} \\
&&+\frac{\pi ^{2}(7-13x)H(0,0,x)x}{6(x-1)^{2}(x+1)}+\frac{2\pi ^{2}H(1,-1,x)x%
}{x^{2}-1}+\frac{\pi ^{2}(5-3x)H(1,0,x)x}{3(x-1)^{2}(x+1)} \\
&&+\frac{2\pi ^{2}H(1,1,x)x}{3-3x^{2}}+\frac{2(13x-7)H(3,0,x)x}{%
(x-1)^{2}(x+1)}+\frac{36(5x-3)H(-2,-1,0,x)x}{(x-1)^{2}(x+1)} \\
&&+\frac{24(3-5x)H(-2,0,0,x)x}{(x-1)^{2}(x+1)}+\frac{12(3-5x)H(-2,1,0,x)x}{%
(x-1)^{2}(x+1)}+\frac{144H(-1,-2,0,x)x}{x^{2}-1} \\
&&-\frac{48H(-1,2,0,x)x}{x^{2}-1}+\frac{12(5-3x)H(1,-2,0,x)x}{(x-1)^{2}(x+1)}%
+\frac{4(3x-5)H(1,2,0,x)x}{(x-1)^{2}(x+1)} \\
&&+\frac{12(3-5x)H(2,-1,0,x)x}{(x-1)^{2}(x+1)}+\frac{2(27x-17)H(2,0,0,x)x}{%
(x-1)^{2}(x+1)}+\frac{4(5x-3)H(2,1,0,x)x}{(x-1)^{2}(x+1)} \\
&&-\frac{216H(-1,-1,-1,0,x)x}{x^{2}-1}+\frac{144H(-1,-1,0,0,x)x}{x^{2}-1}+%
\frac{72H(-1,-1,1,0,x)x}{x^{2}-1} \\
&&-\frac{60H(-1,0,0,0,x)x}{x^{2}-1}+\frac{72H(-1,1,-1,0,x)x}{x^{2}-1}-\frac{%
48H(-1,1,0,0,x)x}{x^{2}-1} \\
&&-\frac{24H(-1,1,1,0,x)x}{x^{2}-1}+\frac{(29x-15)H(0,0,0,0,x)x}{%
(x-1)^{2}(x+1)}+\frac{72H(1,-1,-1,0,x)x}{x^{2}-1} \\
&&-\frac{48H(1,-1,0,0,x)x}{x^{2}-1}-\frac{24H(1,-1,1,0,x)x}{x^{2}-1}+\frac{%
2(7x-13)H(1,0,0,0,x)x}{(x-1)^{2}(x+1)} \\
&&-\frac{24H(1,1,-1,0,x)x}{x^{2}-1}+\frac{4(5x-3)H(1,1,0,0,x)x}{%
(x-1)^{2}(x+1)}+\frac{8H(1,1,1,0,x)x}{x^{2}-1} \\
&&+\frac{\pi ^{4}(35-61x)x}{360(x-1)^{2}(x+1)}+\frac{66H(-1,x)\zeta (3)x}{%
x^{2}-1}+\frac{(18-34x)H(0,x)\zeta (3)x}{(x-1)^{2}(x+1)} \\
&&+\frac{4(7-4x)H(1,x)\zeta (3)x}{(x-1)^{2}(x+1)} \\
b_{2}^{(2)} &=&\frac{16x\zeta (3)H(0,x)}{(x-1)^{2}}-\frac{12x\zeta (3)H(1,x)%
}{(x-1)^{2}}-\frac{2\pi ^{2}xH(-2,x)}{(x-1)^{2}}+\frac{2\pi ^{2}xH(2,x)}{%
3(x-1)^{2}} \\
&&+\frac{36xH(-3,0,x)}{(x-1)^{2}}+\frac{\pi ^{2}xH(0,0,x)}{(x-1)^{2}}-\frac{%
2\pi ^{2}xH(1,0,x)}{3(x-1)^{2}}-\frac{12xH(3,0,x)}{(x-1)^{2}} \\
&&-\frac{72xH(-2,-1,0,x)}{(x-1)^{2}}+\frac{48xH(-2,0,0,x)}{(x-1)^{2}}+\frac{%
24xH(-2,1,0,x)}{(x-1)^{2}} \\
&&-\frac{24xH(1,-2,0,x)}{(x-1)^{2}}+\frac{8xH(1,2,0,x)}{(x-1)^{2}}+\frac{%
24xH(2,-1,0,x)}{(x-1)^{2}}-\frac{20xH(2,0,0,x)}{(x-1)^{2}} \\
&&-\frac{8xH(2,1,0,x)}{(x-1)^{2}}-\frac{14xH(0,0,0,0,x)}{(x-1)^{2}}+\frac{%
12xH(1,0,0,0,x)}{(x-1)^{2}}-\frac{8xH(1,1,0,0,x)}{(x-1)^{2}}+\frac{13\pi
^{4}x}{180(x-1)^{2}}.
\end{eqnarray*}
This pair of MIs belongs to the general class {\bf{J011}}, the $\varepsilon$ expansion of which was studied systematically in \cite{Davydychev:2000na,Davydychev:2003mv}. In the framework of differential equations approach they were calculated in different basis across the literature \cite%
{Bonciani:2003te,Anastasiou:2006hc,Czakon:2004wm}. We use the same basis as
in \cite{Anastasiou:2006hc} thus we verified their results to $\mathcal{O}%
(\epsilon^{2})$ including.

\subsection{Four propagator topology, type \it{a}}

This topology includes the Master integral depicted in Fig. \ref%
{T4a_appendix}.
\begin{figure}[h]
\centering\includegraphics[scale=0.5]{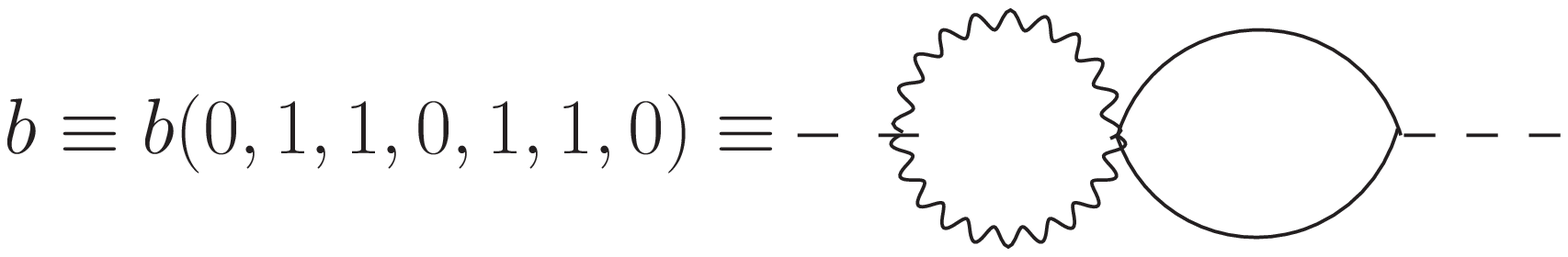}\newline
\caption{The four propagator MI, type $a$.}
\label{T4a_appendix}
\end{figure}
It is a product of a massive (see \cite{Davydychev:2000na}) and a massless bubble. The former one can be
easily extracted from $T3c$ and the latter one can be calculated directly.
So we used this trick instead of solving the corresponding differential
equation. The solution for current orders reads%
\begin{eqnarray*}
b^{(-2)} &=&1 \\
b^{(-1)} &=&-\frac{2H(0,x)}{x-1}+2H(1,x)+4 \\
b^{(0)} &=&8H(1,x)+\left( \frac{4}{x-1}+2\right) H(-1,0,x)+4H(1,1,x)-\frac{%
8H(0,x)}{x-1} \\
&&-\frac{4H(2,x)}{x-1}-\frac{2(x+2)H(0,0,x)}{x-1}-\frac{4H(1,0,x)}{x-1}+%
\frac{\pi ^{2}}{3(x-1)}+12 \\
b^{(1)} &=&-\frac{\pi ^{2}(x+1)\text{$H$}(-1,x)}{3(x-1)}+\frac{\left( \pi
^{2}(x+2)-72\right) \text{$H$}(0,x)}{3(x-1)} \\
&&+\left( \frac{2\pi ^{2}}{3(x-1)}+24\right) \text{$H$}(1,x)-\frac{16\text{$%
H $}(2,x)}{x-1}-\frac{4(x+2)\text{$H$}(3,x)}{x-1} \\
&&+\left( \frac{8}{x-1}+4\right) \text{$H$}(-2,0,x)+\frac{8(x+1)\text{$H$}%
(-1,0,x)}{x-1} \\
&&+\left( \frac{8}{x-1}+4\right) \text{$H$}(-1,2,x)-\frac{8(x+2)\text{$H$}%
(0,0,x)}{x-1}-\frac{16\text{$H$}(1,0,x)}{x-1} \\
&&+16\text{$H$}(1,1,x)-\frac{8\text{$H$}(1,2,x)}{x-1}-\frac{4(x+2)\text{$H$}%
(2,0,x)}{x-1}-\frac{8\text{$H$}(2,1,x)}{x-1} \\
&&-\frac{4(x+1)\text{$H$}(-1,-1,0,x)}{x-1}+\frac{6(x+1)\text{$H$}(-1,0,0,x)}{%
x-1} \\
&&+\left( \frac{8}{x-1}+4\right) \text{$H$}(-1,1,0,x)+\left( -\frac{14}{x-1}%
-6\right) \text{$H$}(0,0,0,x) \\
&&+\left( \frac{8}{x-1}+4\right) \text{$H$}(1,-1,0,x)-\frac{4(x+2)\text{$H$}%
(1,0,0,x)}{x-1} \\
&&-\frac{8\text{$H$}(1,1,0,x)}{x-1}+8\text{$H$}(1,1,1,x)+\frac{4\left(
3(8x+\zeta (3)-8)+\pi ^{2}\right) }{3(x-1)}.
\end{eqnarray*}
Making a check with \cite{Bonciani:2003hc} up to $\mathcal{O}(\epsilon^{1})$
lead to full agreement.

\subsection{Four propagator topology, type \it{b}}

\begin{figure}[h]
\centering\includegraphics[scale=0.5]{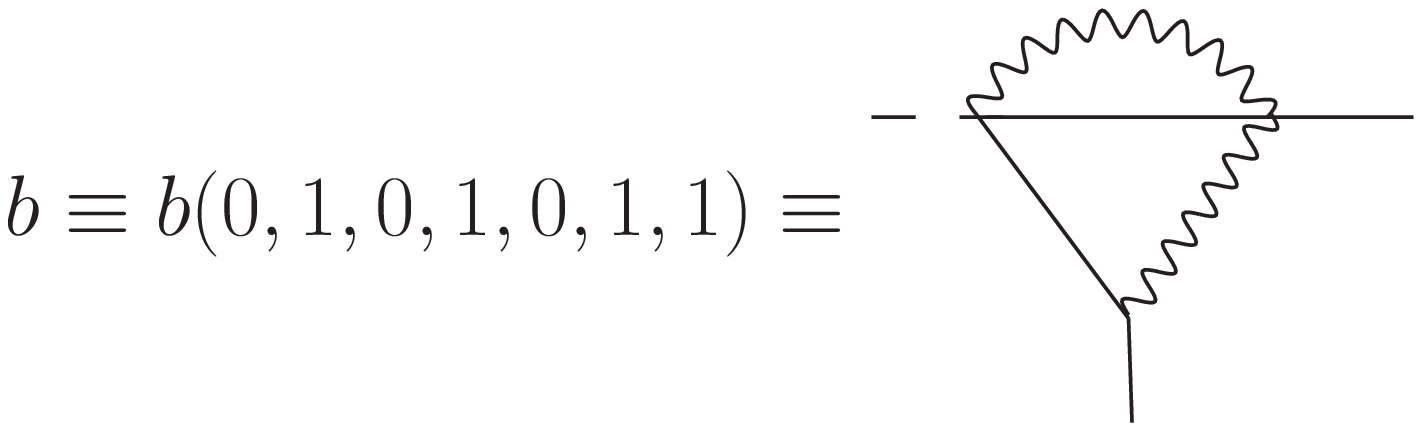}\newline
\caption{The four propagator MI, type $b$.}
\label{T4b_appendix}
\end{figure}
The solution for the one MI depicted in Fig. \ref{T4b_appendix} is (the
integration constants were fixed at $x=1$)
\begin{eqnarray*}
b^{(-2)} &=&\frac{1}{2} \\
b^{(-1)} &=&\frac{5}{2} \\
b^{(0)} &=&\frac{2\pi ^{2}xH(0,x)}{3-3x^{2}}-\frac{4xH(0,0,0,x)}{x^{2}-1}%
-H(0,0,x)+\frac{19}{2} \\
b^{(1)} &=&\frac{\left( 24x\zeta (3)+\pi ^{2}((x-4)x-1)\right) H(0,x)}{%
6\left( x^{2}-1\right) }+\frac{24xH(-3,0,x)}{x^{2}-1}+\frac{4\pi
^{2}xH(-1,0,x)}{3-3x^{2}} \\
&&+\frac{\left( -5x^{2}+2\pi ^{2}x+5\right) H(0,0,x)}{x^{2}-1}+\frac{4\pi
^{2}xH(1,0,x)}{3\left( x^{2}-1\right) }-\frac{8xH(3,0,x)}{x^{2}-1} \\
&&+\frac{(3-x(3x+4))H(0,0,0,x)}{x^{2}-1}+\frac{8xH(2,0,0,x)}{x^{2}-1}-\frac{%
8xH(-1,0,0,0,x)}{x^{2}-1} \\
&&-\frac{4xH(0,0,0,0,x)}{x^{2}-1}+\frac{8xH(1,0,0,0,x)}{x^{2}-1}%
+6H(-2,0,x)-2H(2,0,x) \\
&&+2H(1,0,0,x)+\frac{45\left( x^{2}-1\right) (6\zeta (3)+65)+26\pi ^{4}x}{%
90\left( x^{2}-1\right) }.
\end{eqnarray*}
A total correspondence with \cite{Bonciani:2003te} to $\mathcal{O}%
(\epsilon^{1})$ was found.

\subsection{Four propagator topology, type \it{c}}

\begin{figure}[h]
\centering\includegraphics[scale=0.5]{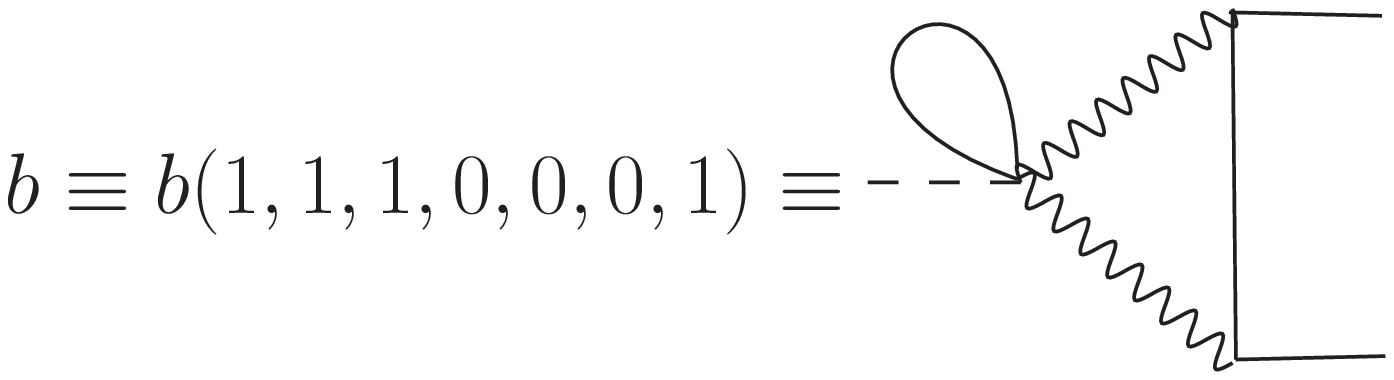}\newline
\caption{The four propagator MI, type $c$.}
\label{T4c_appendix}
\end{figure}

The third four propagator topology contains one Master integral depicted in
Fig. \ref{T4c_appendix}
which has the following solution (integration constants fixed at $x=-1$)%
\begin{eqnarray*}
b^{(-1)} &=&\frac{2xH(2,x)}{x^{2}-1}+\frac{xH(0,0,x)}{x^{2}-1}+\frac{2\pi
^{2}x}{3\left( x^{2}-1\right) } \\
b^{(0)} &=&\frac{4\pi ^{2}xH(-1,x)}{3\left( x^{2}-1\right) }+\frac{\pi
^{2}xH(0,x)}{6-6x^{2}}+\frac{4\pi ^{2}xH(1,x)}{3\left( x^{2}-1\right) }+%
\frac{2xH(2,x)}{x^{2}-1}+\frac{2xH(3,x)}{x^{2}-1}+\frac{4xH(-1,2,x)}{x^{2}-1}
\\
&&+\frac{xH(0,0,x)}{x^{2}-1}+\frac{4xH(1,2,x)}{x^{2}-1}+\frac{2xH(2,0,x)}{%
x^{2}-1}+\frac{4xH(2,1,x)}{x^{2}-1}+\frac{2xH(-1,0,0,x)}{x^{2}-1} \\
&&+\frac{xH(0,0,0,x)}{x^{2}-1}+\frac{2xH(1,0,0,x)}{x^{2}-1}+\frac{x\left(
15\zeta (3)+2\pi ^{2}\right) }{3\left( x^{2}-1\right) } \\
b^{(1)} &=&\frac{2x\left( 15\zeta (3)+2\pi ^{2}\right) H(-1,x)}{3\left(
x^{2}-1\right) }+\frac{2x\left( 15\zeta (3)+2\pi ^{2}\right) H(1,x)}{3\left(
x^{2}-1\right) }+\frac{x\left( 12\zeta (3)+\pi ^{2}\right) H(0,x)}{6-6x^{2}}
\\
&&-\frac{\left( \pi ^{2}-6\right) xH(2,x)}{3\left( x^{2}-1\right) }+\frac{%
2xH(3,x)}{x^{2}-1}+\frac{2xH(4,x)}{x^{2}-1}+\frac{8\pi ^{2}xH(-1,-1,x)}{%
3\left( x^{2}-1\right) }+\frac{\pi ^{2}xH(-1,0,x)}{3-3x^{2}} \\
&&+\frac{8\pi ^{2}xH(-1,1,x)}{3\left( x^{2}-1\right) }+\frac{4xH(-1,2,x)}{%
x^{2}-1}+\frac{4xH(-1,3,x)}{x^{2}-1}-\frac{\left( \pi ^{2}-6\right) xH(0,0,x)%
}{6\left( x^{2}-1\right) } \\
&&+\frac{8\pi ^{2}xH(1,-1,x)}{3\left( x^{2}-1\right) }+\frac{\pi
^{2}xH(1,0,x)}{3-3x^{2}}+\frac{8\pi ^{2}xH(1,1,x)}{3\left( x^{2}-1\right) }+%
\frac{4xH(1,2,x)}{x^{2}-1}+\frac{4xH(1,3,x)}{x^{2}-1} \\
&&+\frac{2xH(2,0,x)}{x^{2}-1}+\frac{4xH(2,1,x)}{x^{2}-1}+\frac{4xH(2,2,x)}{%
x^{2}-1}+\frac{2xH(3,0,x)}{x^{2}-1}+\frac{4xH(3,1,x)}{x^{2}-1} \\
&&+\frac{8xH(-1,-1,2,x)}{x^{2}-1}+\frac{2xH(-1,0,0,x)}{x^{2}-1}+\frac{%
8xH(-1,1,2,x)}{x^{2}-1}+\frac{4xH(-1,2,0,x)}{x^{2}-1} \\
&&+\frac{8xH(-1,2,1,x)}{x^{2}-1}+\frac{xH(0,0,0,x)}{x^{2}-1}+\frac{%
8xH(1,-1,2,x)}{x^{2}-1}+\frac{2xH(1,0,0,x)}{x^{2}-1}+\frac{8xH(1,1,2,x)}{%
x^{2}-1} \\
&&+\frac{4xH(1,2,0,x)}{x^{2}-1}+\frac{8xH(1,2,1,x)}{x^{2}-1}+\frac{%
2xH(2,0,0,x)}{x^{2}-1}+\frac{4xH(2,1,0,x)}{x^{2}-1}+\frac{8xH(2,1,1,x)}{%
x^{2}-1} \\
&&+\frac{4xH(-1,-1,0,0,x)}{x^{2}-1}+\frac{2xH(-1,0,0,0,x)}{x^{2}-1}+\frac{%
4xH(-1,1,0,0,x)}{x^{2}-1}+\frac{xH(0,0,0,0,x)}{x^{2}-1} \\
&&+\frac{4xH(1,-1,0,0,x)}{x^{2}-1}+\frac{2xH(1,0,0,0,x)}{x^{2}-1}+\frac{%
4xH(1,1,0,0,x)}{x^{2}-1}+\frac{x\left( 1800\zeta (3)+240\pi ^{2}+59\pi
^{4}\right) }{360\left( x^{2}-1\right) }.
\end{eqnarray*}
We verified the result of \cite{Bonciani:2003hc} to $\mathcal{O}%
(\epsilon^{1})$.

\subsection{Four propagator topology, type \it{d}}

\begin{figure}[h]
\centering\includegraphics[scale=0.5]{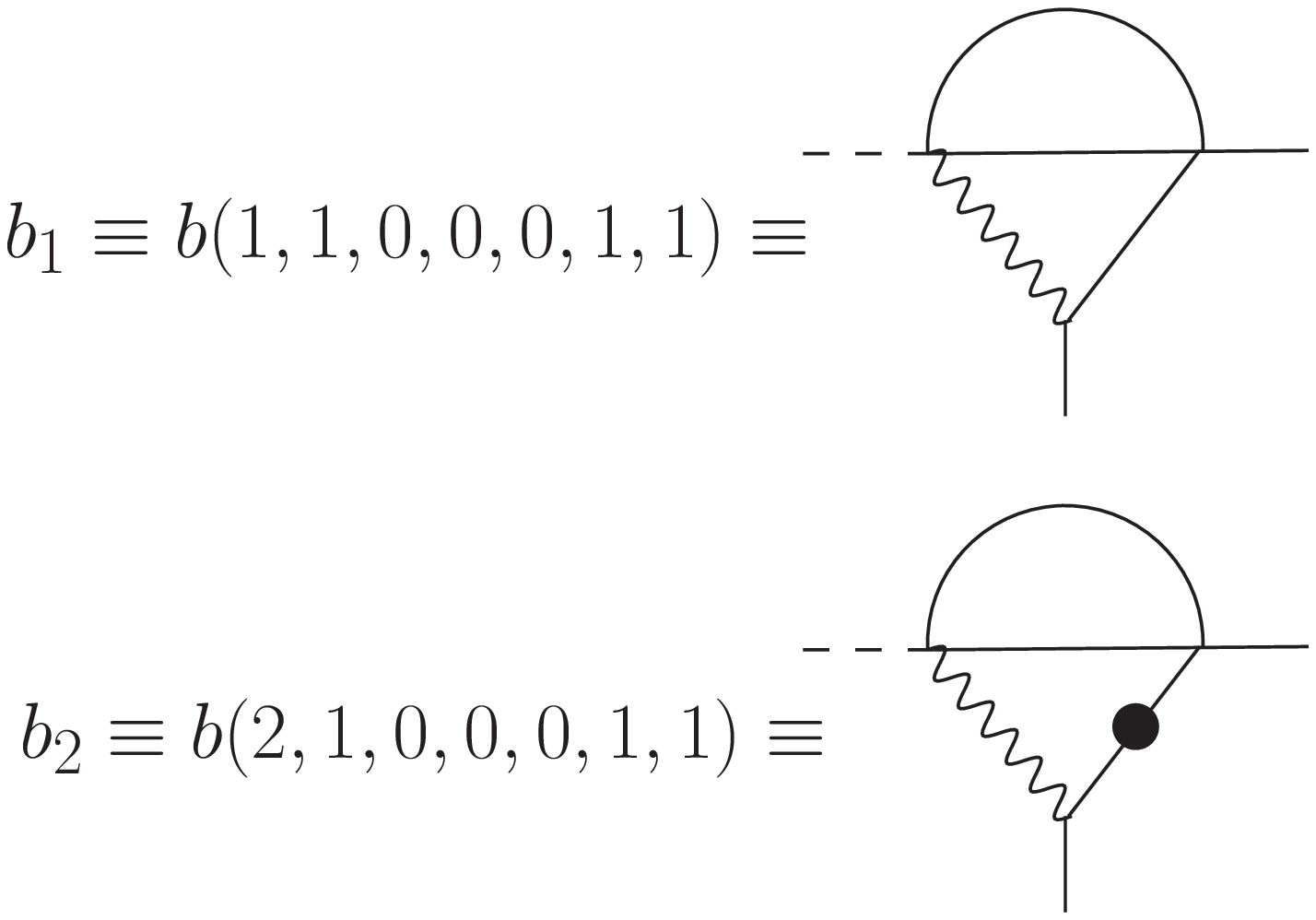}\newline
\caption{The basis of MI for the four propagator topology, type $d$.}
\label{T4d_appendix}
\end{figure}

Our choice for the two Mater integrals in this topology is depicted in Fig. %
\ref{T4d_appendix}.
Here we tried two points $x=\pm 1$ for fixing the integration constants and
both gave the same result. The expansion of the solution in powers of $%
\epsilon $ is%
\begin{eqnarray*}
b_{1}^{(-2)} &=&\frac{1}{2} \\
b_{2}^{(-2)} &=&\frac{1}{2} \\
b_{1}^{(-1)} &=&\frac{5}{2} \\
b_{2}^{(-1)} &=&\frac{(x+1)H(0,x)}{2-2x}+1 \\
b_{1}^{(0)} &=&\frac{2\pi ^{2}xH(0,x)}{3-3x^{2}}-\frac{4xH(0,0,0,x)}{x^{2}-1}%
-H(0,0,x)+\frac{1}{6}\left( 57-2\pi ^{2}\right) \\
b_{2}^{(0)} &=&\frac{(x+1)H(0,x)}{1-x}+\left( \frac{6}{x-1}+3\right)
H(-1,0,x)+\frac{(7x+1)H(0,0,x)}{2-2x} \\
&&+\frac{(x+1)H(1,0,x)}{1-x}-\frac{\pi ^{2}(x-3)}{12(x-1)}+2 \\
b_{1}^{(1)} &=&\frac{4\pi ^{2}xH(-2,x)}{x^{2}-1}+\frac{\pi
^{2}(1-x(9x+4))H(0,x)}{6\left( x^{2}-1\right) }+\frac{24xH(-3,0,x)}{x^{2}-1}+%
\frac{4\pi ^{2}xH(-1,0,x)}{3-3x^{2}} \\
&&+\frac{4\pi ^{2}xH(1,0,x)}{3\left( x^{2}-1\right) }-\frac{8xH(3,0,x)}{%
x^{2}-1}+\frac{8xH(-2,0,0,x)}{x^{2}-1}-\frac{8xH(-1,0,0,0,x)}{x^{2}-1} \\
&&-\frac{12xH(0,0,0,0,x)}{x^{2}-1}+\frac{8xH(1,0,0,0,x)}{x^{2}-1}+2\pi
^{2}H(-1,x)+6H(-2,0,x)-5H(0,0,x) \\
&&-2H(2,0,x)+4H(-1,0,0,x)+\left( \frac{2}{x+1}-\frac{6}{x-1}-11\right)
H(0,0,0,x)-2H(1,0,0,x) \\
&&-\frac{45\left( x^{2}-1\right) (6\zeta (3)-65)+150\pi ^{2}\left(
x^{2}-1\right) +\pi ^{4}x}{90\left( x^{2}-1\right) } \\
b_{2}^{(1)} &=&\frac{\left( \pi ^{2}((16-3x)x+3)-24(x+1)^{2}\right) H(0,x)}{%
12\left( x^{2}-1\right) }+\frac{(3x+1)(9x+1)H(0,0,0,x)}{2-2x^{2}} \\
&&+\frac{\pi ^{2}(x-3)H(-1,x)}{2(x-1)}+\frac{\pi ^{2}(x+1)H(1,x)}{6(x-1)}%
+\left( \frac{24}{x-1}+21\right) H(-2,0,x) \\
&&+\frac{6(x+1)H(-1,0,x)}{x-1}+\left( -\frac{8}{x-1}-7\right) H(0,0,x)-\frac{%
2(x+1)H(1,0,x)}{x-1} \\
&&+\left( -\frac{8}{x-1}-7\right) H(2,0,x)-\frac{18(x+1)H(-1,-1,0,x)}{x-1}%
+\left( \frac{24}{x-1}+14\right) H(-1,0,0,x) \\
&&+\frac{6(x+1)H(-1,1,0,x)}{x-1}+\frac{6(x+1)H(1,-1,0,x)}{x-1}+\left( -\frac{%
8}{x-1}-3\right) H(1,0,0,x) \\
&&-\frac{2(x+1)H(1,1,0,x)}{x-1}+\frac{24(x+\zeta (3)-1)+42x\zeta (3)-\pi
^{2}(x-3)}{6(x-1)}.
\end{eqnarray*}
These Master integrals appear in \cite{Bonciani:2003te} in the same basis.
Making a check leads to complete agreement up to $\mathcal{O}(\epsilon^{1})$.

\subsection{Four propagator topology, type \it{e}}

\begin{figure}[h]
\centering\includegraphics[scale=0.5]{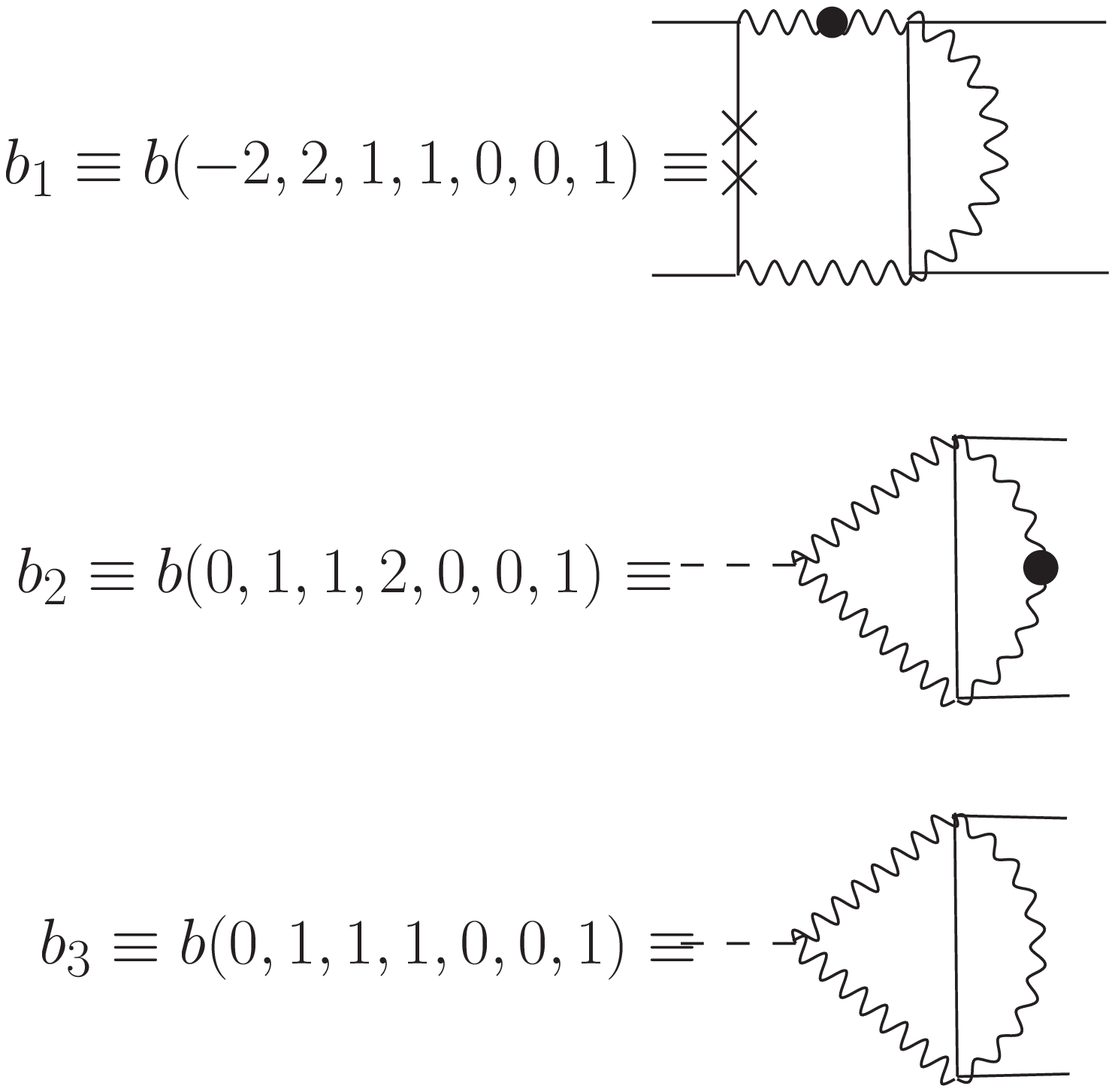}\newline
\caption{New basis of MIs for the four propagator topology, type $e$.}
\label{T4e_appendix}
\end{figure}

This topology includes three coupled Master integrals $b(-1,1,1,1,0,0,1)$, $%
b(0,1,1,1,0,-1,1)$, $b(0,1,1,1,0,0,1)$. We make a transformation to a more
suitable basis depicted in Fig. \ref{T4e_appendix}.
In this basis the system decouples and the solution may be found as in
previous cases (using the point $x=-1$ to fix the unknown integration
constants)
\begin{eqnarray*}
b_{1}^{(-2)} &=&\frac{x^{2}+1}{2x} \\
b_{2}^{(-2)} &=&0 \\
b_{3}^{(-2)} &=&\frac{1}{2} \\
b_{1}^{(-1)} &=&\left( x+\frac{1}{x}-1\right) H(0,x)+2\left( x+\frac{1}{x}%
-1\right) H(1,x)+\frac{7x}{4}+\frac{7}{4x}-1 \\
b_{2}^{(-1)} &=&-\frac{2xH(2,x)}{x^{2}-1}+\frac{xH(0,0,x)}{1-x^{2}}+\frac{%
2\pi ^{2}x}{3-3x^{2}} \\
b_{3}^{(-1)} &=&H(0,x)+2H(1,x)+\frac{5}{2} \\
b_{1}^{(0)} &=&\frac{\left( 2x^{3}-2x+4\right) H(2,x)}{x^{2}+x}+\frac{\left(
x^{3}-x+2\right) H(0,0,x)}{x^{2}+x}+\left( \frac{7x}{2}+\frac{7}{2x}%
-3\right) H(0,x) \\
&&+\left( 7x+\frac{7}{x}-6\right) H(1,x)+\left( 3x+\frac{3}{x}-4\right)
H(1,0,x)+\left( 6x+\frac{6}{x}-8\right) H(1,1,x) \\
&&+\frac{135-x\left( 3((7-45x)x+7)+4\pi ^{2}(x(4x-3)+1)\right) }{24x(x+1)} \\
b_{2}^{(0)} &=&\frac{4\pi ^{2}xH(-1,x)}{3-3x^{2}}+\frac{\pi ^{2}xH(0,x)}{%
6\left( x^{2}-1\right) }+\frac{8\pi ^{2}xH(1,x)}{3-3x^{2}}-\frac{4xH(3,x)}{%
x^{2}-1}-\frac{4xH(-1,2,x)}{x^{2}-1} \\
&&-\frac{8xH(1,2,x)}{x^{2}-1}-\frac{2xH(2,0,x)}{x^{2}-1}-\frac{4xH(2,1,x)}{%
x^{2}-1}-\frac{2xH(-1,0,0,x)}{x^{2}-1} \\
&&-\frac{2xH(0,0,0,x)}{x^{2}-1}-\frac{4xH(1,0,0,x)}{x^{2}-1}-\frac{9x\zeta
(3)}{x^{2}-1} \\
b_{3}^{(0)} &=&\frac{\left( \left( \pi ^{2}-15x\right) x+15\right) H(0,x)}{%
3-3x^{2}}-\frac{2xH(3,x)}{x^{2}-1}-\frac{2xH(2,0,x)}{x^{2}-1}-\frac{%
4xH(2,1,x)}{x^{2}-1} \\
&&+\frac{xH(0,0,0,x)}{1-x^{2}}+10H(1,x)+\left( \frac{2}{x+1}+2\right)
H(2,x)+\left( \frac{1}{x+1}+1\right) H(0,0,x) \\
&&+3H(1,0,x)+6H(1,1,x)+\frac{x\left( -4\pi ^{2}(x-1)+57x+12\zeta (3)\right)
-57}{6\left( x^{2}-1\right) } \\
b_{1}^{(1)} &=&\frac{(x(x((3-2x)x+3)-7)+4)H(0,0,0,x)}{x-x^{3}}-\frac{2\pi
^{2}(x-1)\left( x^{2}+1\right) H(-1,x)}{3x(x+1)} \\
&&+\frac{\left( x\left( 3(45x-38)x^{2}+2\pi ^{2}(x((x-3)x-1)+1)+114\right)
-135\right) H(0,x)}{12x\left( x^{2}-1\right) } \\
&&+\frac{2(2x-3)\left( x^{3}-x+1\right) H(2,0,x)}{x\left( x^{2}-1\right) }+%
\frac{4(2x-3)\left( x^{3}-x+1\right) H(2,1,x)}{x\left( x^{2}-1\right) } \\
&&+\frac{\left( \pi ^{2}(x((9-11x)x-7)+5)+3(x(x(45x+7)+7)+45)\right) H(1,x)}{%
6x(x+1)} \\
&&+\left( 7x-\frac{4(x+5)}{x+1}+\frac{14}{x}\right) H(2,x)+\left( 4x+\frac{1%
}{1-x}-\frac{9}{x+1}+\frac{8}{x}-6\right) H(3,x) \\
&&+\left( -2x-\frac{8}{x+1}+\frac{2}{x}+4\right) H(-1,2,x)+\left( \frac{7x}{2%
}-\frac{8}{x+1}+\frac{7}{x}-2\right) H(0,0,x) \\
&&+\frac{3}{2}\left( 7x+\frac{7}{x}-8\right) H(1,0,x)+3\left( 7x+\frac{7}{x}%
-8\right) H(1,1,x) \\
&&+\left( 6x-\frac{8(x+3)}{x+1}+\frac{14}{x}\right) H(1,2,x)+\left( -x-\frac{%
4}{x+1}+\frac{1}{x}+2\right) H(-1,0,0,x) \\
&&+\left( 3x-\frac{8}{x+1}+\frac{7}{x}-4\right) H(1,0,0,x)+2\left( 5x+\frac{5%
}{x}-8\right) H(1,1,0,x) \\
&&+4\left( 5x+\frac{5}{x}-8\right) H(1,1,1,x)-\frac{1}{48x\left(
x^{2}-1\right) }((x-1)(x(-837x^{2}+543(x+1) \\
&&+8\pi ^{2}(7x(2x-1)+11))-837)+48(x(x(x(13x-24)+7)+6)-4)\zeta (3))
\end{eqnarray*}

\
\begin{eqnarray*}
b_{2}^{(1)} &=&-\frac{18x\zeta (3)H(-1,x)}{x^{2}-1}+\frac{3x\zeta (3)H(0,x)}{%
x^{2}-1}-\frac{32x\zeta (3)H(1,x)}{x^{2}-1}+\frac{4\pi ^{2}xH(-2,x)}{3-3x^{2}%
}+\frac{7\pi ^{2}xH(2,x)}{3-3x^{2}} \\
&&-\frac{8xH(4,x)}{x^{2}-1}-\frac{4xH(-2,2,x)}{x^{2}-1}+\frac{8\pi
^{2}xH(-1,-1,x)}{3-3x^{2}}+\frac{\pi ^{2}xH(-1,0,x)}{3\left( x^{2}-1\right) }%
+\frac{16\pi ^{2}xH(-1,1,x)}{3-3x^{2}} \\
&&-\frac{8xH(-1,3,x)}{x^{2}-1}+\frac{16\pi ^{2}xH(1,-1,x)}{3-3x^{2}}+\frac{%
32\pi ^{2}xH(1,1,x)}{3-3x^{2}}-\frac{20xH(1,3,x)}{x^{2}-1}-\frac{12xH(2,2,x)%
}{x^{2}-1} \\
&&-\frac{6xH(3,0,x)}{x^{2}-1}-\frac{12xH(3,1,x)}{x^{2}-1}-\frac{2xH(-2,0,0,x)%
}{x^{2}-1}-\frac{8xH(-1,-1,2,x)}{x^{2}-1}-\frac{16xH(-1,1,2,x)}{x^{2}-1} \\
&&-\frac{4xH(-1,2,0,x)}{x^{2}-1}-\frac{8xH(-1,2,1,x)}{x^{2}-1}-\frac{%
16xH(1,-1,2,x)}{x^{2}-1}-\frac{32xH(1,1,2,x)}{x^{2}-1}-\frac{12xH(1,2,0,x)}{%
x^{2}-1} \\
&&-\frac{24xH(1,2,1,x)}{x^{2}-1}-\frac{6xH(2,0,0,x)}{x^{2}-1}-\frac{%
4xH(2,1,0,x)}{x^{2}-1}-\frac{8xH(2,1,1,x)}{x^{2}-1}-\frac{4xH(-1,-1,0,0,x)}{%
x^{2}-1} \\
&&-\frac{4xH(-1,0,0,0,x)}{x^{2}-1}-\frac{8xH(-1,1,0,0,x)}{x^{2}-1}-\frac{%
4xH(0,0,0,0,x)}{x^{2}-1}-\frac{8xH(1,-1,0,0,x)}{x^{2}-1} \\
&&-\frac{10xH(1,0,0,0,x)}{x^{2}-1}-\frac{16xH(1,1,0,0,x)}{x^{2}-1}-\frac{%
287\pi ^{4}x}{360\left( x^{2}-1\right) } \\
b_{3}^{(1)} &=&-\frac{2\left( \pi ^{2}(x-1)^{2}-6x\zeta (3)\right) H(-1,x)}{%
3\left( x^{2}-1\right) }+\frac{\left( x\left( \pi ^{2}(x-3)+114x+18\zeta
(3)\right) -114\right) H(0,x)}{6\left( x^{2}-1\right) } \\
&&+\frac{\left( x\left( 30(x+1)+\pi ^{2}\right) -60\right) H(2,x)}{3\left(
x^{2}-1\right) }+\frac{2\left( 2x^{2}+x-4\right) H(3,x)}{x^{2}-1}-\frac{%
6xH(4,x)}{x^{2}-1}+\frac{2\pi ^{2}xH(-1,0,x)}{3-3x^{2}} \\
&&-\frac{4xH(-1,3,x)}{x^{2}-1}+\frac{\left( x\left( \pi ^{2}-30(x+1)\right)
+60\right) H(0,0,x)}{6-6x^{2}}+\frac{\left( 4x^{2}-6\right) H(2,0,x)}{x^{2}-1%
} \\
&&+\frac{4\left( 2x^{2}-3\right) H(2,1,x)}{x^{2}-1}-\frac{12xH(2,2,x)}{%
x^{2}-1}-\frac{4xH(3,0,x)}{x^{2}-1}-\frac{8xH(3,1,x)}{x^{2}-1} \\
&&-\frac{4xH(-1,2,0,x)}{x^{2}-1}-\frac{8xH(-1,2,1,x)}{x^{2}-1}+\frac{\left(
2x^{2}+x-4\right) H(0,0,0,x)}{x^{2}-1}-\frac{6xH(2,0,0,x)}{x^{2}-1} \\
&&-\frac{12xH(2,1,0,x)}{x^{2}-1}-\frac{24xH(2,1,1,x)}{x^{2}-1}-\frac{%
2xH(-1,0,0,0,x)}{x^{2}-1} \\
&&-\frac{3xH(0,0,0,0,x)}{x^{2}-1}+\frac{\left( \pi
^{2}(5-11x)+228(x+1)\right) H(1,x)}{6(x+1)} \\
&&+\left( \frac{4}{x+1}-2\right) H(-1,2,x)+15H(1,0,x)+30H(1,1,x)+\left(
\frac{8}{x+1}+6\right) H(1,2,x) \\
&&+\left( \frac{2}{x+1}-1\right) H(-1,0,0,x)+\left( \frac{4}{x+1}+3\right)
H(1,0,0,x)+10H(1,1,0,x)+20H(1,1,1,x) \\
&&+\frac{2925\left( x^{2}-1\right) +90((11-13x)x+4)\zeta (3)-300\pi
^{2}(x-1)x+16\pi ^{4}x}{90\left( x^{2}-1\right) }.
\end{eqnarray*}%
The last two MIs $b_{2}$ and $b_{3}$ can be found in \cite{Bonciani:2003hc}
in the same basis while for the first one we use a different basis. The
former two were checked to $\mathcal{O}(\epsilon ^{1})$ and the latter one
was left uncontrolled.

\subsection{Five propagator topology, type \it{a}}

\begin{figure}[h]
\centering\includegraphics[scale=0.5]{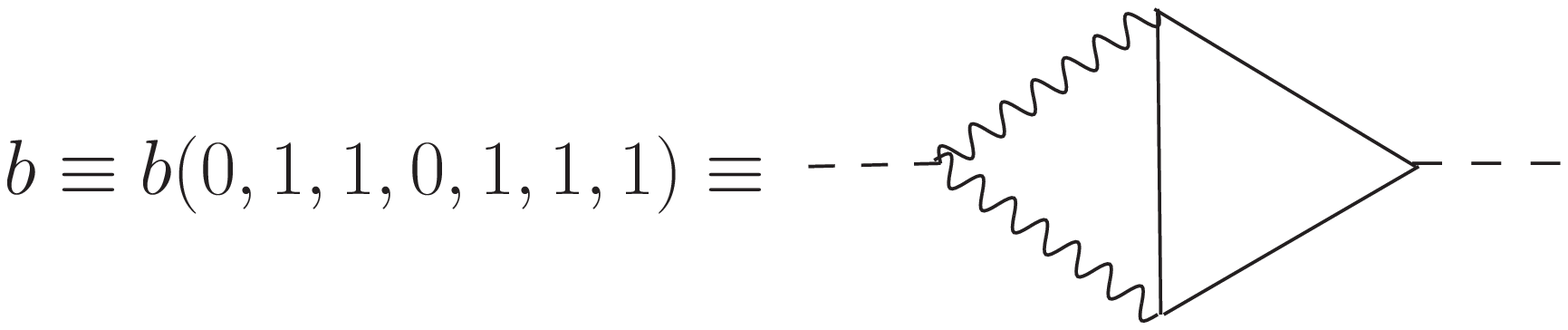}\newline
\caption{The five propagator MI, type $a$.}
\label{T5a_appendix}
\end{figure}

The current topology is formed by the MI depicted in Fig. \ref{T5a_appendix}%
. It corresponds to {\bf{F10101}} of \cite{Davydychev:2000na,Davydychev:2003mv} where the $\varepsilon$ expansion was given.
This MI has the following expansion using $x=1$ for fixing the integration
constants
\begin{eqnarray*}
b^{(0)} &=&-\frac{2xH(3,x)}{(x-1)^{2}}-\frac{2xH(2,0,x)}{(x-1)^{2}}+\frac{%
4xH(1,0,0,x)}{(x-1)^{2}}-\frac{6x\zeta (3)}{(x-1)^{2}} \\
b^{(1)} &=&-\frac{12x\zeta (3)H(0,x)}{(x-1)^{2}}-\frac{24x\zeta (3)H(1,x)}{%
(x-1)^{2}}+\frac{\pi ^{2}xH(2,x)}{3(x-1)^{2}}-\frac{4xH(3,x)}{(x-1)^{2}}-%
\frac{8xH(4,x)}{(x-1)^{2}} \\
&&-\frac{10xH(-3,0,x)}{(x-1)^{2}}+\frac{4xH(-2,2,x)}{(x-1)^{2}}-\frac{2\pi
^{2}xH(1,0,x)}{3(x-1)^{2}}-\frac{4xH(1,3,x)}{(x-1)^{2}} \\
&&-\frac{4xH(2,0,x)}{(x-1)^{2}}-\frac{4xH(2,2,x)}{(x-1)^{2}}-\frac{4xH(3,0,x)%
}{(x-1)^{2}}-\frac{4xH(3,1,x)}{(x-1)^{2}}+\frac{4xH(-2,0,0,x)}{(x-1)^{2}} \\
&&+\frac{4xH(-2,1,0,x)}{(x-1)^{2}}-\frac{24xH(1,-2,0,x)}{(x-1)^{2}}+\frac{%
8xH(1,0,0,x)}{(x-1)^{2}}+\frac{4xH(1,2,0,x)}{(x-1)^{2}} \\
&&+\frac{4xH(2,-1,0,x)}{(x-1)^{2}}-\frac{6xH(2,0,0,x)}{(x-1)^{2}}-\frac{%
4xH(2,1,0,x)}{(x-1)^{2}}+\frac{12xH(1,0,0,0,x)}{(x-1)^{2}} \\
&&-\frac{x\left( 120\zeta (3)+\pi ^{4}\right) }{10(x-1)^{2}}.
\end{eqnarray*}
Verifying the result with \cite{Bonciani:2003hc} to $\mathcal{O}%
(\epsilon^{1})$ gave full agreement.

\subsection{Five propagator topology, type \it{b}}

We have chosen the two Master integrals as depicted in Fig. \ref%
{T5b_appendix} (the AIR original basis was $b(0,1,1,1,0,1,1)$ and $%
b(0,1,1,1,-1,1,1)$).
\begin{figure}[h]
\centering\includegraphics[scale=0.5]{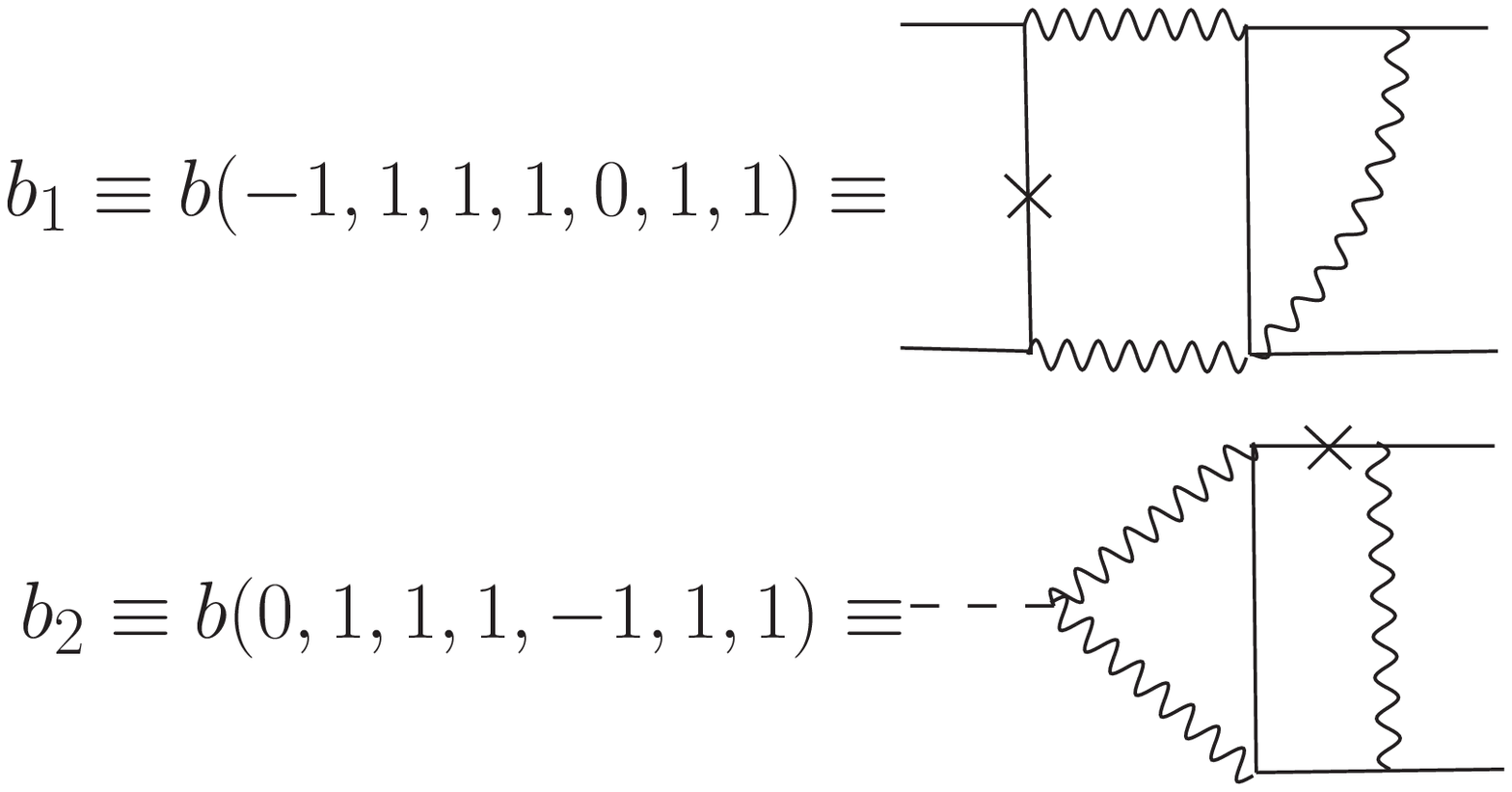}\newline
\caption{Suitable basis of MIs for the five propagator topology, type $b$.}
\label{T5b_appendix}
\end{figure}
This choice has been done in order to decouple the system of corresponding
differential equations. The appropriate point for fixing the integration
constants was $x=-1$. Then it has the solution%
\begin{eqnarray*}
b_{1}^{(-2)} &=&\frac{1}{2} \\
b_{2}^{(-2)} &=&\frac{1}{2} \\
b_{1}^{(-1)} &=&\frac{5}{2} \\
b_{2}^{(-1)} &=&H(0,x)+2H(1,x)+\frac{5}{2} \\
b_{1}^{(0)} &=&\frac{3x^{2}H(0,0,0,x)}{x^{2}-1}+\frac{\left( \pi ^{2}\left(
3x^{2}-1\right) +12x(x+1)\right) H(0,x)}{6\left( x^{2}-1\right) }-\frac{%
2H(3,x)}{x^{2}-1} \\
&&+\frac{\left( x\left( -2x^{2}+x-2\right) -1\right) H(0,0,x)}{(x-1)^{2}(x+1)%
}+\frac{\left( x^{2}+1\right) H(2,0,x)}{1-x^{2}} \\
&&-\frac{2\left( x^{2}+1\right) H(2,1,x)}{x^{2}-1}+2H(1,x)+\frac{2H(2,x)}{x+1%
}+H(1,0,x)+2H(1,1,x) \\
&&+2H(1,0,0,x)+\frac{x^{2}(57-12\zeta (3))+\pi ^{2}((4-3x)x-1)+24\zeta (3)-57%
}{6\left( x^{2}-1\right) } \\
b_{2}^{(0)} &=&\frac{\left( 3(x-1)(4(x-1)\zeta (3)+5x+5)+\pi
^{2}(x(3x-1)-1)\right) H(0,x)}{3\left( x^{2}-1\right) } \\
&&+\frac{2(x-2)H(3,x)}{x^{2}-1}-\frac{2((x-1)x+1)H(2,0,x)}{x^{2}-1}-\frac{%
4((x-1)x+1)H(2,1,x)}{x^{2}-1} \\
&&+\frac{3x(2x-1)H(0,0,0,x)}{x^{2}-1}+\left( \frac{2(x-1)\zeta (3)}{x+1}%
+10\right) H(1,x) \\
&&+\left( \frac{10}{x+1}-2\right) H(2,x)+\left( \frac{4}{x+1}-2\right) H(4,x)
\\
&&+\frac{\left( -\left( 24+\pi ^{2}\right) x+\pi ^{2}+6\right) H(0,0,x)}{%
6(x+1)}+\frac{\left( \pi ^{2}(x-1)+9(x+1)\right) H(1,0,x)}{3(x+1)} \\
&&+6H(1,1,x)+\left( \frac{4}{x+1}-2\right) H(1,3,x)+\left( \frac{2}{x+1}%
-1\right) H(3,0,x) \\
&&+\left( \frac{4}{x+1}-2\right) H(3,1,x)+4H(1,0,0,x) \\
&&+\left( \frac{4}{x+1}-2\right) H(1,2,0,x)+\left( \frac{8}{x+1}-4\right)
H(1,2,1,x) \\
&&+\left( \frac{4}{x+1}-2\right) H(2,0,0,x)+\left( 3-\frac{6}{x+1}\right)
H(1,0,0,0,x) \\
&&+\frac{-72\left( 2x^{2}+x-4\right) \zeta (3)+342\left( x^{2}-1\right) +\pi
^{4}(x-1)^{2}-12\pi ^{2}(7x-3)(x-1)}{36\left( x^{2}-1\right) }
\end{eqnarray*}

\
\begin{eqnarray*}
b_{1}^{(1)} &=&\frac{H(0,0,0,0,x)x^{2}}{x^{2}-1}+\frac{2}{3}\pi ^{2}H(-2,x)+%
\frac{\left( 12(x-1)(x+6)+\pi ^{2}\left( 9x^{2}-7\right) \right) H(2,x)}{%
6\left( x^{2}-1\right) } \\
&&+\frac{2(x(x+3)-5)H(3,x)}{x^{2}-1}+\frac{2\left( x^{2}-4\right) H(4,x)}{%
x^{2}-1}-\frac{6\left( 3x^{2}+1\right) H(-3,0,x)}{x^{2}-1} \\
&&+\frac{12((x-1)x+1)H(-2,0,x)}{(x-1)^{2}}+2H(-2,2,x) \\
&&+\frac{\left( \pi ^{2}\left( x^{2}+1\right) -18(x+1)^{2}\right) H(-1,0,x)}{%
3\left( x^{2}-1\right) } \\
&&+\left( \frac{4}{x+1}-2\right) H(-1,2,x)-\frac{2\left( x^{2}+1\right)
H(-1,3,x)}{x^{2}-1} \\
&&-\frac{\left( \pi ^{2}(x-1)\left( 10x^{2}+3\right)
+6(x(x(2x-9)+14)+5)\right) H(0,0,x)}{6(x-1)^{2}(x+1)} \\
&&+\frac{\left( 3(x+1)(9x-5)-\pi ^{2}\left( x^{2}+3\right) \right) H(1,0,x)}{%
3\left( x^{2}-1\right) }+14H(1,1,x) \\
&&+\left( 2+\frac{8}{x+1}\right) H(1,2,x)+2H(1,3,x)+\frac{%
(2-2x(2(x-1)x+5))H(2,0,x)}{(x-1)^{2}(x+1)} \\
&&+\frac{4(2x-3)H(2,1,x)}{x^{2}-1}-\frac{2\left( x^{2}+5\right) H(2,2,x)}{%
x^{2}-1}+\frac{\left( 7x^{2}-3\right) H(3,0,x)}{x^{2}-1} \\
&&+\frac{2\left( x^{2}-5\right) H(3,1,x)}{x^{2}-1}+H(-2,0,0,x)+\left( \frac{2%
}{x+1}-1\right) H(-1,0,0,x) \\
&&-\frac{2\left( x^{2}+1\right) H(-1,2,0,x)}{x^{2}-1}-\frac{4\left(
x^{2}+1\right) H(-1,2,1,x)}{x^{2}-1} \\
&&+\frac{\left( x^{3}-10x-3\right) H(0,0,0,x)}{(x-1)^{2}(x+1)}-12H(1,-2,0,x)
\\
&&+\left( \frac{4}{x+1}+9+\frac{4}{x-1}+\frac{4}{(x-1)^{2}}\right)
H(1,0,0,x)+6H(1,1,0,x) \\
&&+12H(1,1,1,x)+4H(1,2,0,x)+\frac{\left( 5x^{2}+9\right) H(2,0,0,x)}{1-x^{2}}
\\
&&-\frac{6\left( x^{2}+1\right) H(2,1,0,x)}{x^{2}-1}-\frac{12\left(
x^{2}+1\right) H(2,1,1,x)}{x^{2}-1}+\frac{3\left( x^{2}+1\right)
H(-1,0,0,0,x)}{x^{2}-1} \\
&&+\frac{\left( 5x^{2}-13\right) H(1,0,0,0,x)}{x^{2}-1}+\frac{H(1,x)\left(
\pi ^{2}(7-9x)+12(x+1)(7-6\zeta (3))\right) }{6(x+1)} \\
&&+\frac{H(-1,x)\left( 6\left( x^{2}+1\right) \zeta (3)-2\pi
^{2}(x-1)^{2}\right) }{3\left( x^{2}-1\right) } \\
&&+\frac{H(0,x)\left( 84x\left( x^{2}-1\right) +\pi
^{2}(x+1)(x(10x-13)+5)-6(x-1)\left( 4x^{2}-3\right) \zeta (3)\right) }{%
6(x-1)^{2}(x+1)} \\
&&-\frac{1}{90(x-1)^{2}(x+1)}\left( -2925(x+1)(x-1)^{2}+\pi ^{4}\left(
27x^{2}-17\right) (x-1)\right. \\
&&+\left. 15\pi ^{2}(x(17x-18)+5)(x-1)+90(x(x(9x-22)-3)+4)\zeta (3)\right)
\end{eqnarray*}

\begin{eqnarray*}
b_{2}^{(1)} &=&-\frac{2\pi ^{2}(x-1)H(-3,x)}{3(x+1)}-\frac{\left( 7\pi
^{2}(x-1)^{2}+12(x-2)(4x-5)\right) H(3,x)}{6\left( x^{2}-1\right) } \\
&&+\left( \frac{1}{x+1}+8-\frac{3}{x-1}\right) H(4,x)+\left( \frac{16}{x+1}%
-8\right) H(5,x) \\
&&+\left( \frac{12}{x+1}-6\right) H(-4,0,x)-\frac{12(x(3x-2)+1)H(-3,0,x)}{%
x^{2}-1} \\
&&+\left( \frac{4}{x+1}-2\right) H(-3,2,x)+\frac{\left( \pi
^{2}(x-1)+54(x+1)\right) H(-2,0,x)}{3(x+1)} \\
&&+4H(-2,2,x)+\left( \frac{4}{x+1}-2\right) H(-2,3,x)+\left( \frac{20}{x+1}%
-10\right) H(-1,2,x) \\
&&-\frac{4((x-1)x+1)H(-1,3,x)}{x^{2}-1}+\left( \frac{8}{x+1}-4\right)
H(-1,4,x) \\
&&+\frac{\left( \pi ^{2}(x-1)-30(x-3)\right) H(1,2,x)}{3(x+1)}+\frac{%
8xH(1,3,x)}{x+1} \\
&&+\left( \frac{20}{x+1}-10\right) H(1,4,x)+\left( \frac{17+2\pi ^{2}}{x+1}%
-\pi ^{2}-10+\frac{1}{1-x}\right) H(2,0,x) \\
&&-\frac{4(2(x-4)x+7)H(2,1,x)}{x^{2}-1}-\frac{4((x-3)x+5)H(2,2,x)}{x^{2}-1}
\\
&&+\left( \frac{4}{x+1}-2\right) H(2,3,x)+\frac{(8x(2x-1)-4)H(3,0,x)}{x^{2}-1%
} \\
&&+\frac{8\left( x^{2}-2\right) H(3,1,x)}{x^{2}-1}+\left( \frac{20}{x+1}%
-10\right) H(3,2,x)+\left( \frac{6}{x+1}-3\right) H(4,0,x) \\
&&+\left( \frac{20}{x+1}-10\right) H(4,1,x)+\left( \frac{2}{x+1}-1\right)
H(-3,0,0,x)+2H(-2,0,0,x) \\
&&+\left( \frac{4}{x+1}-2\right) H(-2,2,0,x)+\left( \frac{8}{x+1}-4\right)
H(-2,2,1,x) \\
&&-\frac{\left( 15+\pi ^{2}\right) (x-1)H(-1,0,0,x)}{3(x+1)}+\frac{2\pi
^{2}(x-1)H(-1,1,0,x)}{3(x+1)} \\
&&+\left( \frac{8}{x+1}-4\right) H(-1,1,3,x)-\frac{4((x-1)x+1)H(-1,2,0,x)}{%
x^{2}-1} \\
&&-\frac{8((x-1)x+1)H(-1,2,1,x)}{x^{2}-1}+\left( \frac{4}{x+1}-2\right)
H(-1,3,0,x) \\
&&+\left( \frac{8}{x+1}-4\right) H(-1,3,1,x)-\frac{\left( \pi
^{2}(x-1)^{2}+26x^{2}-50x+2\right) H(0,0,0,x)}{2\left( x^{2}-1\right) } \\
&&+\left( \frac{48}{x+1}-24\right) H(1,-3,0,x)-24H(1,-2,0,x)+\frac{2\pi
^{2}(x-1)H(1,-1,0,x)}{3(x+1)} \\
&&+\left( \frac{8}{x+1}-4\right) H(1,-1,3,x)+\left( -\frac{5\pi ^{2}}{2}+1+%
\frac{5\left( 4+\pi ^{2}\right) }{x+1}\right) H(1,0,0,x) \\
&&+\left( 10-\frac{2\pi ^{2}(x-1)}{3(x+1)}\right)
H(1,1,0,x)+20H(1,1,1,x)+\left( \frac{8}{x+1}-4\right) H(1,1,3,x) \\
&&+\left( 12-\frac{8}{x+1}\right) H(1,2,0,x)+\frac{8(x-1)H(1,2,1,x)}{x+1}%
+\left( \frac{24}{x+1}-12\right) H(1,2,2,x) \\
&&+\left( 2-\frac{4}{x+1}\right) H(1,3,0,x)+\left( \frac{24}{x+1}-12\right)
H(1,3,1,x)+\frac{12(x-1)H(2,-2,0,x)}{x+1} \\
&&+\frac{(-6(x-1)x-14)H(2,0,0,x)}{x^{2}-1}-\frac{12((x-1)x+1)H(2,1,0,x)}{%
x^{2}-1} \\
&&-\frac{24((x-1)x+1)H(2,1,1,x)}{x^{2}-1}+\left( \frac{8}{x+1}-4\right)
H(2,2,0,x)+\left( \frac{18}{x+1}-9\right) H(3,0,0,x) \\
&&+\left( \frac{12}{x+1}-6\right) H(3,1,0,x)+\left( \frac{24}{x+1}-12\right)
H(3,1,1,x)+\left( 3-\frac{6}{x+1}\right) H(-2,0,0,0,x) \\
&&+\frac{6((x-1)x+1)H(-1,0,0,0,x)}{x^{2}-1}+\left( \frac{8}{x+1}-4\right)
H(-1,1,2,0,x) \\
&&+\left( \frac{16}{x+1}-8\right) H(-1,1,2,1,x)+\left( \frac{8}{x+1}%
-4\right) H(-1,2,0,0,x)+\frac{x(2x-1)H(0,0,0,0,x)}{x^{2}-1} \\
&&+\left( \frac{8}{x+1}-4\right) H(1,-1,2,0,x)+\left( \frac{16}{x+1}%
-8\right) H(1,-1,2,1,x) \\
&&+\frac{4(x(x+5)-8)H(1,0,0,0,x)}{x^{2}-1}+\left( \frac{8}{x+1}-4\right)
H(1,1,2,0,x)+\left( \frac{16}{x+1}-8\right) H(1,1,2,1,x) \\
&&+\left( \frac{36}{x+1}-18\right) H(1,2,0,0,x)+\left( \frac{24}{x+1}%
-12\right) H(1,2,1,0,x)+\left( \frac{48}{x+1}-24\right) H(1,2,1,1,x) \\
&&+\left( \frac{26}{x+1}-13\right) H(2,0,0,0,x)+\frac{6(x-1)H(-1,1,0,0,0,x)}{%
x+1}+\frac{6(x-1)H(1,-1,0,0,0,x)}{x+1} \\
&&+\frac{(x-1)H(1,0,0,0,0,x)}{x+1}+\left( \frac{4}{x+1}-2\right)
H(1,1,0,0,0,x) \\
&&+\frac{2H(-1,0,x)\left( 12\zeta (3)(x-1)^{2}+\pi ^{2}((x-1)x+1)\right) }{%
3\left( x^{2}-1\right) } \\
&&+\frac{H(1,0,x)\left( 21\zeta (3)(x-1)^{2}-4\pi ^{2}((x-2)x+2)+45\left(
x^{2}-1\right) \right) }{3\left( x^{2}-1\right) } \\
&&+H(-2,x)\left( \frac{2\zeta (3)(x-1)}{x+1}+\frac{4\pi ^{2}}{3}\right)
+H(1,1,x)\left( \frac{4\zeta (3)(x-1)}{x+1}+30\right) \\
&&+\frac{H(1,x)\left( 3\pi ^{2}(37-43x)-\pi ^{4}(x-1)+684(x+1)-72(7x+5)\zeta
(3)\right) }{18(x+1)} \\
&&+\frac{H(-1,x)\left( \pi ^{2}\left( -60+\pi ^{2}\right)
(x-1)^{2}+72((x-1)x+1)\zeta (3)\right) }{18\left( x^{2}-1\right) } \\
&&+\frac{H(0,x)\left( 17\pi ^{4}(x-1)^{2}+1710\left( x^{2}-1\right) +15\pi
^{2}(x(2x+13)-5)-90(x(16x-17)+2)\zeta (3)\right) }{90\left( x^{2}-1\right) }
\\
&&+\frac{H(0,0,x)\left( \pi ^{2}(9(1-2x)x-4)+6(x-1)(3\zeta
(3)(x-1)-8x+5)\right) }{6\left( x^{2}-1\right) } \\
&&+\frac{H(2,x)\left( \pi ^{2}(x(9x-1)-7)+6(x-1)(6\zeta
(3)(x-1)+x+14)\right) }{3\left( x^{2}-1\right) } \\
&&+\frac{1}{90\left( x^{2}-1\right) }\left( \pi ^{4}((20-59x)x+29)+15\pi
^{2}(x-1)(9\zeta (3)(x-1)-46x+6)\right. \\
&&+\left. 45\left( -6\zeta (5)(x-1)^{2}+65x^{2}+\left(
-44x^{2}+78x-30\right) \zeta (3)-65\right) \right) \\
&&+\frac{4(x-1)H(-1,1,x)\zeta (3)}{x+1}+\frac{4(x-1)H(1,-1,x)\zeta (3)}{x+1}.
\end{eqnarray*}%
This couple of MIs can be found in \cite{Czakon:2004tg}. We performed a
transformation to the original basis and made a check only for $%
b(0,1,1,1,0,1,1)$ which exactly appears also in \cite{Czakon:2004tg}. The
result is in correspondence up to $\mathcal{O}(\epsilon^{0})$. However, this
verifies both of our MIs since the transformation mixes them to $%
b(0,1,1,1,0,1,1)$. The above paper does not offer higher orders so we add
them to the list of existing MIs.

\bibliographystyle{utphys}
\bibliography{clanek}

\end{document}